\documentclass[useAMS,usenatbib]{mn2e}
\usepackage{natbib}
\usepackage{amsmath}
\usepackage{graphicx}
\usepackage{multirow}
\usepackage{url}
\numberwithin{equation}{section}
\title[ISM chemistry \& cooling - I. Optically thin regime]{Non-equilibrium chemistry and cooling in the diffuse interstellar medium - I. Optically thin regime}
\author[A. J. Richings, J. Schaye and B. D. Oppenheimer]{A. J. Richings$^{1}$, J. Schaye$^{1}$ and B. D. Oppenheimer$^{2}$\\
$^{1}$Leiden Observatory, Leiden University, PO Box 9513, 2300 RA Leiden, the Netherlands\\
$^{2}$Center for Astrophysics and Space Astronomy, Department of Astrophysical and Planetary Sciences, University of Colorado, \\
389 UCB, Boulder, CO 80309, USA}

\begin{document}

\date{Accepted 2014 March 13. Received 2014 March 12; in original form 2014 January 19}

\pagerange{\pageref{firstpage}--\pageref{lastpage}} \pubyear{2014}

\maketitle

\label{firstpage}

\begin{abstract}

An accurate treatment of the multiphase interstellar medium (ISM) in hydrodynamic galaxy simulations requires that we follow not only the thermal evolution of the gas, but also the evolution of its chemical state, including its molecular chemistry, without assuming chemical (including ionisation) equilibrium. We present a reaction network that can be used to solve for this thermo-chemical evolution. Our model follows the evolution of all ionisation states of the 11 elements that dominate the cooling rate, along with important molecules such as H$_{2}$ and CO, and the intermediate molecular species that are involved in their formation (20 molecules in total). We include chemical reactions on dust grains, thermal processes involving dust, cosmic ray ionisation and heating and photochemical reactions. We focus on conditions typical for the diffuse ISM, with densities of $10^{-2} \, \text{cm}^{-3} \la n_{\rm{H}} \la 10^{4} \, \text{cm}^{-3}$ and temperatures of $10^{2} \, \text{K} \la T \la 10^{4} \, \text{K}$, and we consider a range of radiation fields, including no UV radiation. In this paper we consider only gas that is optically thin, while paper II considers gas that becomes shielded from the radiation field. We verify the accuracy of our model by comparing chemical abundances and cooling functions in chemical equilibrium with the photoionisation code \textsc{Cloudy}. We identify the major coolants in diffuse interstellar gas to be C\textsc{ii}, Si\textsc{ii} and Fe\textsc{ii}, along with O\textsc{i} and H$_{2}$ at densities $n_{\rm{H}} \ga 10^{2} \, \text{cm}^{-3}$. Finally, we investigate the impact of non-equilibrium chemistry on the cooling functions of isochorically or isobarically cooling gas. We find that, at $T < 10^{4}$ K, recombination lags increase the electron abundance above its equilibrium value at a given temperature, which can enhance the cooling rate by up to two orders of magnitude. The cooling gas also shows lower H$_{2}$ abundances than in equilibrium, by up to an order of magnitude. 

\end{abstract}


\section{Introduction}

Understanding the chemical evolution of the interstellar medium (ISM) is imperative for the accurate modelling of the radiative cooling rate of interstellar gas, and hence for the dynamics of the ISM. However, a full treatment of the ISM chemistry can involve hundreds of chemical species and thousands of reactions (e.g. the \textsc{umist} database for astrochemistry\footnote{\url{http://www.udfa.net/}}; \citealt{leteuff00}) and the computational cost of such a complex treatment renders it impractical for use within large-scale 3D hydrodynamical simulations of galaxy formation. This problem is further compounded by the fact that chemical rate equations are typically stiff and so must be integrated implicitly, with a computational cost that scales with the cube of the number of species involved. 

To avoid these issues, existing cosmological hydrodynamical simulations typically make use of simplifying assumptions to model the gas cooling rate. For example, the cosmological simulations run as part of the OverWhelmingly Large Simulations project (OWLS) \citep{schaye10} use pre-computed cooling functions tabulated by temperature, density and abundances of individual elements, which were calculated using \textsc{Cloudy} \citep{ferland98,ferland13} assuming ionisation equilibrium in the presence of the UV background of \citet{haardt01} \citep[see][]{wiersma09}. However, such an approach has a number of disadvantages. Firstly, it assumes that the gas is in ionisation equilibrium, which may not be valid when the cooling or dynamical time-scale of the gas is shorter than its chemical time-scale \citep[e.g.][]{kafatos73,gnat07,oppenheimer13a,vasiliev13} or in the presence of a time variable radiation field \citep[e.g.][]{oppenheimer13b}. Secondly, one needs to rely on simple assumptions about the UV radiation field that is present. For example, many authors tabulate the cooling rate assuming that the gas is in collisional ionisation equilibrium i.e. they neglect the UV radiation altogether (e.g. \citealt{cox69}; \citealt{sutherland93}; \citealt{smith08}). Other studies such as \citet{wiersma09} assume the presence of a uniform UV background, thus neglecting local sources as well as shielding effects. \citet{derijcke13} account for the self shielding of gas by exponentially suppressing the hydrogen ionising part of the UV background spectrum at high H\textsc{i} densities above a threshold $n_{\rm{HI}} = 0.007 \, \text{cm}^{-3}$, while \citet{vogelsberger13} use the fitting function of \citet{rahmati13} for the hydrogen ionisation rate as a function of density. Such simplifying assumptions are used because correctly accounting for the UV radiation field would require additional dimensions in the pre-computed tables of cooling rates, and reproducing the spectral shape of the radiation potentially requires tabulating in many frequency bins, which would greatly increase the size of the tables (although a recent study by \citealt{gnedin12} suggests that this can be achieved with just 3 or 4 frequency bins).

An alternative approach is to solve the ISM chemistry within hydrodynamical simulations using a greatly simplified reaction network that captures the most important features of the chemical evolution while avoiding the complexity of more complete networks. Such a model was used by \citet{glover07} to investigate low-metallicity gas, and \citet{glover10} used a similar model including the CO chemistry to study the formation of molecular clouds in the turbulent ISM. \citet{grassi11} have developed a model of the ISM that they implemented in numerical simulations by first evaluating their model on a large grid of input parameters and then using this to train an artificial neural network to behave like their gas model, and \citet{grassi13} present a package for embedding chemistry in hydrodynamical simulations. There are also methods to optimise a chemical network by selecting only the species and/or reactions from the network that are relevant to the particular physical conditions that one is interested in \citep[e.g.][]{tupper02,wiebe03,grassi12}.

We present a new chemical and thermal model of the ISM that covers a wide range of physical conditions, and is accurate for temperatures $10^{2}$ K $\leq T \leq 10^{9}$ K and densities $n_{\rm{H}} \leq 10^{4} \, \text{cm}^{-3}$. At higher densities and lower temperatures additional molecular species not included in our model may become important. We intend to apply this model to hydrodynamic galaxy simulations in which it is important that we are able to resolve the Jeans mass of the gas. Colder gas has a smaller Jeans mass ($M_{J} \propto T^{3/2}$), making it difficult to resolve gas at very low temperatures. We therefore choose not to consider gas below $10^{2}$ K, which approximately corresponds to the smallest Jeans mass that can be resolved in current simulations of galaxy formation.

We are primarily interested in the regime of the diffuse ISM, with typical densities of $10^{-2} \, \text{cm}^{-3} \la n_{\rm{H}} \la 10^{4} \, \text{cm}^{-3}$ and temperatures of $10^{2} \, \text{K} \la T \la 10^{4} \, \text{K}$. Our model enables us to calculate the radiative cooling rate in the diffuse ISM without assuming ionisation or chemical equilibrium. Additionally, we will also be able to study the formation of some of the simplest molecules in the ISM, particularly H$_{2}$ and CO. Molecular hydrogen is known to be an important coolant below $10^{4}$ K in primordial (e.g. \citealt{saslaw67}; \citealt{peebles68}; \citealt{lepp84}; \citealt{puy93}) and low-metallicity gas (e.g. \citealt{omukai05}; \citealt{santoro06}; \citealt{jappsen07}). 

By following the evolution of the CO abundance in our model, we will be able to simulate mock observations of CO emission, which will help us to compare our simulations to observations. This will also allow us to investigate how the $X_{CO}$ factor, which relates the observable CO emission to the abundance of molecular hydrogen, evolves in our simulations. Additionally, cooling from heavy molecular species such as CO, H$_{2}$O and OH, which are all included in our network, can also contribute to the cooling rate of diffuse interstellar gas (e.g. \citealt{neufeld93}; \citealt{neufeld95}).

Beyond this regime of the diffuse ISM, we are also interested in gas that is at higher temperatures, above $10^{4}$ K, and/or at lower densities, below $10^{-2} \, \text{cm}^{-3}$. For such gas we include the species and reactions from the chemical model of \citet{oppenheimer13a} for metal-enriched gas in the intergalactic medium (IGM), which includes higher ionisation states. This enables us to self-consistently follow the full chemical evolution of gas as it cools from the IGM and the circumgalactic medium (CGM) onto the ISM, and also to accurately model interstellar gas that is irradiated by a strong UV radiation field from local sources. 

In this paper we shall focus on photoionised gas in the optically thin regime. The chemistry and cooling properties of shielded gas, in which the incident radiation field becomes attenuated by dust and by the gas itself, will be important for the cold phase of the ISM, in particular for molecular clouds. We will present results for such shielded conditions, along with the methods that we use to calculate the attenuation of the photochemical reactions by dust and gas, in a companion paper to this work (hereafter paper II).

Throughout this paper we use the default solar abundances used by \textsc{Cloudy}, version 13.01 (see for example table 1 in \citealt{wiersma09}). In particular, we take the solar metallicity to be $Z_{\odot} = 0.0129$. This paper is organised as follows. In sections~\ref{ChemicalModel} and \ref{ThermalModel} we describe the details of our chemical and thermal models respectively (readers who are not interested in the details of our methods may skip these two sections). In section~\ref{ResultsI} we compare the abundances and cooling functions predicted by our model in chemical equilibrium with \textsc{Cloudy} to confirm its validity in the range of physical conditions that we are interested in, and we highlight the most important processes and the dominant coolants in this regime. In section~\ref{NonEqSect} we investigate what impact non-equilibrium chemistry can have on the cooling of interstellar gas, and we discuss our results and conclusions in section~\ref{conclusions}.

\section{The chemical model}\label{ChemicalModel}

To reproduce the gas cooling rate in the diffuse ISM, we will need to ensure that we include all of the most important coolants for the range of physical conditions that we are interested in. In non-primordial gas the cooling rate below $10^{4}$ K is typically dominated by fine structure line emission from metals (e.g. \citealt{maio07}). \citet{glover07} considered the contribution of various neutral and singly ionised metal species to the cooling rate at temperatures $50 K < T < 10^{4}$ K and densities $10^{-3} \, \text{cm}^{-3} < n < 10^{2}$ cm$^{-3}$ at a metallicity of 0.1 $Z_{\odot}$. They identified the most important coolants to be C\textsc{i}, C\textsc{ii}, O\textsc{i} and Si\textsc{ii}, contributing at least $25\%$ to the cooling rate somewhere within these ranges, in addition to inverse Compton cooling (at high redshift) and Ly$\alpha$ cooling. They also included cooling from molecular hydrogen, which dominates below $10^{4}$ K in primordial gas and remains important in low-metallicity gas. We therefore take the reaction network used by \citet{glover07} as the basis for our chemical model, supplemented with additional reactions for the primordial chemistry, and more recent rate coefficients for some existing reactions, taken from \citet{glover08} and others. \citet{glover07} follow 18 chemical species including the coolants described above and those species that have a significant impact on their abundances. They also include cooling from the HD molecule, which can become important in primordial and very low metallicity gas at low temperatures $T < 200$ K and densities $n_{\rm{H}} > 10^{4} \, \text{cm}^{-3}$ \citep{flower00,omukai05,glover06}. However, as we are interested in the diffuse ISM rather than dense molecular clouds, we do not explore temperatures below $10^{2}$ K, nor do we consider densities above $10^{4} \, \text{cm}^{-3}$. We therefore decide to neglect HD cooling, which allows us to remove the deuterium chemistry from the reaction network. 

In addition to tracking the coolants described above, we are also interested in following the abundance of CO in our model. Modelling the CO chemistry can become very complicated as a large number of intermediate molecular species are involved in its formation and destruction. \citet{glover12} compare a number of simplified models that aim to approximate the CO chemistry in giant molecular clouds. We include the molecular CO network of \citet{glover10} (the most complex model considered by \citealt{glover12}, although still a significant simplification compared to the full \textsc{umist} database), with some modifications as described in section~\ref{COmodel}. We demonstrate in section~\ref{ISMmolecules} that this model produces equilibrium CO fractions that are in good agreement with those calculated using version 13.01 of \textsc{Cloudy} \citep{ferland13} in the physical regimes that we are interested in here, i.e. those relevant to the diffuse ISM. We also include cooling from the molecular species CO, H$_{2}$O and OH in our thermal model.

Finally, we extend our chemical network to include the higher ionisation states that are relevant in the circumgalactic medium, the warm and hot ISM, and in regions with a strong interstellar radiation field. We combine the chemical reactions involving molecules and low ionisation states described above with the model of \citet{oppenheimer13a}, which includes all ionisation states of the 11 elements\footnote{H, He, C, N, O, Ne, Mg, Si, S, Ca \& Fe} tabulated by \citet{wiersma09} and used in the cosmological simulations of the OWLS project \citep{schaye10}. \citet{oppenheimer13a} have tabulated the temperature dependence of the rate coefficients for collisional ionisation, radiative and di-electronic recombination and charge transfer reactions, based on the rates used by \textsc{Cloudy}, which we make use of in our model, although we recalculate the photoionisation rates (including Auger ionisation) so that we can apply this model to any general UV radiation field, rather than just the extragalactic UV background models used by \citet{oppenheimer13a} (see section~\ref{photoion}). Also, while \citet{oppenheimer13a} only include the charge transfer ionisation and recombination of metals by hydrogen and helium, we supplemented these with a small number of charge transfer reactions between metal species, which we take from the \textsc{umist} database.

We now have a chemical model that follows 157 species. These are the molecules H$_{2}$, H$_{2}^{+}$, H$_{3}^{+}$, OH, H${_2}$O, C$_{2}$, O$_{2}$, HCO$^{+}$, CH, CH$_{2}$, CH$_{3}^{+}$, CO, CH$^{+}$, CH$_{2}^{+}$, OH$^{+}$, H$_{2}$O$^{+}$, H$_{3}$O$^{+}$, CO$^{+}$, HOC$^{+}$ and O$_{2}^{+}$, along with electrons and all ionisations states of H, He, C, N, O, Ne, Si, Mg, S, Ca and Fe (including H$^{-}$, C$^{-}$ and O$^{-}$). The 907 reactions included in our model are summarised in table~\ref{reaction_table1} in appendix~\ref{reactions_summary}. We describe some of these reactions, and the rates that we use for them, in more detail in sections~\ref{photoion} to \ref{COmodel} below. 

To confirm the validity of our chemical model, we compare it to the photoionisation code \textsc{Cloudy}, version 13.01\footnote{\url{http://nublado.org/}} \citep{ferland13}, which simulates the ionisation and molecular state of gas in chemical (including ionisation) equilibrium, along with the thermal state and the level populations of each species, and hence predicts the spectrum of the gas. \textsc{Cloudy} uses a more extensive chemical network than our model, including all ionisation states of the 30 lightest elements, and a molecular network that incorporates 83 molecules. \textsc{Cloudy} includes a detailed treatment of the microphysical processes involved with molecular hydrogen, including the level populations of 1893 rovibrational states \citep[for details of the microphysics of the H$_{2}$ molecule implemented in \textsc{Cloudy}, see][]{shaw05}. \textsc{Cloudy} also follows the radiative transfer of radiation through the gas, although we only consider optically thin gas in this paper, so we compare one-zone calculations in \textsc{Cloudy} to our model in chemical equilibrium. \textsc{Cloudy} is applicable to a wide range of physical conditions, e.g. densities up to $n_{\rm{H}} \sim 10^{15} \, \rm{cm}^{-3}$ and temperatures up to $\sim 10^{10}$ K. See \citet{rollig07} for a comparison of several PDR codes, including \textsc{Cloudy}. 

\subsection{Numerical implementation}\label{numImplementation}

We follow the non-equilibrium abundances of the chemical species in our network by integrating the rate equations from the initial conditions using the backward difference formula method and Newton iteration, implemented in \textsc{Cvode} (a part of the \textsc{Sundials}\footnote{\url{https://computation.llnl.gov/casc/sundials/main.html}} suite of non-linear differential/algebraic equation solvers). We use a relative tolerance of $10^{-6}$ and an absolute tolerance of $10^{-13}$. The abundances are also subject to the following constraint equations:

\begin{align}
x_{e} &= \sum_{i,n} n x_{i^{n+}} - \sum_{j} x_{j^{-}}, \label{chargeNeutralityEqn} \\
x_{i^{0}} &= x_{i, \rm{tot}} - \sum_{n} x_{i^{n+}} - x_{i^{-}} - \sum_{mol} s_{i, mol} x_{mol},
\end{align}
where $s_{i, mol}$ is the number of atoms of element $i$ in the molecular species $mol$, and the abundance $x_{i^{n+}}$ of ion species $i^{n+}$ is defined with respect to the total number density of hydrogen, $n_{\rm{H_{tot}}}$, i.e. $x_{i^{n+}} \equiv n_{i^{n+}} / n_{\rm{H_{tot}}}$. Be aware that these abundances are number fractions, not mass fractions. \citet{glover07} use these constraint equations to reduce the number of rate equations that need to be integrated. However, following \citet{oppenheimer13a}, we track all of the species from their rate equations and use the above constraint equations as an independent check on the accuracy of the \textsc{Cvode} solver. If the sum of species of a given element differs from its constraint by more than 1\%, then these abundances are renormalised by multiplying the abundances of each species by the ratio of their constraint to their sum. Similarly, if the electron abundance differs from the sum of charged species as given in equation~\ref{chargeNeutralityEqn}, then it is also renormalised.

\subsection{Photochemical reactions}\label{photoion}

\subsubsection{Photoionisation}

The optically thin photoionisation rate $\Gamma$ for an incident spectrum with an intensity per unit solid angle per unit frequency $J_{\nu}$ is given by the equation:

\begin{equation}\label{photo_eqn}
\Gamma_{\rm{thin}} = \int_{\nu_{0}}^{\infty} \frac{4 \pi J_{\nu} \sigma_{\nu}}{h \nu} d\nu \, ,
\end{equation}
where $\nu_{0}$ is the ionisation threshold frequency of the ion. 

Equation~\ref{photo_eqn} does not include the effect of secondary ionisations, which can be significant if the ionising radiation is dominated by X-rays \citep[e.g.][]{abel97}. We use the values for the number of secondary ionisations of hydrogen per primary ionisation tabulated by \citet{furlanetto10} as a function of primary electron energy and ionised hydrogen fraction $x_{\rm{HII}}$ (see section~\ref{CR_ion_section}) to calculate how important secondary ionisations would be for the photoionisation of hydrogen by the interstellar radiation field of \citet{black87} and the redshift zero extragalactic UV background of \citet{haardt01}. We found that, for these two spectra, the secondary ionisation rate of hydrogen was just $0.02 \%$ and $0.08 \%$ of the primary ionisation rate respectively at an H\textsc{ii} abundance $x_{\rm{HII}} = 10^{-4}$, and even less at higher H\textsc{ii} abundances. We therefore choose to neglect secondary ionsations from UV photoionisation in our model, although we find that we do need to consider them for cosmic rays (see section~\ref{CR_ion_section}).

We can replace the frequency dependent cross section $\sigma_{\nu}$ with an average cross section $\overline{\sigma}$ using the optically thin grey approximation. Then the photoionisation rate of species $i$ is given by the equation:

\begin{equation}\label{grey_photo_eqn}
\Gamma_{i, \rm{thin}} = \overline{\sigma_{i}} \int_{\nu_{0, \rm{H}}}^{\infty} \frac{4 \pi J_{\nu}}{h \nu} d\nu,
\end{equation}
where:

\begin{equation}\label{grey_x_section}
\overline{\sigma_{i}} = \frac{\int_{\nu_{0, i}}^{\infty} \frac{4 \pi J_{\nu} \sigma_{\nu, i}}{h \nu}d\nu}{\int_{\nu_{0, \rm{H}}}^{\infty} \frac{4 \pi J_{\nu}}{h \nu}d\nu} \, .
\end{equation}

The integral on the right hand side of equation~\ref{grey_photo_eqn} gives the number of hydrogen ionising photons per unit area and time. Once we have specified the spectral shape of the incident UV radiation field through $J_{\nu}$, we can calculate the average cross sections $\overline{\sigma_{i}}$ using equation~\ref{grey_x_section}. We use the frequency dependent cross sections $\sigma_{\nu, i}$ from \citet{verner95} and \citet{verner96}, as used by \textsc{Cloudy}. The number of hydrogen ionising photons per unit area and time, which determines the intensity of the UV radiation field, is an input parameter to our model, hence we can calculate the photoionisation rates using equation~\ref{grey_photo_eqn}. 

The photoionisation of inner shell electrons can lead to the ejection of multiple electrons by a single photon (Auger ionisation). We therefore multiply the cross sections calculated using equation~\ref{grey_x_section} for each subshell of every ion by the electron vacancy distribution probabilities from \citet{kaastra93}, and then sum over all subshells to obtain the Auger ionisation rates for each species. 

\subsubsection{Photodissociation}

Molecular hydrogen in the ground electronic state can be dissociated by photons in the Lyman Werner band ($11.2 \, \text{eV} < h \nu < 13.6 \, \text{eV}$) via the two step Solomon process. The absorbed photon excites the H$_{2}$ into an electronically and vibrationally excited state, and when it decays back to the ground electronic state there is a probability that it will dissociate. An accurate treatment of the Solomon process would require us to follow the level populations of the rovibrational states of H$_{2}$, which would be computationally expensive. However, there are approximations that we can use to estimate the photodissociation rate. \citet{abel97} argue that photodissociation of H$_{2}$ occurs mainly via absorptions in the very narrow energy band $12.24 \, \text{eV} < h \nu < 13.51 \, \text{eV}$. Therefore, if the UV spectrum is approximately constant in this range, the photodissociation rate will be proportional to the spectral intensity $J_{\nu}$ evaluated at $h\nu = 12.87$ eV (corresponding to the vibrational state $v = 13$). They then derive this rate to be:

\begin{equation}\label{H2dissocEqn}
\Gamma_{\rm{H_{2}, thin}} = 1.38 \times 10^{9} \, \text{s}^{-1} \, \left( \frac{J_{\nu}(h \overline{\nu} = 12.87 \, \text{eV})}{\text{erg s}^{-1} \text{cm}^{-2} \text{Hz}^{-1} \text{sr}^{-1}} \right) .
\end{equation}
This rate is also used by \citet{glover07} (their equation 49).

To investigate the accuracy of this approximation, we calculated the photodissociation rate in \textsc{Cloudy} using its `big H2' model for the molecular hydrogen, in which the level populations of 1893 rovibrational states of H$_{2}$ are followed and dissociation from these states via the Solomon process is calculated self-consistently (see \citealt{shaw05} for a description of how the microphysics of molecular hydrogen is implemented in \textsc{Cloudy}). We considered three different UV spectra, the interstellar radiation field of \citet{black87}, the redshift zero extragalactic UV background of \citet{haardt01} and a black body spectrum with a temperature $10^{5}$ K, and compared the photodissociation rates estimated by equation~\ref{H2dissocEqn} with those from \textsc{Cloudy} for the range of densities $1 \, \text{cm}^{-3} \leq n_{\rm{H}} \leq 10^{4} \, \text{cm}^{-3}$. We found that the rates from \textsc{Cloudy} were generally higher than those estimated using equation~\ref{H2dissocEqn}, by a factor $\sim 2$. Furthermore, the dissociation rates calculated by \textsc{Cloudy} did not scale exactly with $J_{\nu}(h \overline{\nu} = 12.87 \, \text{eV})$. We considered different ways to parameterise the dependence of the photodissociation rate on the UV spectrum, and we found that the most robust method was to assume that it scales with the number density of photons in the energy band $12.24 \, \text{eV} < h \nu < 13.51 \, \text{eV}$, $n_{12.24-13.51 \rm{eV}}$. We normalised this relation to the average rate calculated by \textsc{Cloudy} over the density range $1 \, \text{cm}^{-3} \leq n_{\rm{H}} \leq 10^{4} \, \text{cm}^{-3}$ in the presence of the \citet{black87} interstellar radiation field. The optically thin photodissociation rate of H$_{2}$ is then:

\begin{equation}
\Gamma_{\rm{H_{2}, thin}} = 7.5 \times 10^{-11} \, \text{s}^{-1} \left( \frac{n_{12.24-13.51 \rm{eV}}}{2.256 \times 10^{-4} \, \text{cm}^{-3}} \right) .
\end{equation}

There remains a dependence of the photodissociation rate on the density that we are unable to capture in this approximation, as we do not follow the level populations of the rovibrational states of H$_{2}$. This introduces errors of up to $\sim 25 \%$ over the density range $1 \, \text{cm}^{-3} \leq n_{\rm{H}} \leq 10^{4} \, \text{cm}^{-3}$.

For the molecular species from the CO network we use the photoionisation and photodissociation rates given by \citet{vandishoeck06} and \citet{visser09} where available, or \citet{glover10} otherwise, as calculated for the interstellar radiation field of \citet{draine78}. This radiation field has a field strength $G_{0} = 1.7$, where the dimensionless parameter $G_{0}$ is defined as the ratio of the energy density in the UV radiation field in the energy range 6 eV to 13.6 eV with respect to the \citet{habing68} field:

\begin{equation}\label{G_eqn}
G_{0} \equiv \frac{u_{6 - 13.6 \text{eV}}}{5.29 \times 10^{-14} \, \text{erg cm}^{-3}} .
\end{equation}
We assume that the photoionisation and photodissociation rates of these molecular species are proportional to $G_{0}$, so we multiply these rates calculated for the \citet{draine78} interstellar radiation field by $G_{0} / 1.7$.

The above photochemical rates are valid in the optically thin regime. In paper II we will consider the impact of shielding on these rates.

\subsection{Cosmic ray ionisation}\label{CR_ion_section}
The ionisation rate of different species due to cosmic rays depends on the spectrum of the cosmic rays. This spectrum is, however, still uncertain both at high redshifts and even in the local ISM. Following the approach of \citet{glover07}, we allow the primary ionisation rate of atomic hydrogen due to cosmic rays, $\zeta_{\rm{HI}}$, to be a free parameter in our model. For example, the default value used by \textsc{Cloudy} for the galactic background is $\zeta_{\rm{HI}} = 2.5 \times 10^{-17} \text{s}^{-1}$ \citep{williams98}. We then scale the primary ionisation rates of other species linearly with this parameter. \citet{glover07} assume that the ratios of the ionisation rates of the other elements with respect to hydrogen are equal to the ratios of the values given in the \textsc{umist} database for astrochemistry \citep{leteuff00} where these rates are available. For Si\textsc{i} and Si\textsc{ii}, whose cosmic ray ionisation rates are not included in this database, they obtain the cosmic ray ionisation rate with respect to $\zeta_{\rm{HI}}$ using the method from \citet{langer78}, who use the following equation from \citet{silk70}:

\begin{equation}
\zeta_{i} = \overline{\xi_{i}} \left( \frac{\chi_{\rm{H}}}{\chi_{i}} \right) \zeta_{\rm{HI}},
\end{equation}
where $\overline{\xi_{i}}$ is the effective number of electrons in the outer shell and $\chi_{\rm{H}}$ and $\chi_i$ are the ionisation energies of hydrogen and the ion species $i$ respectively. $\overline{\xi_{i}}$ can be calculated using the equations in \citet{lotz67}:

\begin{equation}
\overline{\xi} = \overline{\chi} \sum_{j=1}^{2} (\xi_j / \chi_j),
\end{equation}
where the summation is over the two outermost (sub-) shells, with $\xi_j$ electrons and ionisation potential $\chi_j$. $\overline{\chi}$ is a weighted average of these ionisation potentials, given by:

\begin{equation}
\ln \overline{\chi} = \frac{\sum_{j=1}^{2} (\xi_j / \chi_j) \ln \chi_j}{\sum_{j=1}^{2} (\xi_j / \chi_j)}.
\end{equation}

Many of the species that we have added from the model of \citet{oppenheimer13a} are not included in the \textsc{umist} database, so for these we also use the above equations to calculate their cosmic ray ionisation rates in terms of $\zeta_{\rm{HI}}$. We note that cosmic ray ionisation is only important for species in low ionisation states, as the ionisation rates calculated above become very small for higher ionisation states.

In addition to the primary ionisations described above, the highly energetic ejected electrons can ionise further atoms. \citet{furlanetto10} used Monte Carlo simulations to calculate the fraction of the primary electron energy $E$ that is deposited as heat, in collisional ionisation of H\textsc{i}, He\textsc{i} or He\textsc{ii}, and in collisional excitation of H\textsc{i}. They tabulate these energy deposition fractions as functions of $E$ and of the ionisation fraction, $x_{i} = x_{\rm{HII}}$. We therefore calculate the ionisation rates of H\textsc{i} and He\textsc{i} due to secondary ionisations by interpolating the tables of \citet{furlanetto10} as a function of the ionised hydrogen fraction $x_{\rm{HII}}$. For the primary electron energy we take a typical mean value $E = 35$ eV \citep{spitzer78,wolfire95}. At this energy, the secondary ionisation rate of He\textsc{ii} is zero, as it is below the He\textsc{ii} ionisation energy of 54.4 eV. 

Note that this introduces a dependence on the ionisation fraction, as Coulomb interactions can reduce the energy of the primary electron if the electron abundance is high, thereby reducing the number of secondary ionisations. If we only include primary ionisations, then we miss this dependence, which we find does have a small but noticeable effect on the ionisation balance at low densities and low ionisation fractions. 

We were unable to find equivalent calculations for the secondary ionisation rates of metal species in the literature, so we only include primary cosmic ray ionisations for these species.

\citet{glover10} also include a number of ionisation and dissociation reactions of molecules from cosmic ray induced UV emission, which we have also included in our model. These are summarised in table~\ref{reaction_table1} in appendix~\ref{reactions_summary}, and are labelled as `$\gamma_{cr}$'.

\subsection{Dust grain physics}\label{dust_physics}

Our model includes a number of reactions that occur on the surface of dust grains. The most important of these is the formation of molecular hydrogen on dust, which typically dominates over the gas phase reactions except in dust free or very low metallicity environments. For this reaction we use the rate from \citet{cazaux02}, as given by their equation 18:

\begin{equation}\label{H2_dust_rate}
R_{\rm{H_{2}}} = \frac{1}{2} n_{\rm{HI}} v_{\rm{H}}(T_{\rm{gas}}) n_{\rm{d}} \sigma_{\rm{d}} \epsilon_{\rm{H_{2}}}(T_{\rm{dust}}) S_{\rm{H}}(T_{\rm{gas}}, T_{\rm{dust}}) \; \text{cm}^{-3} \, \text{s}^{-1},
\end{equation}
where $n_{\rm{HI}}$ is the number density of neutral hydrogen, $v_{\rm{H}}(T_{\rm{gas}})$ is the thermal velocity of the gas at temperature $T_{\rm{gas}}$ and $n_{\rm{d}} \sigma_{\rm{d}}$ is the total cross sectional area of dust grains per unit volume. Following \citet{krumholz11}, we use a cross sectional area of $10^{-21} Z / Z_{\odot} \, \text{cm}^{2}$ per H nucleus, which is intermediate between the values from the models of \citet{weingartner01a} for the Milky Way (with a ratio of visual extinction to reddening $R_{v} = 3.1$ or $5.5$), the Large Magellanic Cloud and the Small Magellanic Cloud, and assumes that the dust content of the gas scales linearly with metallicity. The dimensionless recombination efficiency $\epsilon_{\rm{H_{2}}}(T_{\rm{dust}})$ is given in equation 13 of \citet{cazaux02}. For the dimensionless sticking probability, $S_{\rm{H}}(T_{\rm{gas}}, T_{\rm{dust}})$, we use equation 3.7 from \citet{hollenbach79}.

Equation~\ref{H2_dust_rate} depends on the temperatures of both the gas and the dust, so we need to make an assumption about the dust temperature. \citet{glover12} calculate the dust temperature by assuming that it is in thermal equilibrium and then solving the thermal balance equation for the dust grains (their equation A2). However, we find that for the physical conditions that we are interested in, i.e. those relevant to the diffuse ISM, the molecular hydrogen abundance is insensitive to the temperature of the dust (see appendix~\ref{dustPhysicsAppendix}). Therefore, to reduce the computational cost of our model, we shall simply assume a constant dust temperature of 10 K.

Following \citet{glover07}, we include the recombination of H\textsc{ii}, He\textsc{ii}, C\textsc{ii}, O\textsc{ii} and  Si\textsc{ii} on dust grains, along with Fe\textsc{ii}, Mg\textsc{ii}, S\textsc{ii}, Ca\textsc{ii} and Ca\textsc{iii}, which were not present in the model of \citet{glover07}. We use the rate coefficients quoted in table 3 of their paper, where present, which are based on calculations by \citet{hollenbach79} and \citet{weingartner01b}. For the reactions not included in \citet{glover07}, we take the rate coefficients directly from \citet{weingartner01b}. These reactions are typically one or two orders of magnitude smaller than the radiative recombination rates, although the two can become comparable at temperatures above a few thousand Kelvin.

\begin{figure}
\centering
\mbox{
	\includegraphics[width=84mm]{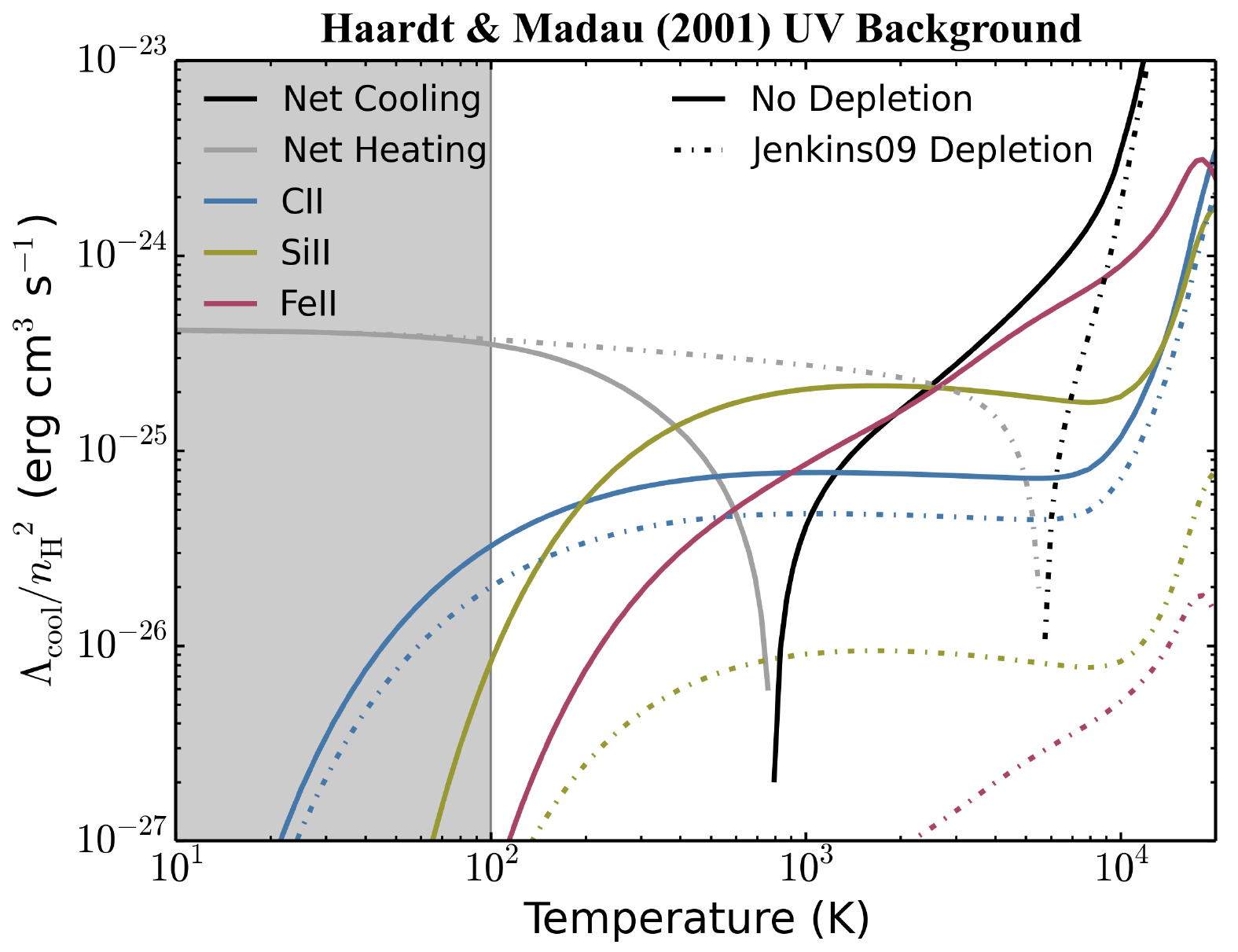}}
\caption{Comparison of the cooling functions of interstellar gas at solar metallicity without depletion of metal atoms onto dust grains \textit{(solid curves)} and including the depletion factors from \citet{jenkins09} for C, N, O, Mg, Si and Fe. These cooling functions were calculated using our model at a density $n_{\rm{H}} = 1 \, \rm{cm}^{-3}$ in the presence of the redshift zero \citet{haardt01} extragalactic UV background. The shaded grey region indicates temperatures below 100 K that are outside the range of temperatures that we are primarily interested in, but we include this regime here for completeness. We see that, in this example, the depletion of metal atoms onto dust grains can increase the thermal equilibrium temperature from $T_{\rm{eq}} \sim 800$ K to $T_{\rm{eq}} \sim 5600$ K.}
\label{depletionsFig}
\end{figure}

There are several additional dust grain processes that we do not currently include in our model. In particular, we do not explicitly follow the creation and destruction of dust, assuming instead that the dust-to-gas ratio simply scales linearly with the metallicity. More importantly for the gas cooling rates, we also do not model the depletion of metal atoms onto dust grains. This can reduce the gas phase abundances of metals by factors of a few. For example, \citet{jenkins09} investigates the variation in the depletion factors (i.e. the fraction of a species remaining in the gas phase) of metals along different sight lines in the local ISM. The depletion factors of carbon and oxygen that they find are typically $\sim 0.6 - 0.8$ and $\sim 0.6 - 1.0$ respectively, while heavily depleted elements such as iron can have depletion factors below one per cent. Thus the rates of metal line cooling, particularly from iron, could be significantly affected by depletion onto dust grains. 

Since these depletion factors are highly uncertain, we do not include them in our model, and we aim to address these issues in a future work. However, in figure~\ref{depletionsFig} we illustrate the impact that metal depletion onto dust grains could potentially have on the cooling function of interstellar gas. We show the cooling rates of gas at a density $n_{\rm{H}} = 1 \, \rm{cm}^{-3}$ in the presence of the redshift zero \citet{haardt01} extragalactic UV background, plotted as a function of temperature. These were calculated using our model with element abundances at solar metallicity (solid curves), and with the solar abundances of C, N, O, Mg, Si and Fe reduced by the depletion factors given in column seven of table 4 in \citet{jenkins09} (dot-dashed curves). We see that, in this example, cooling from C\textsc{ii}, Si\textsc{ii} and Fe\textsc{ii} are important to balance the heating from the photoionisation of H\textsc{i}. However, these elements can be strongly depleted, particularly Si and Fe. Therefore, when we include the depletion factors of \citet{jenkins09}, the thermal equilibrium temperature, at which the total cooling and heating rates are equal, increases from $T_{\rm{eq}} \sim 800$ K to $T_{\rm{eq}} \sim 5600$ K.

\subsection{CO model}\label{COmodel}
We base our model for CO chemistry on the reaction network of \citet{glover10}, as we find that this is able to reproduce the molecular fraction of CO in the diffuse ISM fairly well, as predicted by \textsc{Cloudy}, without requiring the full complexity of more complete networks (see section~\ref{ISMmolecules}). For the photochemical reaction rates they assume the interstellar radiation field of \citet{draine78}, with a radiation field strength (see equation~\ref{G_eqn}) of $G_{0} = 1.7$ in units of the \citet{habing68} field. To generalise these rates to any UV radiation field, we multiply them by $G_{0} / 1.7$. 

Our comparisons with \textsc{Cloudy} showed that it is necessary to include four additional reactions in our network (reactions 111, 283, 303 and 304 in table~\ref{reaction_table1}) that were not included in the model of \citet{glover10}. We also updated the rates for a small number of reactions in \citet{glover10} to improve agreement with \textsc{Cloudy}, and we updated some of the photodissociation rates using the values from \citet{vandishoeck06} and \citet{visser09}; see appendix~\ref{reactions_summary}. 

\section{Thermal processes}\label{ThermalModel}
Once we have calculated the abundance of each species, we can evaluate its contribution to the net cooling rate. The processes that we include in our model are summarised in Table~\ref{cooling_processes}, and some of these are discussed in more detail below.

\begin{table}
\centering
\begin{minipage}{84mm}
\caption{Summary of included thermal processes.}
\centering
\begin{tabular}{lr}
\hline 
Process & References\footnote{1 - \citet{oppenheimer13a}; 2 - \citet{glover07}; 3 - \citet{wolfire03}; 4 - \citet{cen92}; 5 - \citet{glover08}; 6 - \citet{maclow86}; 7 - \citet{glover10}; 8 - \citet{neufeld93}; 9 - \citet{neufeld95}; 10 - \citet{hollenbach79}; 11 - \citet{verner95}; 12 - \citet{verner96}; 13 - \citet{goldsmith78}; 14 - \citet{wolfire95}; 15 - \citet{black77}; 16 - \citet{burton90}; 17 - \citet{launay91}; 18 - \citet{karpas79}} \\
\hline
\multicolumn{2}{c}{\textbf{Cooling Processes}} \\ 
\hline
H excitation & 1 \\
He excitation & 1 \\
He$^{+}$ excitation & 1 \\
H collisional ionisation & 1 \\
He collisional ionisation & 1 \\
He$^{+}$ collisional ionisation & 1 \\
H$^{+}$ recombination & 1 \\
He$^{+}$ recombination\footnote{radiative plus dielectronic} & 1 \\
He$^{++}$ recombination & 1 \\
Grain-surface recombination & 2, 3 \\
Inverse Compton Cooling from the CMB & 2, 4 \\
Bremsstrahlung & 1 \\
Metal line cooling & 1, 2 \\
H$_{2}$ rovibrational cooling & 5 \\
H$_{2}$ collisional dissociation & 2, 6 \\
CO rovibrational cooling & 7, 8, 9 \\
H$_{2}$O rovibrational cooling & 7, 8, 9 \\
OH rotational cooling & 10 \\
\hline
\multicolumn{2}{c}{\textbf{Heating Processes}} \\ 
\hline
Photoheating & 11, 12 \\
Cosmic ray heating\footnote{Assuming 20 eV deposited as heat into the gas per primary ionisation \citep{goldsmith78}} & 2, 13 \\
Dust photoelectric effect & 14 \\
H$_{2}$ photodissociation & 2, 15 \\
H$_{2}$ UV pumping & 2, 16 \\
H$_{2}$ formation; gas phase & 2, 17, 18 \\
H$_{2}$ formation; dust grains & 2, 10 \\
\hline
\vspace{-0.3 in}
\label{cooling_processes}
\end{tabular}
\end{minipage}
\end{table}

\subsection{Metal line cooling}

\citet{oppenheimer13a} have tabulated the radiative cooling rates from the nine metal species that we include in our model (see section~\ref{ChemicalModel}), along with hydrogen and helium, as a function of temperature using \textsc{Cloudy} (they use version 10.00 in their paper, but they have since updated these tables using version 13.01 of \textsc{Cloudy}\footnote{These upated tables can be found on the website: \url{http://noneq.strw.leidenuniv.nl}}). See their section 2.2 for more details of their method. We use these tabulated cooling rates for most of these species. However, they note that these rates assume that the radiative cooling is dominated by electron-ion collisions and that this assumption can break down at temperatures $T \la 10^{3}$ K for some species, for example O\textsc{i}. 

We indeed find that their tabulated cooling rates for O\textsc{i} and C\textsc{i} are unable to reproduce \textsc{Cloudy}'s cooling curves at low temperatures, so for these two species we calculate the radiative cooling rates following the same method as \citet{glover07}. They consider only the three lowest fine-structure energy levels of each species, although for C\textsc{i} we found it necessary to extend their method to include the nine lowest energy levels, using transition probabilities from the NIST Atomic Spectra Database (ver. 5.0)\footnote{\url{http://physics.nist.gov/asd}} and effective collision strengths from \citet{pequignot76} and \citet{dunseath97}. We tabulate these cooling rates in 55 temperature bins from $10$ K to $10^{4}$ K and 30 bins each in H\textsc{i}, H\textsc{ii} and electron densities from $10^{-8} \, \text{cm}^{-3}$ to $10^{5} \, \text{cm}^{-3}$. By tabulating in four dimensions in this way we are able to follow the radiative cooling rates of these species in regimes that are dominated by collisions with H\textsc{i} or H\textsc{ii}, as well as when electron-ion collisions dominate. We use these new cooling rates for O\textsc{i} and C\textsc{i} at temperatures below $10^{4}$ K, and revert back to the cooling rates tabulated by \citet{oppenheimer13a} at higher temperatures. 

We also found that for some low-ionisation metal species there was an extra density dependence in the cooling rates that was not captured in the tables of \citet{oppenheimer13a}, most notably for C\textsc{ii}, N\textsc{ii}, Si\textsc{ii} and Fe\textsc{ii}. This led to the cooling rates from these species being overestimated by up to an order of magnitude at the highest densities that we consider here ($n_{\rm{H}} \sim 10^{4} \, \text{cm}^{-3}$, much higher than those considered by \citealt{oppenheimer13a}) compared to those calculated by \textsc{Cloudy}. For these four species we therefore used the \textsc{Chianti} database version 7.1\footnote{\url{http://www.chiantidatabase.org/chianti.html}} \citep{dere97,landi13} to calculate line emissivities and hence cooling rates per ion, which we tabulate as functions of temperature in 80 bins from $10$ K to $10^{5}$ K and electron density in 130 bins from $10^{-8} \, \text{cm}^{-3}$ to $10^{5} \, \text{cm}^{-3}$. 

\subsection{H$_{2}$ rovibrational cooling}\label{H2_rovib_cooling}

\begin{figure*}
\centering
\mbox{
	\includegraphics[width=145mm]{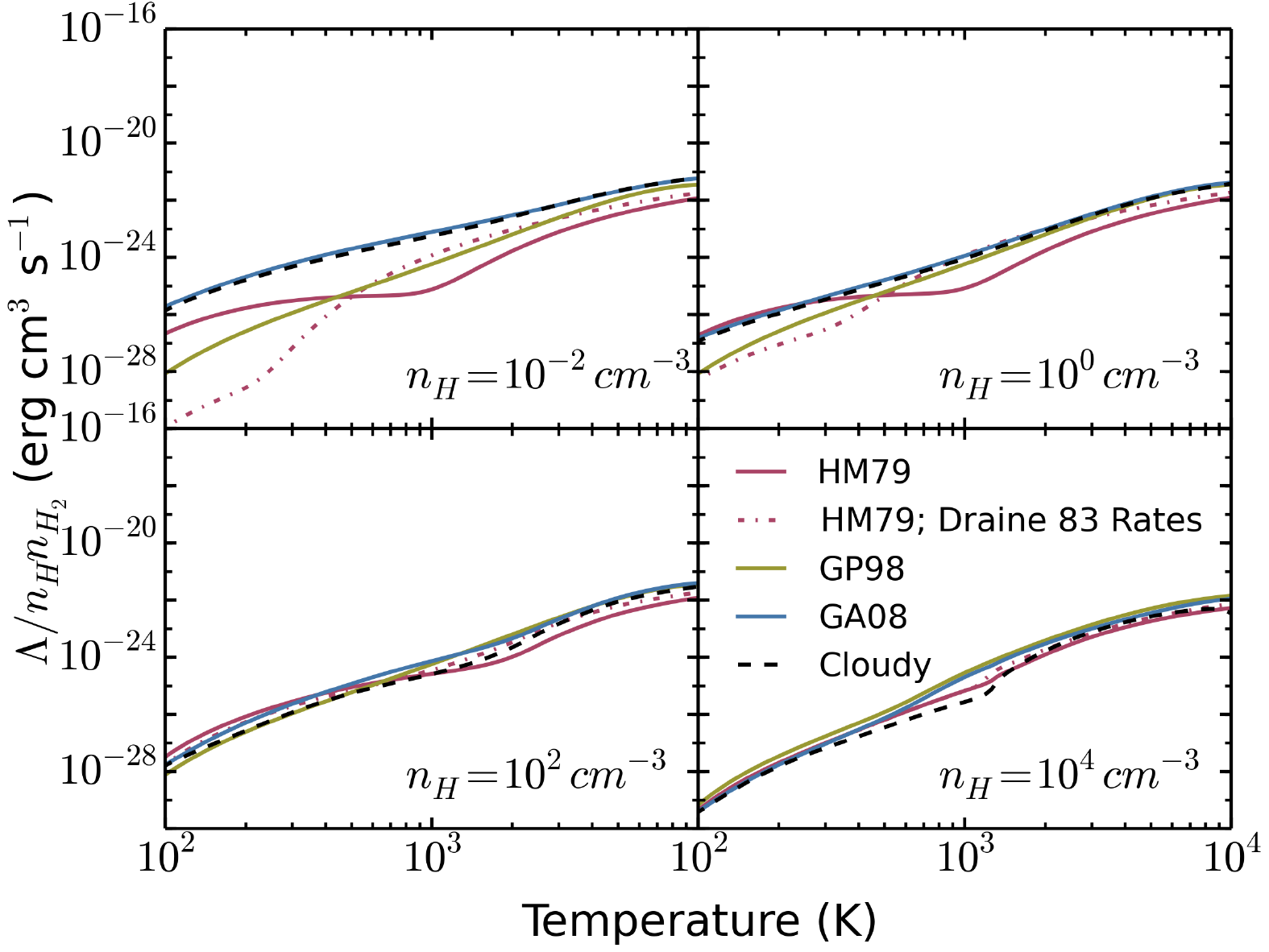}}
\caption{Comparison of various approximations for the H$_{2}$ cooling rate: \citet{hollenbach79} \textit{(HM79; solid red curve)}, \citet{hollenbach79} with the collisional excitation rates from \citet{draine83} \textit{(dot-dashed red curve)}, \citet{galli98} \textit{(GP98; solid yellow curve)} and \citet{glover08} \textit{(GA08; solid blue curve)}. These are compared to the H$_{2}$ cooling rates calculated by \textsc{Cloudy} \textit{(dashed black curve)}, which follows the level populations of several thousand rovibrational levels. The rovibrational cooling rate per H$_{2}$ molecule is plotted as a function of temperature, calculated in the absence of UV at various densities. We see that overall the GA08 cooling function gives the best match to the \textsc{Cloudy} cooling rates, so we shall use the GA08 cooling rates in our model.}
\label{H2Cooling}
\end{figure*}

Cooling from molecular hydrogen can become important in the ISM, particularly in primordial and low-metallicity gas. The primary mechanism by which H$_{2}$ cools the gas is transitions between its rotationally and vibrationally excited energy levels. To treat this properly, we would need to solve for the populations of these levels and hence obtain the rate at which they radiate away energy, but this can become computationally expensive due to the large number of levels involved. A number of studies have looked at this problem and have suggested analytic fits to the H$_{2}$ rovibrational cooling function. We have considered the analytic fits proposed by \citet{hollenbach79}, \citet{galli98} and \citet{glover08}, which we compare to the cooling rates predicted by \textsc{Cloudy} using its `big H2' model to accurately follow all of the rotational and vibrational energy levels of the H$_{2}$ molecule. We also considered using the \citet{hollenbach79} cooling function but using the collisional rates from \citet{draine83}. These cooling functions typically express the H$_{2}$ cooling in terms of temperature and the densities of H\textsc{i} and H$_{2}$, although \citet{glover08} also include its dependence upon the densities of H\textsc{ii}, He\textsc{i} and electrons, which arises from the collisional excitation of molecular hydrogen by these species. When calculating these cooling rates we have used the abundances calculated by \textsc{Cloudy} to compute the collisional excitation rates, so that any discrepancies are due to differences in the analytic fits to the cooling function rather than to differences in the solution of the chemical model. These comparisons are shown in figure~\ref{H2Cooling}. 

We see that the cooling function from \citet{glover08} shows the closest agreement with the cooling rates computed by \textsc{Cloudy}, so we shall use this in our model. For the low density limit of the cooling function we use equation 40 from \citet{glover08} with the fitting coefficients listed in their table 8, for a fixed ratio of ortho- to para-H$_{2}$ of 3:1. This is then combined with the local thermodynamic equilibrium (LTE) cooling rate using their equation 39 to obtain the final H$_{2}$ cooling function. We calculate the LTE cooling rate using the radiative de-excitation rates of the rovibrational transitions in the ground electronic state taken from \citet{wolniewicz98}. 

\subsection{CO, H$_{2}$O and OH cooling}
We include cooling from CO and H$_{2}$O molecules using the same prescriptions as \citet{glover10} (see their section 2.3.2). \citet{glover10} base these prescriptions on the cooling rates calculated by \citet{neufeld93} and \citet{neufeld95} but extend their treatment to account for collisions with electrons and atomic hydrogen, in addition to molecular hydrogen. Following \citet{glover10}, we make use of the tabulated fit parameters given in \citet{neufeld93} and \citet{neufeld95}, which are given as functions of temperature and the effective column density per unit velocity, $\tilde{N}_{m}$. If the gas flow has no special symmetry and if the large velocity gradient approximation is valid, then \citep{neufeld93}:
\begin{equation}\label{effNeq}
\tilde{N}_{m} = \frac{n_{m}}{\lvert \underline{\nabla} \cdot \underline{v} \rvert},
\end{equation}
where $n_{m}$ is the number density of species $m$ and $\underline{\nabla} \cdot \underline{v}$ is the divergence of the velocity field. For static gas this is equivalent to:
\begin{equation}
\tilde{N}_{m} = \frac{N_{m}}{\sigma},
\end{equation}
where $N_{m}$ is the column density of species $m$ and $\sigma$ is the thermal velocity dispersion:
\begin{equation}
\sigma = \sqrt{\frac{3 k_{B} T}{\mu m_{p}}}.
\end{equation}
We use the static gas case in this work. However, when we use this in a hydrodynamic simulation, we could drop this assumption and calculate $\tilde{N}_{m}$ directly from equation~\ref{effNeq}. We also include rotational cooling from the OH molecule using the `universal' cooling function for molecules with dipole moments from \citet{hollenbach79} (their equations 6.21 and 6.22), using the molecular cooling parameters for OH summarised in their table 3. This function gives the cooling rate from OH in terms of the temperature and density along with the OH column density and dust extinction of the gas cell.

\subsection{Photoheating}\label{photoheat_section}

We include the photoheating of H\textsc{i}, He\textsc{i}, He\textsc{ii} and all metal atoms and ions in our model. The optically thin photoheating rate per unit volume from the photoionisation of species $i$ is $n_{i} \mathcal{E}_{i, \rm{thin}}$, where:

\begin{equation}\label{photoheating_eq}
\mathcal{E}_{i, \rm{thin}} = \int_{\nu_{0, i}}^{\infty} \frac{4 \pi J_{\nu} \sigma_{\nu, i}}{h \nu} (h \nu - h \nu_{0, i}) d\nu.
\end{equation}
We can express this as:

\begin{equation}
\mathcal{E}_{i, \rm{thin}} = \Gamma_{i, \rm{thin}} \left< \epsilon_{i, \rm{thin}} \right>,
\end{equation}
where $\Gamma_{i, \rm{thin}}$ is the photoionisation rate of species $i$, given by equation~\ref{grey_photo_eqn}, and $\epsilon_{i, \rm{thin}}$ is the average excess energy of ionising photons:

\begin{equation}\label{epsilonEqn}
\left< \epsilon_{i, \rm{thin}} \right> = \frac{\int_{\nu_{0, i}}^{\infty} \frac{4 \pi J_{\nu} \sigma_{\nu, i}}{h \nu} (h \nu - h \nu_{0, i}) d\nu}{\int_{\nu_{0, i}}^{\infty} \frac{4 \pi J_{\nu} \sigma_{\nu, i}}{h \nu} d\nu}.
\end{equation}

For each species $\left< \epsilon_{i, \rm{thin}} \right>$ is calculated for the given incident UV spectrum $J_{\nu}$, using the frequency dependent cross sections from \citet{verner95} and \citet{verner96}, and is then provided to our model as an input parameter.

Note that \citet{glover07} include an efficiency factor in equation~\ref{photoheating_eq} for the fraction of the excess photon energy that is converted into heat (as calculated using the results of e.g. \citealt{shull85} or \citealt{dalgarno99}). This factor accounts for the fact that the electrons released by photoionisation reactions can cause secondary ionisations/excitations of H\textsc{i} and He\textsc{i}. The energy used in these secondary reactions will subsequently be radiated away and thus does not contribute to heating the gas. However, as explained in section~\ref{photoion}, we do not need to include the effects of secondary ionisation, as they only enhance the primary photoionisation rate by less than $0.1\%$ for the UV radiation fields that we consider here and are thus unimportant. We therefore take this efficiency factor to be equal to 1 in equation~\ref{photoheating_eq}.

The above equations are applicable to the optically thin regime. In paper II we will consider the effect of shielding on the photoheating rates. 

\subsection{Photoelectric heating}
Another potentially important source of heating is photoelectric emission from dust grains. When a UV photon is absorbed by a dust grain it can release an electron, and the excess energy absorbed by the electron is quickly thermalised, thereby heating the gas. The photoelectric heating rate is given by \citep{bakes94,wolfire95}:

\begin{equation}
\Gamma = 1.0 \times 10^{-24} \epsilon G_{0} \, \text{erg} \, \text{cm}^{-3} \, \text{s}^{-1},
\end{equation}
where $G_{0}$ is the UV radiation field in units of the \citet{habing68} field (see equation~\ref{G_eqn}) and the heating efficiency $\epsilon$ is given by the following analytic fit for gas temperatures up to $10^{4}$ K:

\begin{align}
\epsilon = & \frac{4.9 \times 10^{-2}}{1.0 + ((G_{0} T^{0.5} / n_{e}) / (1925 \, \text{K}^{0.5} \, \text{cm}^{3}))^{0.73}} \notag \\
& \left. + \frac{3.7 \times 10^{-2} (T / 10^{4} \, \text{K})^{0.7}}{1.0 + ((G_{0} T^{0.5} / n_{e}) / (5000 \, \text{K}^{0.5} \, \text{cm}^{3}))} \right. .
\end{align}

\section{Equilibrium chemistry and cooling in the ISM}\label{ResultsI}

While our chemical model is primarily focussed on the non-equilibrium evolution of chemistry and gas cooling in the ISM, it is important that we confirm the validity of our model by comparing its results in chemical equilibrium with those from existing codes, in particular to test whether we have missed any important reactions in our chemical network. We therefore calculated the equilibrium ionisation balance and cooling rates from our model as a function of temperature from 10 K to $2 \times 10^{4}$ K. We are primarily interested in the temperature range $10^{2} \leq T \leq 10^{4}$ K, corresponding to the transition from a warm to a cold phase in the ISM, but we also look at how our model behaves outside this regime. Abundances and cooling rates were calculated at a range of fixed densities relevant to the diffuse ISM, from $10^{-2} \, \text{cm}^{-3}$ to $10^{4} \, \text{cm}^{-3}$, for various metallicities and using four different assumptions for the UV radiation field: the interstellar radiation field (ISRF) of \citet{black87}, both by itself and also multiplied by a factor of 10; the extragalactic UV background (UVB) of \citet{haardt01} and in the absence of UV radiation (i.e. fully shielded gas). The calculations were started from fully neutral and atomic initial conditions and evolved until they reached chemical equilibrium. The examples including a UV radiation field were assumed to be optically thin, i.e. with no intervening column density to shield the gas. We will consider the effect of shielding in paper II. We used the default solar metal abundances from \textsc{Cloudy} (as used by e.g. \citealt{wiersma09,oppenheimer13a}), and we used a hydrogen cosmic ray ionisation rate of $\zeta_{\rm{HI}} = 2.5 \times 10^{-17} \, \text{s}^{-1}$ \citep{williams98}. 

We compared the predicted equilibrium ionisation fractions and cooling rates from our model to those calculated using version 13.01 of \textsc{Cloudy} \citep{ferland13}. In most of the \textsc{Cloudy} runs the dust temperature was calculated self-consistently assuming thermal balance. However, in the absence of a UV radiation field we found it necessary to fix the dust temperature in \textsc{Cloudy} at 10 K, as used in our model, to prevent unrealistically low dust temperatures. This problem arose because our \textsc{Cloudy} calculations considered an optically thin parcel of gas in which the dust is also optically thin to its own infrared radiation, so in the absence of UV radiation there is very little heating of the dust grains. We generally find that our network agrees well with \textsc{Cloudy}, as we shall demonstrate in the next section. We also use these results to identify which are the most important coolants in the ISM in the various physical regimes explored here. More plots of our results from these tests can be found on our website~\footnote{\url{http://noneqism.strw.leidenuniv.nl}}.

\subsection{Dominant coolants}\label{coolantsSection}

\begin{figure*}
\centering
\mbox{
	\includegraphics[width=164mm]{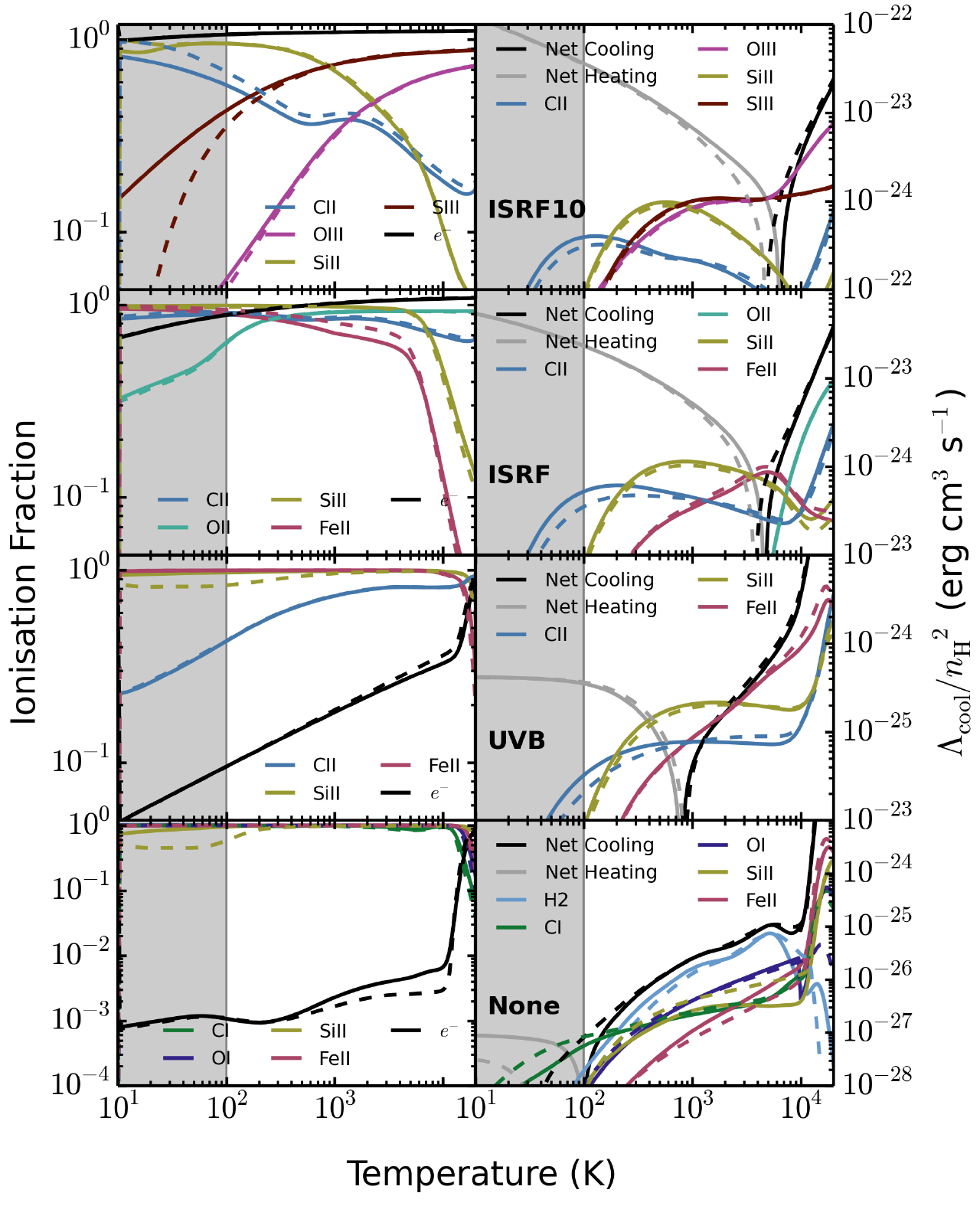}}
\caption{Comparison of the equilibrium ionisation fractions \textit{(left)} and cooling functions \textit{(right)} predicted by our model \textit{(solid lines)} and \textsc{Cloudy} \textit{(dashed lines)}, plotted against temperature. These were calculated at solar metallicity and a density $n_{\rm{H}} = 1 \, \text{cm}^{-3}$ for four different UV radiation fields: the \citet{black87} interstellar radiation field (ISRF) multiplied by a factor 10 \textit{(ISRF10; top)}, the \citet{black87} ISRF \textit{(ISRF; second row)}, the redshift zero \citet{haardt01} extragalactic UV background \textit{(UVB; third row)} and in the absence of UV radiation \textit{(None; bottom)}. In each row we show only those species that contribute at least 20 per cent to the cooling rate anywhere in the temperature range $10^{2} \, \text{K} < T < 10^{4} \, \text{K}$. The panels in the right column show the cooling rates from individual species along with the net cooling rate (total cooling minus heating). The shaded grey region highlights temperatures below 100 K that are outside the range of temperatures that we are primarily interested in, but we include this regime here for completeness.}
\label{ISMcoolants}
\end{figure*}  

In figure~\ref{ISMcoolants} we show the equilibrium ionisation fractions (left panels) and cooling functions (right panels) as a function of temperature at a density $n_{\rm{H}} = 1 \, \text{cm}^{-3}$ for the four different UV radiation fields we consider here. We show temperatures ranging from 10 K to $2 \times 10^{4}$ K, but have greyed out the region $T < 100$ K as this lies outside the range of temperatures that we are primarily interested in. 

The top row was calculated in the presence of the \citet{black87} ISRF multiplied by a factor of 10, which represents the UV radiation field one might find within a strongly star-forming galaxy. The dominant coolants in this case are, going from high to low temperatures, O\textsc{iii}, S\textsc{iii}, Si\textsc{ii} and C\textsc{ii}, while strong photoheating, primarily from hydrogen, balances the cooling at a thermal equilibrium temperature $T_{\rm{eq}} \sim 5000$ K. 

The second row shows the results for the \citet{black87} ISRF, which is more typical of the local solar neighbourhood. Fe\textsc{ii} contributes significantly to the cooling here, while the more highly ionised species (e.g. O\textsc{iii} and S\textsc{iii}) become less important. Despite the weaker photoheating, the equilibrium temperature is still close to $5000$ K, as the metal line cooling is also lower.

The results in the third row were calculated in the presence of the \citet{haardt01} extragalactic UV background at redshift zero, in which the flux of hydrogen-ionising photons is approximately three orders of magnitude lower than the \citet{black87} field. In this example we find that the cooling is now dominated by Fe\textsc{ii}, Si\textsc{ii} and C\textsc{ii}, and the lower photoheating rate now balances the metal line cooling at $T_{\rm{eq}} \sim 800$ K. 

Finally, in the bottom row of figure~\ref{ISMcoolants} we show the equilibrium state of gas in the absence of UV radiation. The dominant coolant here is molecular hydrogen, although there is still a significant contribution from some metal species, most notably Si\textsc{ii}, O\textsc{i} and C\textsc{i}. Without UV radiation to provide photoheating, the main source of heating comes from cosmic rays, resulting in an equilibrium temperature $T_{\rm{eq}} \sim 100$ K. 

In table~\ref{dominantCoolantsTable} we summarise the species that contribute at least 5\% to the total equilibrium cooling rate in the temperature range $10^{2} \, \text{K} \leq T \leq 10^{4} \, \text{K}$ for solar metallicity and a density $n_{\rm{H}} = 1 \, \text{cm}^{-3}$, in the presence of the four UV radiation fields considered in figure~\ref{ISMcoolants}. For each species we give the minimum and maximum temperatures, $T_{\rm{min}}$ and $T_{\rm{max}}$, for which that species contributes at least 5\% to the total equilibrium cooling rate, the temperature at which the relative contribution of that species to the total equilibrium cooling rate is highest, T$_{\rm{peak}}$, and the relative contribution of that species at $T_{\rm{peak}}$. More tables of dominant coolants for different metallicities and densities, and extending to higher temperatures up to $10^{9}$ K, can be found on our website.

\citet{bertone13} have also investigated the dominant cooling channels in the Universe, although they focus on diffuse gas on cosmological scales, typically with lower densities ($n_{\rm{H_{tot}}} < 0.1 \, \text{cm}^{-3}$) than we have considered. The species that they found most important at these lower densities, in the presence of the \citet{haardt01} extragalactic UV background, were O\textsc{iii}, C\textsc{ii}, C\textsc{iii}, Si\textsc{ii}, Si\textsc{iii}, Fe\textsc{ii} and S\textsc{iii}.

\begin{table}
\centering
\begin{minipage}{84mm}
\caption{Summary of species that contribute at least 5\% to the total equilibrium cooling rate in the temperature range $10^{2} \, \text{K} \leq T \leq 10^{4} \, \text{K}$ at solar metallicity and a density $n_{\rm{H}} = 1 \, \text{cm}^{-3}$, in the presence of four different UV radiation fields: the \citet{black87} interstellar radiation field (ISRF) multiplied by a factor 10 \textit{(ISRF10)}, the \citet{black87} ISRF \textit{(ISRF)}, the redshift zero \citet{haardt01} extragalactic UV background \textit{(UVB)} and in the absence of UV radiation \textit{(None)}. The four sections in the table correspond to the four rows of figure~\ref{ISMcoolants}.}
\centering
\begin{tabular}{lcccc}
\hline
Species & $\log T_{\rm{min}}$\footnote{Minimum temperature at which the species contributes at least 5\% to the total equilibrium cooling rate.} & $\log T_{\rm{max}}$\footnote{Maximum temperature at which the species contributes at least 5\% to the total equilibrium cooling rate.} & $\log T_{\rm{peak}}$\footnote{Temperature at which the relative contribution of the species to the total equilibrium cooling rate is highest.} & Peak\footnote{Relative contribution of the species to the total equilibrium cooling rate at $T_{\rm{peak}}$.} \\
\hline
\multicolumn{5}{c}{\textbf{ISRF10}} \\
\hline
C\textsc{ii} & 2.0 & 3.1 & 2.0 & 44.3\% \\
H\textsc{ii} & 2.0 & 4.0 & 2.0 & 17.9\% \\
N\textsc{ii} & 2.0 & 2.5 & 2.0 & 15.8\% \\
Si\textsc{ii} & 2.0 & 3.7 & 2.5 & 32.9\% \\
S\textsc{iii} & 2.1 & 4.0 & 3.3 & 27.2\% \\
O\textsc{ii} & 3.9 & 4.0 & 4.0 & 7.5\% \\
O\textsc{iii} & 2.0 & 4.0 & 4.0 & 35.7\% \\
\hline
\multicolumn{5}{c}{\textbf{ISRF}} \\
\hline
C\textsc{ii} & 2.0 & 3.9 & 2.0 & 61.8\% \\
N\textsc{ii} & 2.0 & 4.0 & 2.0 & 13.4\% \\
Si\textsc{ii} & 2.0 & 4.0 & 2.7 & 42.4\% \\
Fe\textsc{ii} & 2.4 & 4.0 & 3.7 & 23.9\% \\
H\textsc{ii} & 2.0 & 4.0 & 3.7 & 15.7\% \\
S\textsc{iii} & 2.5 & 4.0 & 3.7 & 14.1\% \\
O\textsc{ii} & 3.8 & 4.0 & 4.0 & 22.4\% \\
S\textsc{ii} & 3.8 & 4.0 & 4.0 & 9.2\% \\
\hline
\multicolumn{5}{c}{\textbf{UVB}} \\
\hline
C\textsc{ii} & 2.0 & 3.9 & 2.0 & 72.5\% \\
Si\textsc{ii} & 2.0 & 4.0 & 2.7 & 54.9\% \\
Fe\textsc{ii} & 2.3 & 4.0 & 3.7 & 47.7\% \\
Ca\textsc{ii} & 3.8 & 4.0 & 3.9 & 7.7\% \\
S\textsc{ii} & 3.7 & 4.0 & 3.9 & 12.5\% \\
H\textsc{i} & 3.9 & 4.0 & 4.0 & 18.5\% \\
Mg\textsc{ii} & 3.9 & 4.0 & 4.0 & 8.4\% \\
\hline
\multicolumn{5}{c}{\textbf{None}} \\
\hline
C\textsc{i} & 2.0 & 4.0 & 2.0 & 60.6\% \\
Si\textsc{ii} & 2.0 & 3.5 & 2.3 & 15.4\% \\
H$_{2}$ & 2.0 & 4.0 & 3.7 & 70.2\% \\
O\textsc{i} & 2.0 & 4.0 & 3.9 & 30.7\% \\
Fe\textsc{ii} & 3.2 & 4.0 & 4.0 & 19.0\% \\
H\textsc{i} & 3.9 & 4.0 & 4.0 & 16.9\% \\
\hline
\vspace{-0.1in}
\label{dominantCoolantsTable}
\end{tabular}
\end{minipage}
\end{table}

Comparing our equilibrium abundances and cooling functions (solid lines) to \textsc{Cloudy} (dashed lines) in figure~\ref{ISMcoolants}, we find that they are in good agreement. Below 100 K the agreement between our model and \textsc{Cloudy} tends to be poorer for all UV radiation fields, but most noticeably in the absence of UV. This is partly because we include fewer molecular species in our network than \textsc{Cloudy} does. However, we see in figure~\ref{ISMcoolants} that many species are still in good agreement at these low temperatures.

Figure~\ref{ISMcoolants} only shows our results for a density $n_{\rm{H}} = 1 \, \text{cm}^{-3}$, but we also considered densities in the range $10^{-2} \leq n_{\rm{H}} \leq 10^{4} \, \text{cm}^{-3}$. At lower densities photoionisations and cosmic ray ionisations become relatively more important, as they scale linearly with density while two body collisional processes scale with the density squared. In the presence of a strong UV radiation field, higher ionisation species such as S\textsc{iv} and Fe\textsc{vi} contribute significantly to the cooling function in these cases, although heating is also relatively stronger. Conversely, at higher densities the importance of the UV radiation and cosmic rays becomes relatively weaker, with the cooling dominated by low ionisation state species such as O\textsc{i}, Si\textsc{ii} and Fe\textsc{ii}, along with molecular hydrogen. More plots comparing the ionisation fractions and cooling functions from our model and from \textsc{Cloudy} can be found on our website. 

\subsection{Molecular abundances}\label{ISMmolecules}

\begin{figure*}
\centering
\mbox{
	\includegraphics[width=135mm]{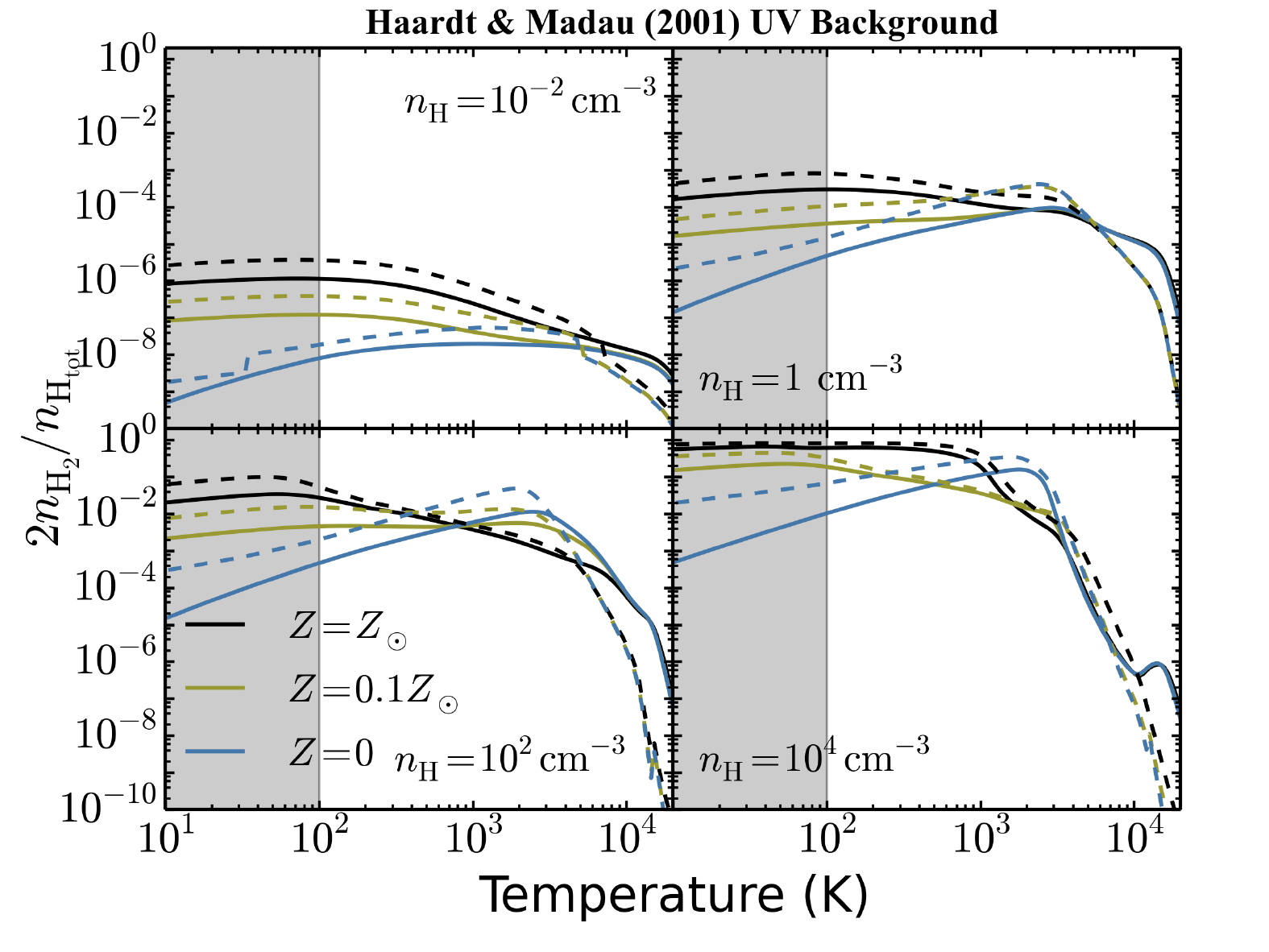}}
\mbox{
	\includegraphics[width=135mm]{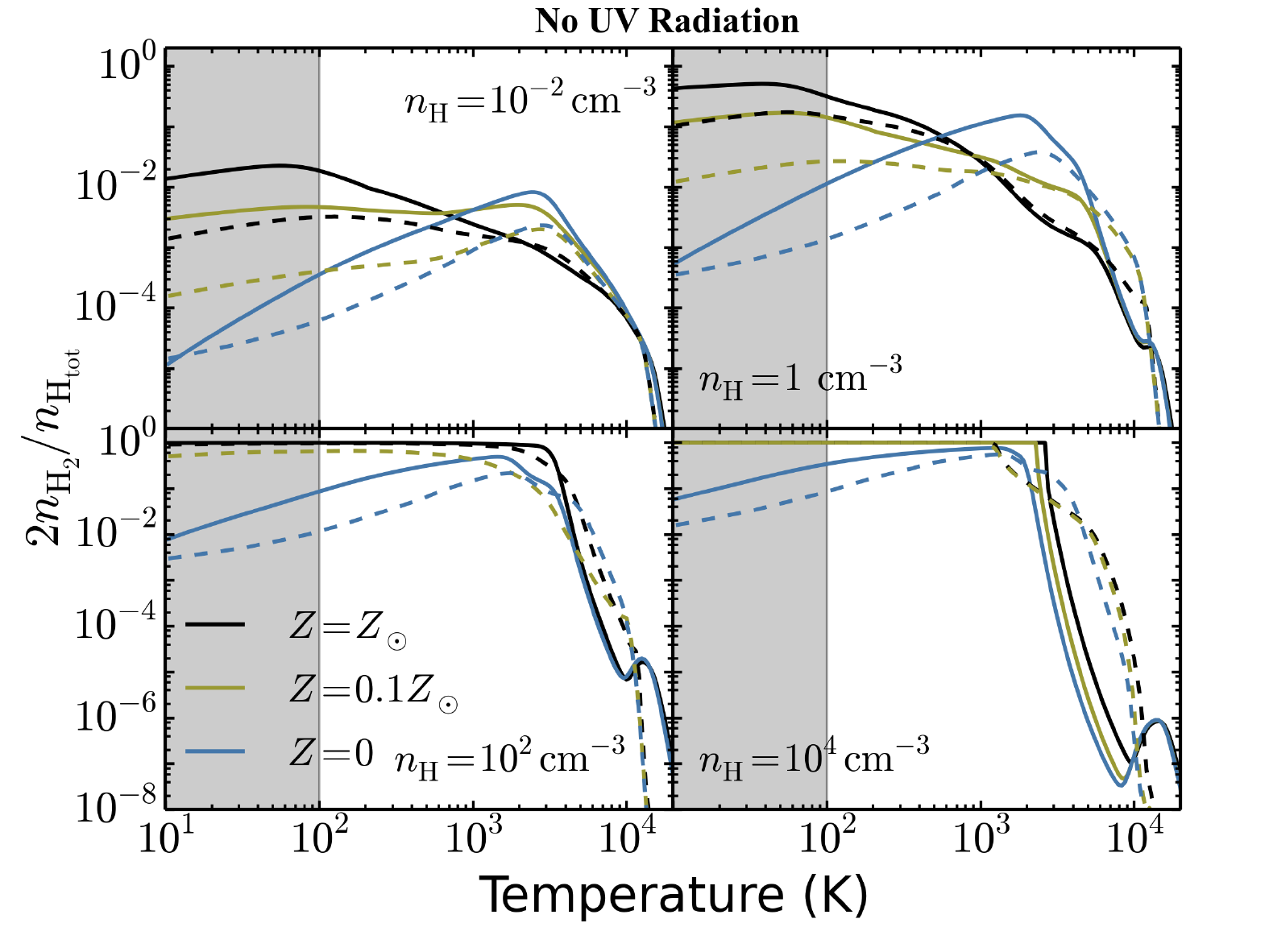}}
\caption{Equilibrium molecular fractions of H$_{2}$ predicted by our chemical model \textit{(solid lines)} and \textsc{Cloudy} \textit{(dashed lines)}. These were calculated in the presence of the redshift zero \citet{haardt01} extragalactic UV background \textit{(top four panels)} and in the absence of UV radiation \textit{(bottom four panels)} at solar metallicity and densities $10^{-2} \, \text{cm}^{-3} \leq n_{\rm{H}} \leq 10^{4} \, \text{cm}^{-3}$. The shaded grey region highlights temperatures below 100 K that are outside the range of temperatures that we are primarily interested in, but we include this regime here for completeness. In the presence of the UV background the greatest discrepancies are found in primordial gas \textit{(blue curves)}, and are mainly due to uncertainties in the formation rate of H$_{2}$ via H$^{-}$. In the absence of UV we tend to find higher molecular hydrogen abundances than \textsc{Cloudy}, primarily because we use a lower cosmic ray dissociation rate of H$_{2}$. See section~\ref{ISMmolecules} for a more detailed discussion.}
\label{H2_Zfig}
\end{figure*}

\begin{figure*}
\centering
\mbox{
	\includegraphics[width=145mm]{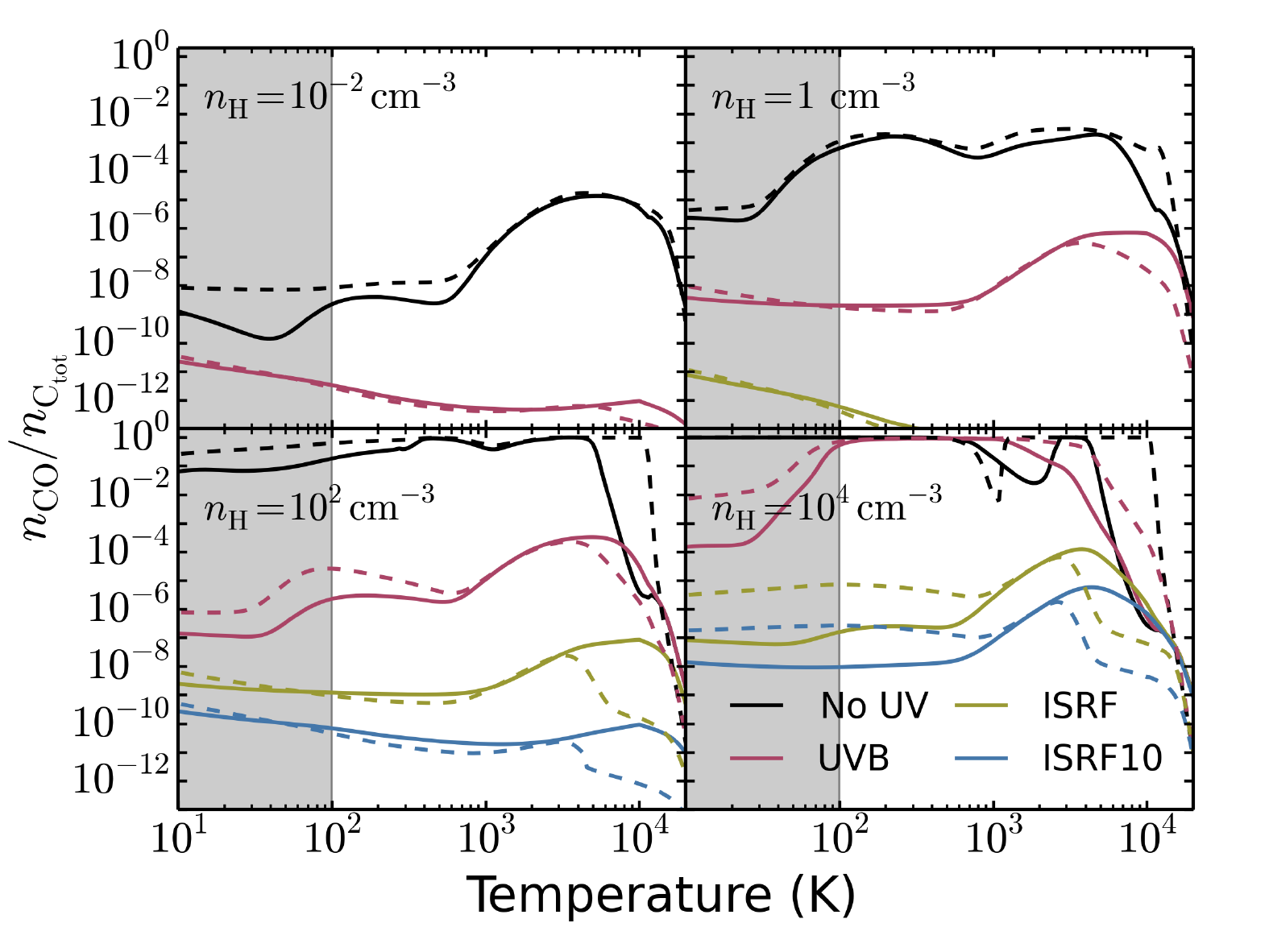}}
\caption{Equilibrium molecular fractions of CO predicted by our chemical model \textit{(solid lines)} and \textsc{Cloudy} \textit{(dashed lines)}. These were calculated at solar metallicity and for densities $10^{-2} \, \text{cm}^{-3} \leq n_{\rm{H}} \leq 10^{4} \, \text{cm}^{-3}$ in the absence of UV \textit{(black, No UV)}, in the presence of the redshift zero \citet{haardt01} extragalactic UV background \textit{(red, UVB)}, the \citet{black87} interstellar radiation field \textit{(yellow, ISRF)} and ten times this interstellar radiation field \textit{(blue, ISRF10)}. The shaded grey region highlights temperatures below 100 K that are outside the range of temperatures that we are primarily interested in, but we include this regime here for completeness. The agreement with \textsc{Cloudy} is good except in gas with both high density and high temperature, which is a combination that will be rare in nature.}
\label{CO_UVfig}
\end{figure*}

In this section we test the accuracy of our molecular network by comparing the equilibrium abundances of H$_{2}$ and CO calculated using our model to those from \textsc{Cloudy}. In figure~\ref{H2_Zfig} we plot the molecular fraction of H$_{2}$ as a function of temperature in the presence of the extragalactic UV background of \citet{haardt01} (top four panels) and in the absence of UV radiation (bottom four panels) for three different metallicities (different colours) and for four different densities (different panels). 

In the presence of the \citet{haardt01} UV background our molecular fractions at solar and $10\%$ solar metallicity match those from \textsc{Cloudy} well. However, we tend to predict somewhat smaller molecular fractions in primordial gas than \textsc{Cloudy}. This is due to uncertainties in the reaction rate of the associative detachment of H$^{-}$, which is the primary formation mechanism of H$_{2}$ at these densities when there is no dust present. \citet{glover06} have demonstrated how uncertainties in this reaction rate can lead to significant uncertainties in the abundance of molecular hydrogen. We have used the rate from \citet{bruhns10}, as given in the \textsc{umist} database, which was measured over a temperature range $30 \, \text{K} < T < 3000$ K and is in much better agreement with theoretical calculations than earlier experimental measurements such as \citet{schmeltekopf67}, \citet{fehsenfeld73} and \citet{martinez09}. However, \textsc{Cloudy} uses the rate coefficients from \citet{launay91}.

In the absence of UV radiation our molecular fractions follow similar trends with temperature as those from \textsc{Cloudy}, although we generally tend to produce higher H$_{2}$ abundances for $T \la 10^{3}$ K. This is primarily due to different dissociation rates of H$_{2}$ by cosmic rays used in \textsc{Cloudy} compared to our model. In particular, \textsc{Cloudy} includes the cosmic ray induced photodissociation of H$_{2}$, which is an order of magnitude higher than the cosmic ray dissociation that we include in our model based on the rates in the \textsc{umist} database.

We find similar trends in the presence of stronger UV radiation fields, although the H$_{2}$ abundances are small in these examples, so we do not show them here.

In figure~\ref{CO_UVfig} we show the molecular fraction of CO as a function of temperature at solar metallicity for four different densities (different panels) and four different UV radiation fields (different colours). We see that in the presence of the two strongest UV radiation fields considered here the CO fraction (i.e. the fraction of carbon that is in CO) remains low ($< 10^{-4}$) even at the highest density shown in figure~\ref{CO_UVfig} ($n_{\rm{H_{tot}}} = 10^{4} \, \text{cm}^{-3}$). In the presence of the \citet{haardt01} extragalactic UV background the CO fraction reaches unity at $n_{\rm{H_{tot}}} \sim 10^{4} \, \text{cm}^{-3}$ for temperatures $10^{2} \, \text{K} \la T \la 10^{3} \, \text{K}$, while in the absence of UV the CO fraction is close to one at these temperatures down to even lower densities, $n_{\rm{H_{tot}}} \sim 10^{2} \, \text{cm}^{-3}$. 

The CO fractions predicted by our model agree well with \textsc{Cloudy}, except at a density $n_{\rm{H}} \sim 10^{4} \, \text{cm}^{-3}$ and temperatures above a few thousand Kelvin where we predict a lower CO abundance than \textsc{Cloudy}. However, since it is unlikely that gas at such high densities would maintain temperatures of a few thousand Kelvin, this is not a problem for us.

\subsection{Impact of cosmic rays on ionisation balance and cooling}

Once the gas becomes shielded from the UV radiation field, its ionisation balance is typically dominated by cosmic ray ionisation at low temperatures. However, the cosmic ray ionisation rate in the ISM is highly uncertain. By modelling the chemistry of 23 low mass molecular cores to reproduce observed molecular ion abundances, \citet{williams98} infer a cosmic ray ionisation rate for H$_{2}$ of $\zeta_{\rm{H_{2}}} = 5 \times 10^{-17} \text{s}^{-1}$, corresponding to an H\textsc{i} ionisation rate from cosmic rays of $\zeta_{\rm{HI}} = 2.5 \times 10^{-17} \text{s}^{-1}$ (this is the default value that we use in our models). However, other authors have measured very different values for the galactic cosmic ray background. For example, \citet{indriolo07} use observations of H$_{3}^{+}$ absorption along twenty galactic sight lines to obtain a cosmic ray ionisation rate $\zeta_{\rm{HI}} = 2 \times 10^{-16} \text{s}^{-1}$, while \citet{mccall03} derive an even higher value of $\zeta_{\rm{HI}} = 1.2 \times 10^{-15} \text{s}^{-1}$ based on laboratory measurements and H$_{3}^{+}$ observations towards $\zeta$ Persei. Furthermore, the cosmic ray ionisation rate may also depend on the galactic environment, for example \citet{suchkov93} estimate a much higher density of cosmic rays in the starburst galaxy M82.

\begin{figure*}
\centering
\mbox{
	\includegraphics[width=168mm]{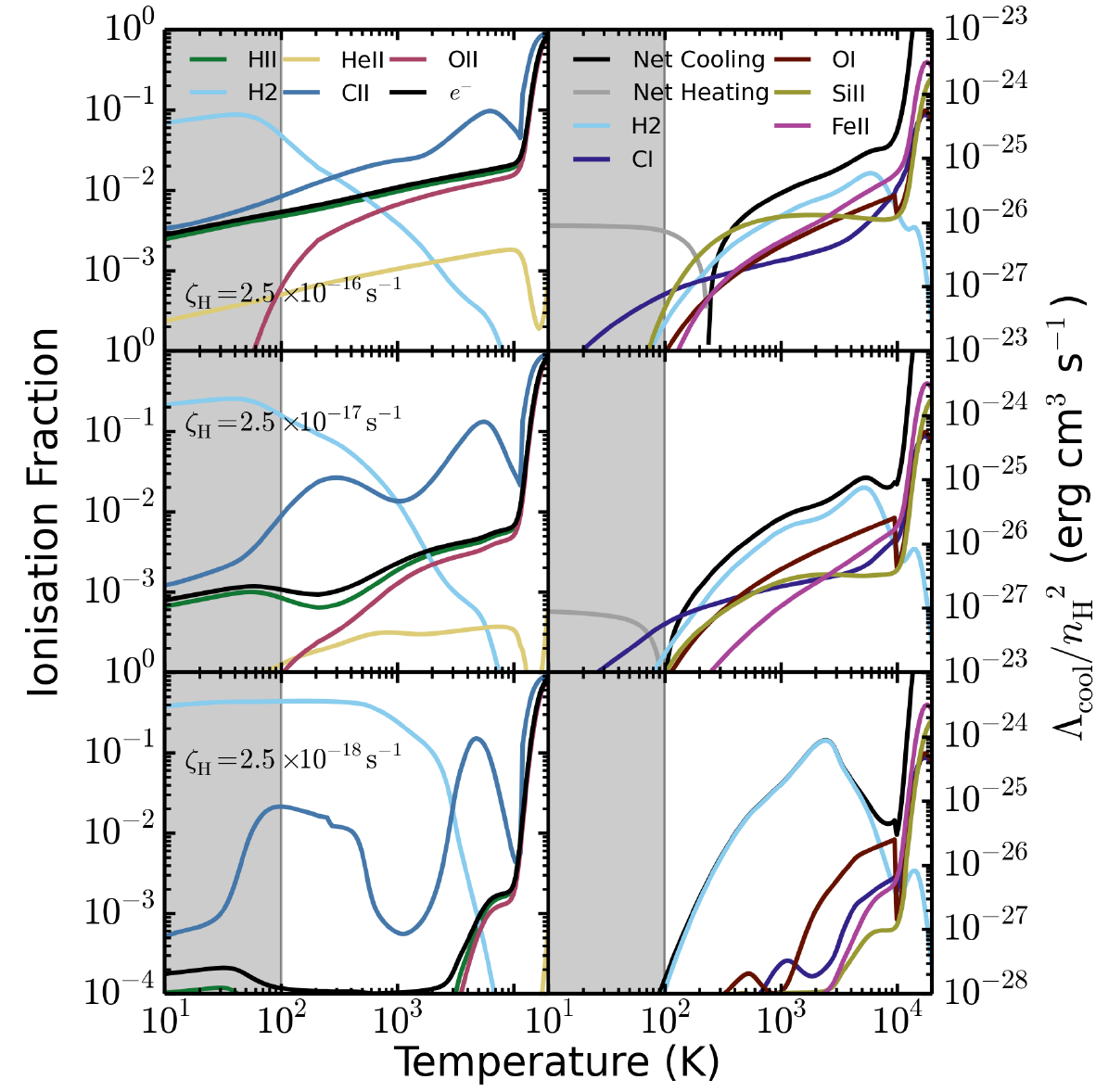}}
\caption{Comparison of the equilibrium ionisation fractions \textit{(left)} and cooling functions \textit{(right)} for three different values of the cosmic ray ionisation rate of hydrogen: $2.5 \times 10^{-16} s^{-1}$ \textit{(top row)}, $2.5 \times 10^{-17} s^{-1}$ \textit{(middle row)} and $2.5 \times 10^{-18} s^{-1}$ \textit{(bottom row)}. These were calculated using our chemical model at solar metallicity and a density $n_{\rm{H}} = 1 \, \text{cm}^{-3}$ in the absence of UV radiation. The shaded grey region highlights temperatures below 100 K that are outside the range of temperatures that we are primarily interested in, but we include this regime here for completeness. The cosmic ray ionisation rate, which is uncertain and may vary with environment, has a significant effect on the abundances and cooling of self shielded gas.}
\label{CRcompfig}
\end{figure*} 

Figure~\ref{CRcompfig} shows the impact of increasing or decreasing our default cosmic ray ionisation rate by a factor of ten on the ionisation balance and cooling function of fully shielded gas at solar metallicity and a density $n_{\rm{H}} = 1 \, \text{cm}^{-3}$. In the left panels we see that the ionisation fractions of singly ionised helium, carbon and oxygen as well as hydrogen, and hence the electron abundance, change dramatically as we vary the cosmic ray ionisation rate, confirming that, for our default value and higher, cosmic rays do indeed dominate the ionisation balance in this fully shielded regime. In the right hand panels we see that the cooling function is also significantly affected, as cosmic rays are the primary source of heating in fully shielded gas at these temperatures and densities, so increasing the cosmic ray ionisation rate by a factor of ten above our default value increases the thermal equilibrium temperature from around 100 K to 300 K. Furthermore, when we decrease it by a factor of ten below our default value, we find that the cooling rate from molecular hydrogen increases and more strongly dominates the cooling function.

\section{Cooling in non-equilibrium interstellar gas}\label{NonEqSect}

\begin{figure*}
\centering
\mbox{
	\includegraphics[width=168mm]{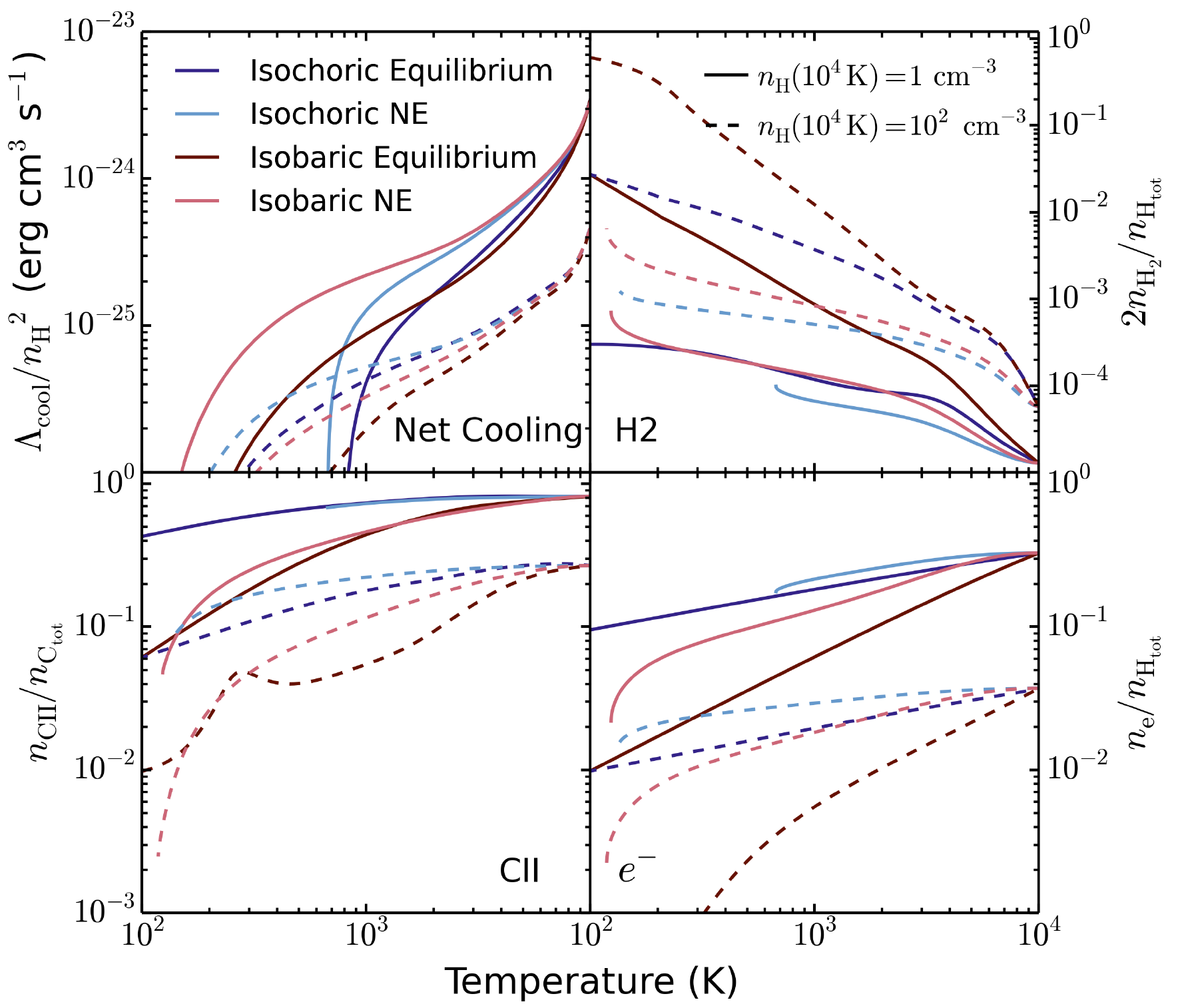}}
\caption{Non-equilibrium cooling functions \textit{(top left panel)} and abundances of H$_{2}$ \textit{(top right)}, C\textsc{ii} \textit{(bottom left)} and electrons \textit{(bottom right)} in gas cooling isochorically or isobarically from $10^{4}$ K, compared to their values in chemical equilibrium. These were calculated using our chemical model in the presence of the redshift zero \citet{haardt01} extragalactic UV background at solar metallicity and densities at $T = 10^{4}$ K of $n_{\rm{H}} (10^{4} K) = 1 \, \text{cm}^{-3}$ \textit{(solid lines)} and $n_{\rm{H}} (10^{4} K) = 10^{2} \, \text{cm}^{-3}$ \textit{(dashed lines)}.}
\label{figNonEqCool}
\end{figure*}  

The cooling functions presented in the previous section were calculated in chemical equilibrium. We now investigate what impact non-equilibrium chemistry can have on the cooling rates of interstellar gas. We consider solar metallicity gas that is cooling either isochorically or isobarically in the presence of the \citet{haardt01} extragalactic UV background for different densities. Isochoric cooling is relevant to applications in hydrodynamic simulations, in which cooling over individual hydrodynamic timesteps is typically done at fixed density. Isochoric cooling is also relevant for rapidly cooling gas where the cooling time is shorter than the dynamical or sound crossing times for self gravitating or externally confined gas respectively. Isobaric cooling is relevant more generally in interstellar gas as the different phases of the ISM are usually in pressure equilibrium, so a parcel of gas that is not subject to heating events such as shocks or stellar feedback would tend to cool at constant pressure. 

We start these calculations from chemical equilibrium and consider two different initial temperatures. Starting from $T = 10^{4}$ K corresponds to gas that is cooling from the warm phase of the ISM to the cold phase, while starting from $T = 10^{6}$ K corresponds to gas that has been strongly heated, for example by a supernova event, and is subsequently cooling back towards the cold phase of the ISM. We compare the resulting net radiative cooling rates of the gas with the equilibrium cooling functions in sections~\ref{LowT_NonEqSect} and \ref{HiT_NonEqSect} below.

\subsection{Cooling from $T = 10^{4}$ K}\label{LowT_NonEqSect} 

\begin{figure}
\centering
\mbox{
	\includegraphics[width=84mm]{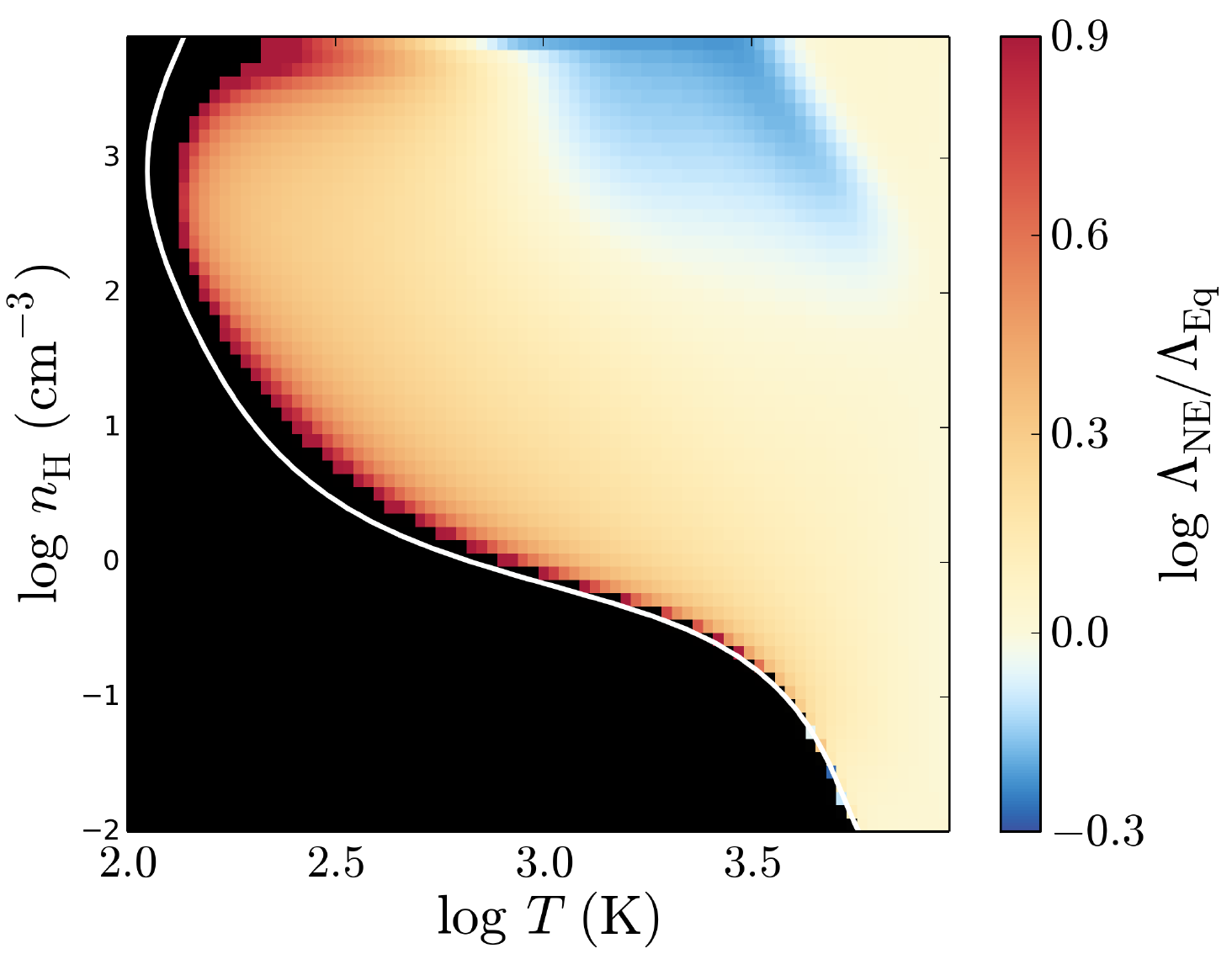}}
\caption{The ratio of the non-equilibrium cooling rate of gas cooling isochorically from chemical equilibrium at $10^{4}$ K, $\Lambda_{\rm{NE}}$, to the cooling rate in chemical equilibrium, $\Lambda_{\rm{Eq}}$, plotted against temperature and density. These were calculated for a solar metallicity gas in the presence of the redshift zero \citet{haardt01} extragalactic UV background. Blue regions indicate where non-equilibrium effects suppress the cooling rate, while red and yellow regions indicate where they enhance the cooling rate. The black region indicates where there is net heating in chemical equilibrium. The white curve shows the minimum temperature to which the isochorically cooling gas is able to cool to before there is net heating in the non-equilibrium cooling function. This is generally below the thermal equilibrium temperature. The non-equilibrium and equilibrium rates differ by up to an order of magnitude.}
\label{figNonEqGrid}
\end{figure}  

The non-equilibrium cooling functions starting from $T = 10^{4}$ K are compared with their corresponding equilibrium cooling functions in the top left panel of figure~\ref{figNonEqCool}, while the remaining panels compare the abundances of H$_{2}$, C\textsc{ii} and electrons. We see that the non-equilibrium gas tends to have higher net cooling rates than in equilibrium. This is because the electron density is decreasing as the gas cools, but it takes a finite time for the electrons to recombine with ions. Thus, if the recombination time is long compared to the cooling time, there will be a recombination lag that results in a higher electron density at a given temperature than we would expect in equilibrium, which is what we see in the bottom right panel of figure~\ref{figNonEqCool}. This recombination lag enhances the radiative cooling rate, which is typically driven by electron-ion collisions. Furthermore, the photoheating rate depends on the neutral fraction, so the recombination lag will also suppress the photoheating of the gas. 

One exception to the above trend can be seen in isochorically cooling gas at a density $n_{\rm{H}} = 10^{2} \, \text{cm}^{-3}$ (the blue dashed curves in figure~\ref{figNonEqCool}). In this example the cooling rate around a temperature $T \sim 4000$ K is slightly lower than in equilibrium. At this temperature the contribution to the equilibrium net cooling rate from molecular hydrogen is comparable to that from Si\textsc{ii} and Fe\textsc{ii}, the other dominant coolants. However, the formation time for molecular hydrogen is long compared to the cooling time-scale here, which results in a lower H$_{2}$ abundance, as can be seen in the top right panel. Hence its contribution to the cooling rate is suppressed in the non-equilibrium cooling function.

The non-equilibrium cooling functions are generally able to cool to lower temperatures than their corresponding equilibrium cooling functions. For example, in the isochoric cooling functions at a density $n_{\rm{H}} = 1 \, \text{cm}^{-3}$ (blue solid curves in the top left panel of figure~\ref{figNonEqCool}), the equilibrium net cooling rate becomes negative (net heating) at $T \sim 850$ K, whereas the isochorically cooling gas cools to $T \sim 700$ K. Thus non-equilibrium effects enable an isochorically or isobarically cooling gas to cool below the thermal equilibrium temperature, before heating back up towards its final thermal and chemical equilibrium. In figure~\ref{figNonEqCool} we only plot the non-equilibrium rates and abundances down to the temperature at which the gas begins to heat up again. 

Furthermore, for some species, such as C\textsc{ii}, the abundances in the examples presented here tend to remain close to equilibrium. For other species, such as H$_{2}$, the abundances in isochorically or isobarically cooling gas can differ by up to an order of magnitude from their equilibrium values. 

To illustrate the density dependence of the non-equilibrium effects on the cooling rate, we show in figure~\ref{figNonEqGrid} the ratio of the non-equilibrium cooling rate of gas cooling isochorically from chemical equilibrium at $10^{4}$ K to the cooling rate in chemical equilibrium, plotted against temperature and density. These were calculated for solar metallicity in the presence of the redshift zero \citet{haardt01} extragalactic UV background (more examples can be found on our website). Blue regions in this plot indicate where the non-equilibrium effects suppress the cooling rate, while red and yellow regions indicate where they enhance the cooling rate. The black region shows where the equilibrium net cooling rate is negative (i.e. there is net heating). The white curve shows the minimum temperature to which the isochorically cooling gas is able to cool before there is net heating in the non-equilibrium cooling function. We see that the gas is generally able to cool below the thermal equilibrium temperature. 

For most densities in figure~\ref{figNonEqGrid}, the enhancement of the cooling rate in non-equilibrium compared to the equilibrium cooling rate increases smoothly from the initial temperature of $10^{4}$ K down to the thermal equilibrium temperature. This enhancement can reach up to a factor $\sim 10$. However, at high densities ($n_{\rm{H}} \ga 10^{2} \, \text{cm}^{-3}$), the cooling rate in non-equilibrium is suppressed at temperatures around a few thousand Kelvin. This corresponds to the region where molecular hydrogen dominates the cooling rate. This figure highlights in which regions of the density-temperature plane these two opposing non-equilibrium effects are important. 

\subsection{Cooling from $T = 10^{6}$ K}\label{HiT_NonEqSect}

\begin{figure*}
\centering
\mbox{
	\includegraphics[width=168mm]{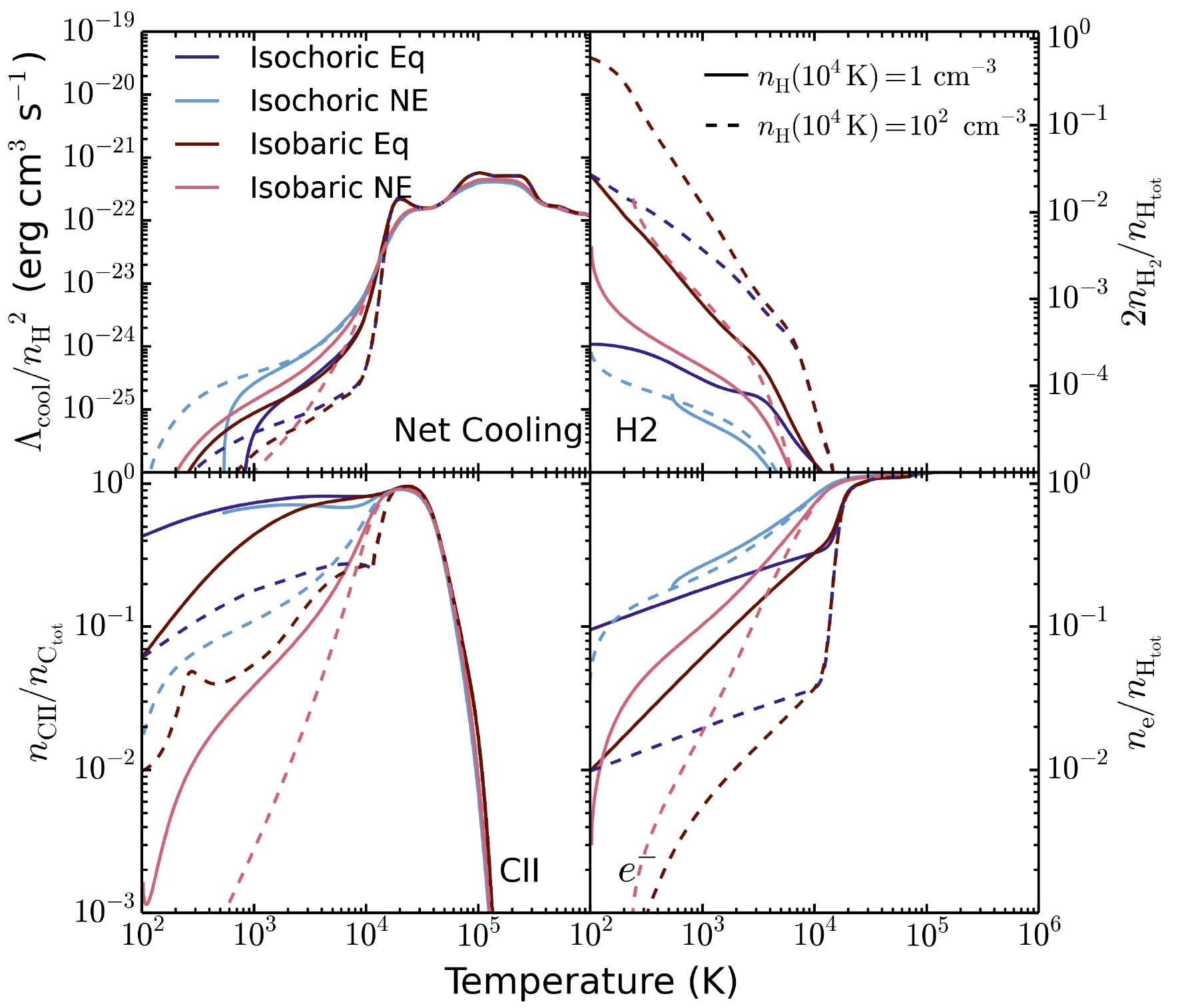}}
\caption{As figure~\ref{figNonEqCool}, but for gas cooling from $10^{6}$ K. Note that for the isobaric cooling curves, the densities given in the legend are valid at $T = 10^{4}$ K.}
\label{figNonEqCoolHiT}
\end{figure*} 

In the top left panel of figure~\ref{figNonEqCoolHiT} we compare the non-equilibrium cooling functions starting from $T = 10^{6}$ K to their corresponding equilibrium cooling functions, while the abundances of H$_{2}$, C\textsc{ii} and electrons are shown in the remaining panels. At high temperatures ($> 10^{4}$ K) we find that non-equilibrium effects tend to decrease the net cooling rates. This agrees with what \citet{oppenheimer13a} found for gas typical of the intergalactic medium, and is caused by the recombination lag allowing species in higher ionisation states to persist down to lower temperatures. The electronic transitions in these high ionisation states are more energetic compared to those in the lower ionisation states, and thus they are more difficult to excite at these lower temperatures. Hence the cooling rates from these more highly ionised species are smaller, and thus the net cooling rate is suppressed with respect to the cooling rate in chemical equilibrium. 

Once the gas temperature drops below $10^{4}$ K, the non-equilibrium cooling rates generally become higher than the equilibrium cooling functions, similar to the trends we saw in figure~\ref{figNonEqCool}. The most significant differences that we see compared to the previous figure are for gas at a density $n_{\rm{H}} = 10^{2} \, \text{cm}^{-3}$. In this example the isochoric non-equilibrium cooling curve in figure~\ref{figNonEqCoolHiT} shows significantly higher cooling rates than in figure~\ref{figNonEqCool}. This is because recombination lags induced as the gas cooled from $10^{6}$ K to $10^{4}$ K still persist and thus the electron density is even higher, as can be seen in the bottom right panel of figure~\ref{figNonEqCoolHiT}, which further enhances the metal line cooling. The isobarically cooling gas at $n_{\rm{H}} = 10^{2} \, \text{cm}^{-3}$ also shows an enhanced net cooling rate just below $10^{4}$ K compared to figure~\ref{figNonEqCool}, for the same reason. However, below $T \sim 2000$ K the cooling rate drops below the corresponding isobaric equilibrium cooling function. This is because the cooling is dominated by H$_{2}$ here, which is suppressed in the non-equilibrium cooling curve due to the long formation time-scale of molecular hydrogen, as can be seen from the top right panel. 

In the bottom left panel of figure~\ref{figNonEqCoolHiT}, the non-equilibrium abundances of C\textsc{ii} remain very close to their equilibrium values down to the peak at $T \sim 2 \times 10^{4}$ K. However, below this temperature the C\textsc{ii} abundance of the cooling gas tends to fall below the equilibrium abundance, by up to an order of magnitude. We saw in figure~\ref{figNonEqCool} that C\textsc{ii} exhibited only a small recombination lag when the gas starts cooling from $10^{4}$ K, which indicates that it has a relatively short recombination time compared to, for example, H\textsc{ii}. Furthermore, the bottom right panel of figure~\ref{figNonEqCoolHiT} demonstrates that the electron abundance below $10^{4}$ K is even higher in this example than in gas that starts cooling from $10^{4}$ K, due to recombination lags of other species that persist from higher temperatures. This suggests that, as the gas cools, the C\textsc{ii} rapidly recombines to a lower abundance than in equilibrium due to the enhanced electron abundance. 

\begin{figure}
\centering
\mbox{
	\includegraphics[width=84mm]{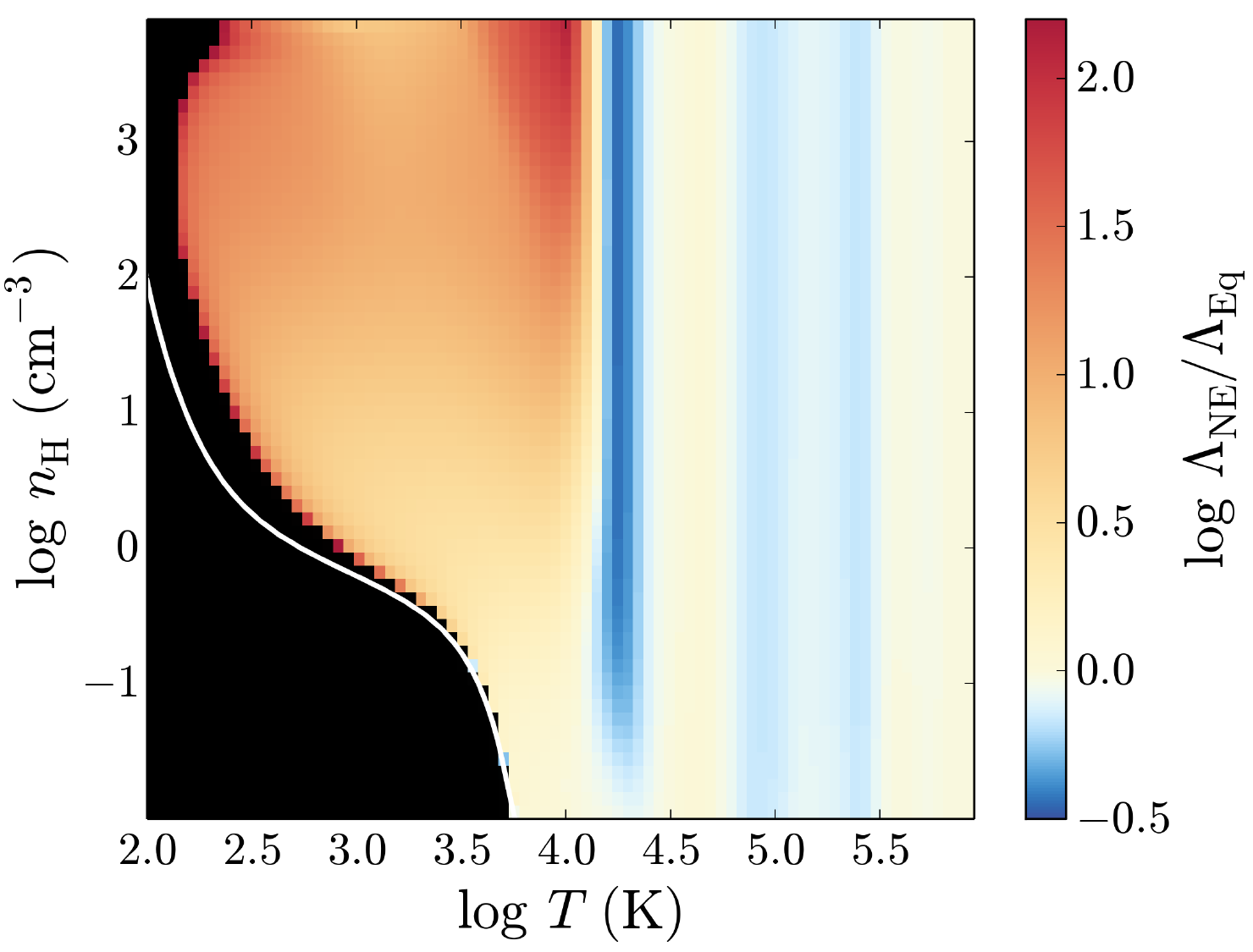}}
\caption{As figure~\ref{figNonEqGrid}, but for gas cooling isochorically from $10^{6}$ K. The non-equilibrium rate above $10^{4}$ K is generally suppressed compared to the equilibrium cooling rate, whereas below 10$^{4}$ K the enhancement of the cooling rate in non-equilibrium can exceed two orders of magnitude and is thus even greater than we found for gas cooling from $10^{4}$ K in figure~\ref{figNonEqGrid}.}
\label{figNonEqGridHiT}
\end{figure}

In figure~\ref{figNonEqGridHiT} we show the ratio of the non-equilibrium cooling rate of gas cooling isochorically from chemical equilibrium at $10^{6}$ K to the cooling rate in chemical equilibrium, plotted against temperature and density, for solar metallicity gas in the presence of the \citet{haardt01} extragalactic UV background. At temperatures $T > 10^{4}$ K, we see that the non-equilibrium cooling rate is suppressed compared to the equilibrium cooling rate, for the reasons discussed above. This suppression becomes approximately independent of density for $n_{\rm{H}} \ga 0.1 \, \text{cm}^{-3}$. Below $10^{4}$ K, the non-equilibrium cooling rate is enhanced due to the higher electron abundance caused by recombination lags. This enhancement is even greater than we saw in figure~\ref{figNonEqGrid} for gas that starts cooling from $10^{4}$ K, with an increase in the non-equilibrium cooling rate of up to a factor $\sim 100$ compared to in equilibrium. Furthermore, this enhancement outweighs the suppression of the non-equilibrium cooling rate at $n_{\rm{H}} \ga 10^{2} \, \text{cm}^{-3}$ and temperatures around a few thousand Kelvin that is caused by the long formation time-scale of molecular hydrogen. 

\section{Conclusions}\label{conclusions}

We have presented a model to follow the non-equilibrium thermal and chemical evolution of interstellar gas that is designed to be incorporated into hydrodynamic simulations of galaxy formation. Our chemical network includes 157 species: the molecules H$_{2}$, H$_{2}^{+}$, H$_{3}^{+}$, OH, H${_2}$O, C$_{2}$, O$_{2}$, HCO$^{+}$, CH, CH$_{2}$, CH$_{3}^{+}$, CO, CH$^{+}$, CH$_{2}^{+}$, OH$^{+}$, H$_{2}$O$^{+}$, H$_{3}$O$^{+}$, CO$^{+}$, HOC$^{+}$ and O$_{2}^{+}$, along with electrons and all ionisations states of the 11 elements that dominate gas cooling (H, He, C, N, O, Ne, Mg, Si, S, Ca and Fe). Our model includes chemical reactions on dust grains, most importantly H$_{2}$ formation, and thermal processes involving dust, such as photoelectric heating (but not yet thermal dust emission). We also include photochemical reactions and cosmic ray ionisations, along with their associated heating rates. 

While we do not include metal depletion on dust grains in our model, as the depletion factors of metals are highly uncertain, we show that the depletion factors measured by \citet{jenkins09} could potentially have a large effect on the cooling function of interstellar gas (see figure~\ref{depletionsFig}).

We have compared the equilibrium ionisation balance and cooling functions predicted by our model with the photoionisation code \textsc{Cloudy} for a range of densities $10^{-2} \leq n_{\rm{H}} \leq 10^{4} \, \text{cm}^{-3}$, temperatures $10 \, \text{K} \leq T \leq 2 \times 10^{4} \, \text{K}$, metallicities $0 \leq Z \leq Z_{\odot}$ and in the presence of four UV radiation fields: the \citet{black87} interstellar radiation field, ten times this interstellar radiation field, the \citet{haardt01} extragalactic UV background and finally in the absence of UV radiation. These physical parameters include the typical conditions in the diffuse ISM that we are interested in, and also extend down to colder gas. The equilibrium abundances and cooling functions predicted by our model generally agree very well with those from \textsc{Cloudy} (see figures~\ref{ISMcoolants}, \ref{H2_Zfig} and \ref{CO_UVfig}), and most remaining differences are due to different choices for uncertain rates. Additional figures showing all of these comparisons can be found on our website~\footnote{\url{http://noneqism.strw.leidenuniv.nl}}.

In our equilibrium cooling functions we find that the cooling rate of interstellar gas is typically dominated by C\textsc{ii}, Si\textsc{ii} and Fe\textsc{ii}, with O\textsc{i} becoming important at high densities ($n_{\rm{H_{tot}}} \ga 10^{2} \, \text{cm}^{-3}$) and when the UV radiation field is weak (for example the redshift zero \citealt{haardt01} extragalactic UV background). We also find that at these high densities and low UV radiation intensities molecular hydrogen dominates the cooling function, typically peaking around $1000 - 2000$ K. See table~\ref{dominantCoolantsTable} for a summary of the dominant coolants at solar metallicity. In primordial gas, molecular hydrogen is the only significant source of cooling below $T \sim 8000$ K in our model. 

We note that \citet{glover07} also investigated which species contribute significantly to the cooling rate of interstellar gas in order to determine which ions to include in their chemical network. They looked at a similar range of physical parameters (although only at a metallicity of $0.1 Z_{\odot}$). We generally agree with their conclusions, although they found that Fe\textsc{ii} only contributed at most $\sim 10\% - 20\%$ to the cooling rate, whereas we find that it is one of the most important coolants in several cases (see for example figure~\ref{ISMcoolants}). 

We also calculate the non-equilibrium abundances and cooling functions of gas that is cooling isochorically or isobarically. We find that non-equilibrium chemistry tends to increase the cooling rates below $10^{4}$ K compared to the cooling rates in chemical (including ionisation) equilibrium, due to a recombination lag that results in a higher electron density at a given temperature (see figures~\ref{figNonEqCool} and \ref{figNonEqGrid}). We find enhancements in the cooling rate of up to an order of magnitude for gas that starts cooling from $10^{4}$ K. This enables the gas to cool below its thermal equilibrium temperature, before subsequently heating back up to thermal and chemical equilibrium. 

The abundance of molecular hydrogen in isochorically or isobarically cooling gas is lower than in equilibrium, by up to an order of magnitude, due to its relatively long formation time-scale compared to the cooling time. This can reduce the cooling rate if H$_{2}$ dominates the equilibrium cooling function, i.e. for high density ($n_{\rm{H_{tot}}} \ga 10^{2} \, \text{cm}^{-3}$) gas at temperatures around a few thousand Kelvin. 

Non-equilibrium effects above $10^{4}$ K generally suppress the cooling rates as higher ionisation species persist to lower temperatures. Such species have lower cooling rates as their electronic transitions are more energetic and thus are more difficult to excite at these lower temperatures. Furthermore, if we consider gas that starts cooling from a temperature $10^{6}$ K, the resulting recombination lag results in a higher electron density once it reaches 10$^{4}$ K compared to equilibrium. Thus the enhancement of the net cooling rate below 10$^{4}$ K due to non-equilibrium effects will be greater if the gas starts cooling from $T \gg 10^{4}$ K than if we start in equilibrium at $\sim 10^{4}$ K. For gas cooling from $10^{6}$ K, we find enhancements of up to two orders of magnitude (see figure~\ref{figNonEqGridHiT}).

Throughout this paper we have focussed on photoionised gas in the optically thin regime, although we also considered a case without any UV radiation. However, gas that is in the cold phase of the ISM, in particular molecular clouds, will become shielded from the UV radiation field, both by dust and by the gas itself, and hence the radiation field will vary with the depth into the cloud. In paper II we shall describe the methods that we use to calculate the attenuated photochemical rates in shielded gas and present the resulting chemistry and cooling properties under such conditions.

\section*{Acknowledgments}
We thank Andreas Pawlik and Ewine van Dishoeck for useful discussions. We gratefully acknowledge support from Marie Curie Training Network CosmoComp (PITN-GA-2009- 238356) and from the European Research Council under the European Union's Seventh Framework Programme (FP7/2007-2013) / ERC Grant agreement 278594-GasAroundGalaxies. 

{}

\appendix

\section{Dust temperature}\label{dustPhysicsAppendix}

\begin{figure}
\centering
\mbox{
	\includegraphics[width=84mm]{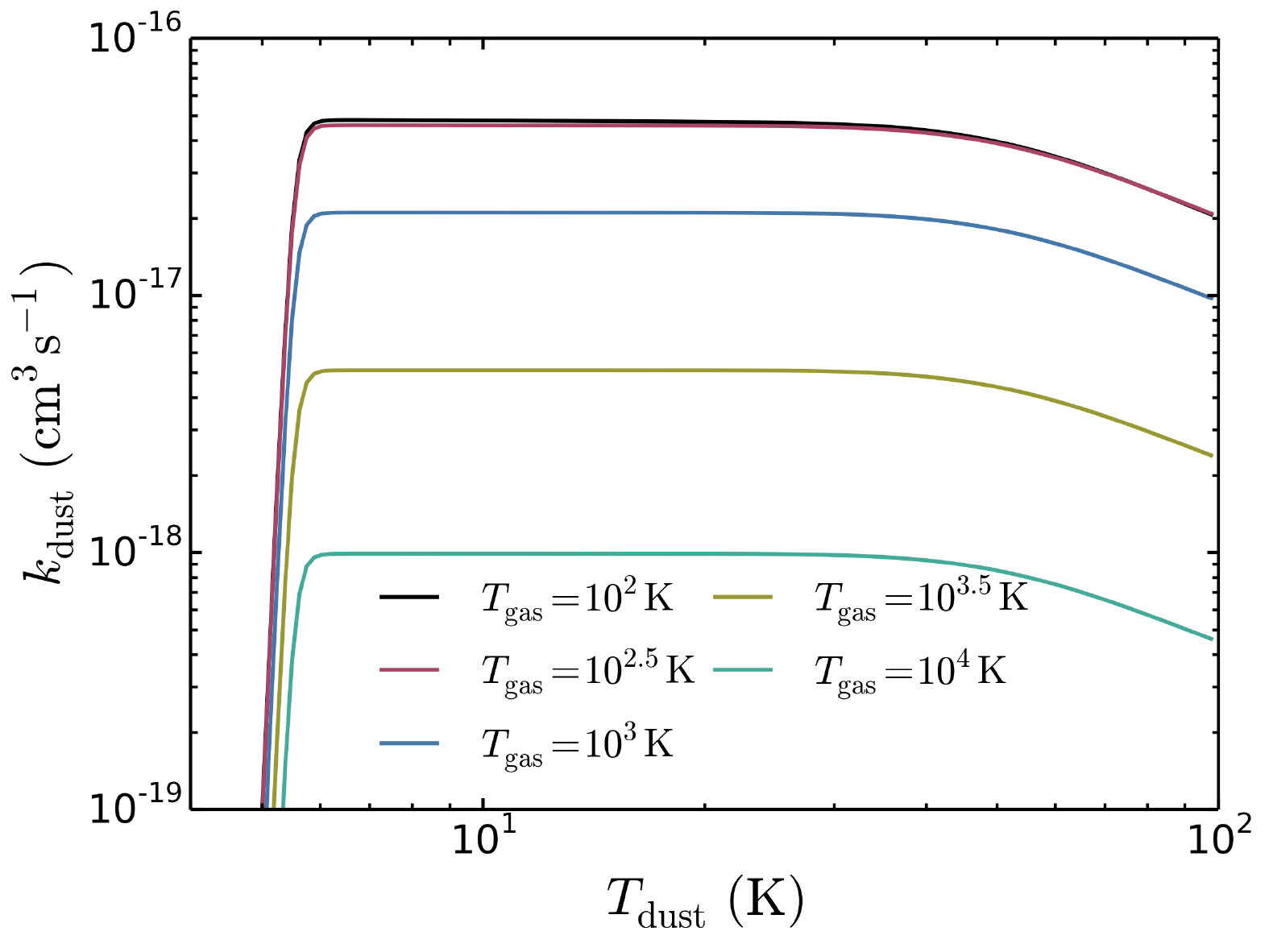}}
\caption{Rate coefficient for the formation of molecular hydrogen on dust grains, calculated from equation~\ref{H2_dust_rate}, plotted against dust temperature for different gas temperatures.}
\label{dust_H2_rate_fig}
\end{figure}

\begin{figure*}
\centering
\mbox{
	\includegraphics[width=168mm]{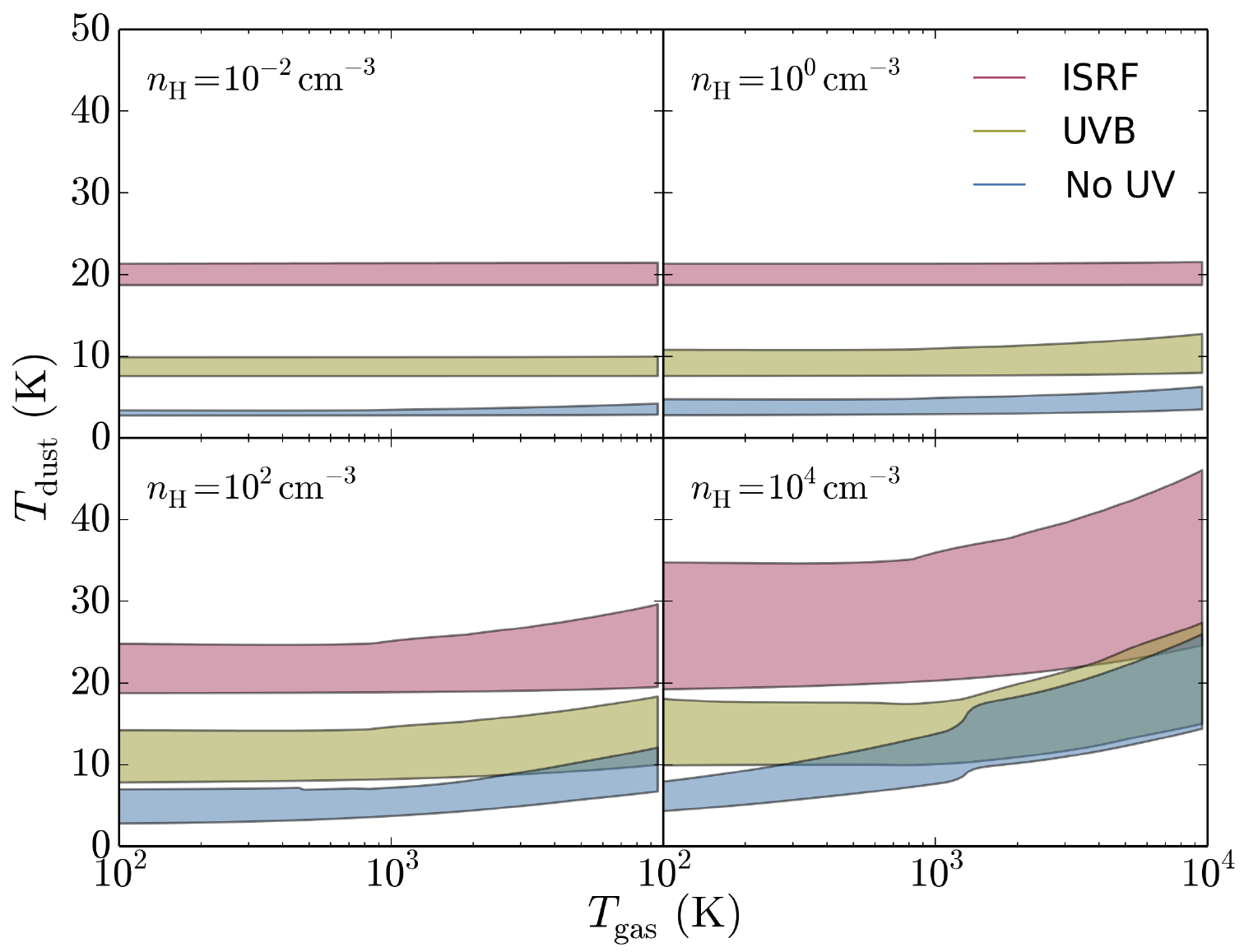}}
\caption{The range of (carbonaceous) dust grain temperatures calculated by \textsc{Cloudy} across the 10 grain size bins, plotted against gas temperature in the presence of the \citet{black87} interstellar radiation field \textit{(red)}, the redshift zero \citet{haardt01} extragalactic UV background \textit{(yellow)} and in the absence of UV radiation \textit{(blue)}.}
\label{dust_temp_fig}
\end{figure*}

The rate of formation of molecular hydrogen on dust grains depends on the temperature of the dust as well as the gas, so we need to make an assumption about the grain temperature. \citet{glover12} calculate the dust temperature by assuming that it is in thermal equilibrium and hence solving the thermal balance equation (their equation A2), but this adds to the computational cost of the model. To investigate the impact that dust temperature has on the molecular hydrogen abundance in our model we plot in figure~\ref{dust_H2_rate_fig} the rate coefficient $k_{\rm{dust}}$ (obtained by dividing equation~\ref{H2_dust_rate} by $n_{\rm{HI}}$) as a function of the dust temperature for different gas temperatures in the regime relevant to the diffuse ISM ($10^{2}$ K to $10^{4}$ K). Below $T_{\rm{dust}} \approx 6$ K the rate falls very rapidly, but we would not expect to find such low dust temperatures in the diffuse ISM. At higher dust temperatures the rate of molecular hydrogen formation on dust grains remains fairly constant up to $50$ K, at which point it begins to decrease again, falling by a factor of just over 2 by $T_{\rm{dust}} = 100$ K. In figure~\ref{dust_temp_fig} we compare the range of dust temperatures calculated by \textsc{Cloudy} using ten grain size bins (from 0.005 to 0.25$\mu$m, assuming the power-law size distribution of \citealt{mathis77}) as a function of the gas temperature, in the presence of different radiation fields. These were calculated for carbonaceous grains, but we found that the temperatures of silicate grains are similar. Even in the presence of the interstellar radiation field of \citet{black87} the highest dust temperatures in \textsc{Cloudy} only just reach $50$ K, and even this temperature is only reached for gas temperatures near $10^{4}$K, where molecular hydrogen becomes unimportant. We therefore would not expect the rate of molecular hydrogen formation on dust grains to be significantly affected by variations in the dust temperature in the range of physical conditions that we are interested in here, i.e. those relevant to the diffuse ISM.

\begin{figure}
\centering
\mbox{
	\includegraphics[width=84mm]{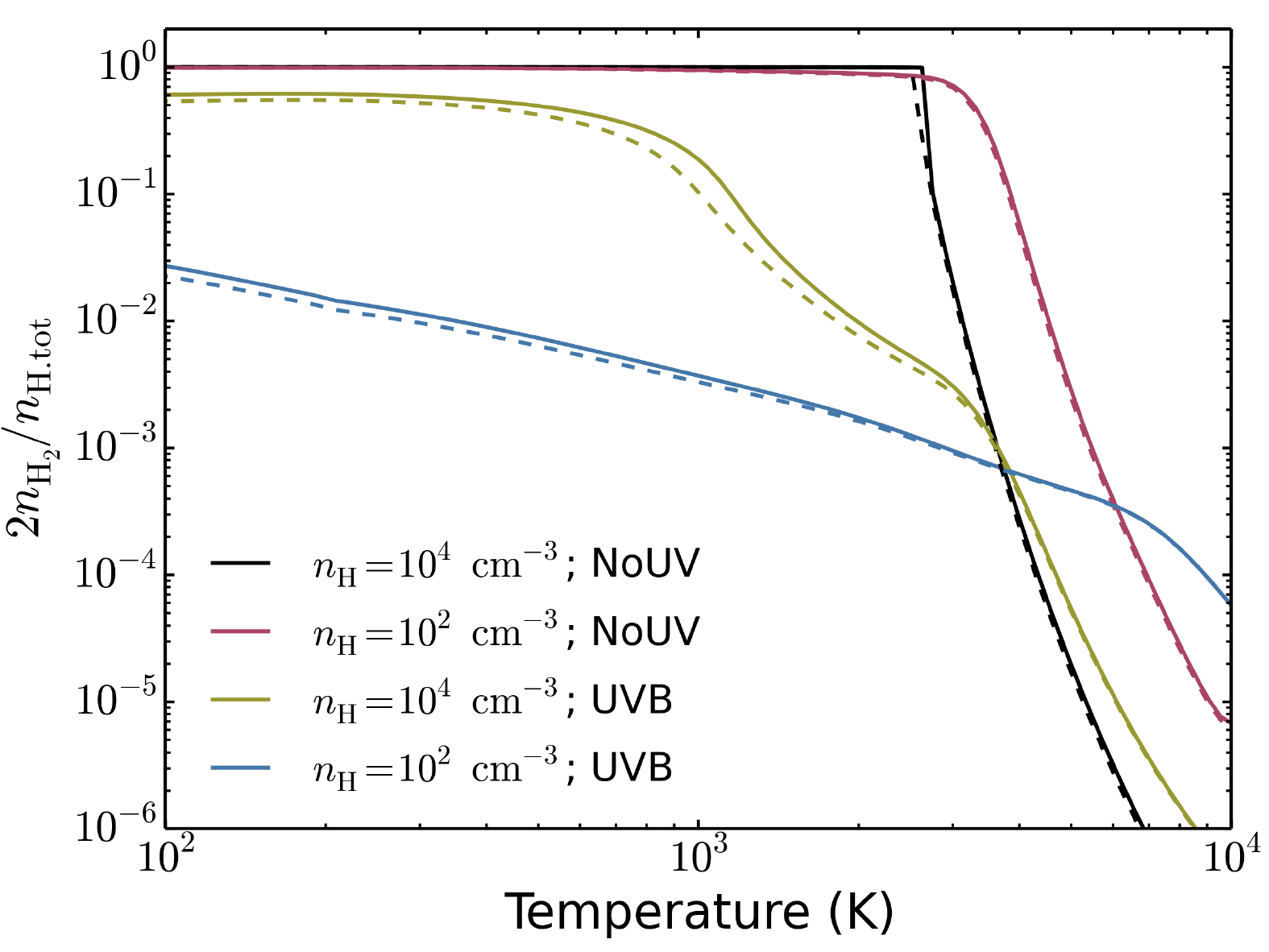}}
\caption{Equilibrium molecular hydrogen fractions as a function of gas temperature, calculated from our model using a constant dust temperature $T_{\rm{dust}} = 10$ K \textit{(solid lines)} and $T_{\rm{dust}} = 50$ K \textit{(dashed lines)}, either in the presence of the redshift zero \citet{haardt01} extragalactic UV background \textit{(UVB)} or in the absence of UV \textit{(NoUV)}.}
\label{H2_dust_effect_fig}
\end{figure}

To confirm the insensitivity to the dust temperature, we compare the molecular hydrogen fractions calculated using our model assuming different fixed dust temperatures. These are shown in figure~\ref{H2_dust_effect_fig}. We find that the change in molecular hydrogen fraction as we increase the dust temperature from 10 K to 50 K is indeed negligible. We therefore choose to fix the dust temperature at a constant 10 K in our model.

\newpage 

\section{Chemical reactions}\label{reactions_summary}

\begin{table}
\centering
\begin{minipage}{84mm}
\caption{Summary of primordial and molecular reactions. \label{reaction_table1}}
\centering
\begin{tabular}{cll}
\hline
Reaction No.\footnote{The reaction numbers match those used in the code.} & Reaction & Reference\footnote{1 - \citet{wishart79}; 2 - \citet{bruhns10}; 3 - \citet{ramaker76}; 4 - \citet{karpas79}; 5 - \citet{moseley70}; 6 - \citet{schneider94}; 7 - \citet{savin04}; 8 - \citet{trevisan02}; 9 - \citet{maclow86}; 10 - \citet{lepp83}; 11 - \citet{martin98}; 12 - \citet{janev87}; 13 - \citet{ferland92}; 14 - \citet{poulaert78}; 15 - \citet{hummer98}; 16 - \citet{aldrovandi73}; 17 - \citet{zygelman89}; 18 - \citet{kimura93}; 19 - \citet{dejong72}; 20 - \citet{shapiro87}; 21 - \citet{vandishoeck88}; 22 - \citet{yan98}; 23 - \citet{wilms00}; 24 - \citet{cazaux02}; 25 - \citet{weingartner01b}; 26 - \citet{williams98}; 27 - \citet{furlanetto10}; 28 - \citet{leteuff00}; 29 - \citet{dalgarno87}; 30 - \citet{schulz67}; 31 - \citet{barlow84}; 32 - \citet{dove87}; 33 - \citet{linder95}; 34 - \citet{kim75}; 35 - \citet{petuchowski89}; 36 - Data from \citet{mccall04}, fit by \citet{woodall07}; 37 - \citet{geppert05}; 38 - \citet{prasad80}; 39 - \citet{nelson99}; 40 - \citet{sidhu92}; 41 - \citet{gerlich92}; 42 - \citet{dalgarno90}; 43 - \citet{singh99}; 44 - \citet{smith02}; 45 - \citet{wagnerredeker85}; 46 - \citet{dean91}; 47 - \citet{harding93}; 48 - \citet{smith04}; 49 - \citet{baulch92}; 50 - \citet{murrell86}; 51 - \citet{warnatz84}; 52 - \citet{frank86}; 53 - Data from \citet{frank84} and \citet{frank88}, fit by \citet{leteuff00}; 54 - \citet{mitchell83}; 55 - \citet{fairbairn69}; 56 - \citet{glover10}; 57 - \citet{natarajan87}; 58 - \citet{tsang86}; 59 - \citet{oldenborg92}; 60 - \citet{carty06}; 61 - \citet{cohen79}; 62 - \citet{azatyan75}; 63 - \citet{adams84}; 64 - \citet{mcewan99}; 65 - \citet{viggiano80}; 66 - \citet{smith77a, smith77b}; 67 - \citet{fehsenfeld76}; 68 - \citet{smith78}; 69 - \citet{adams80}; 70 - \citet{milligan00}; 71 - \citet{dubernet92}; 72 - \citet{jones81}; 73 - \citet{raksit80}; 74 - \citet{kim74}; 75 - \citet{anicich76}; 76 - \citet{adams78}; 77 - \citet{smith92}; 78 - \citet{mauclaire78a, mauclaire78b}; 79 - \citet{adams76a, adams76b}; 80 - \citet{federer84}; 81 - \citet{peart94}; 82 - \citet{mitchell90}; 83 - \citet{larson98}; 84 - \citet{rosen00}; 85 - \citet{jensen00}; 86 - \citet{alge83}; 87 - \citet{rosen98}; 88 - \citet{stancil98}; 89 - \citet{andreazza97}; 90 - \citet{barinovs06}; 91 - \citet{herbst85}; 92 - \citet{field80}; 93 - \citet{orel87}; 94 - \citet{abel02}; 95 - \citet{cohen83}; 96 - \citet{walkauskas75}; 97 - Data from \citet{fairbairn69} and \citet{slack76}, fit by \citet{leteuff00}; 98 - \citet{macgregor73}; 99 - \citet{vandishoeck87}; 100 - \citet{vandishoeck06}; 101 - \citet{vandishoeck84}; 102 - \citet{lee84}; 103 - \citet{sternberg95}; 104 - \citet{visser09}; 105 - \citet{gredel89}; 106 - \citet{maloney96}; 107 - \citet{tielens85}; 108 - \citet{oppenheimer13a}; 109 - \citet{verner95}; 110 - \citet{verner96}; 111 - \citet{kaastra93}; 112 - \citet{lotz67}; 113 - \citet{langer78}; 114 - \citet{scott97}} \\ 
\hline
$1$ & H + e$^{-}$ $\rightarrow$ H$^{-} + \gamma$ & 1 \\
$2$ & H$^{-}$ + H $\rightarrow$ H$_{2}$ + e$^{-}$ & 2 \\
$3$ & H + H$^{+}$ $\rightarrow$ H$_{2}^{+}$ + $\gamma$ & 3 \\
$4$ & H + H$_{2}^{+}$ $\rightarrow$ H$_{2}$ + H$^{+}$ & 4 \\
$5$ & H$^{-}$ + H$^{+}$ $\rightarrow$ H + H & 5 \\
$6$ & H$_{2}^{+}$ + e$^{-}$ $\rightarrow$ H + H & 6 \\
$7$ & H$_{2}$ + H$^{+}$ $\rightarrow$ H$_{2}^{+}$ + H & 7 \\
\hline
\end{tabular}
\end{minipage}
\end{table}

\begin{table}
\centering
\begin{minipage}{84mm}
\contcaption{}
\centering
\begin{tabular}{cll}
\hline
$8$ & H$_{2}$ + e$^{-}$ $\rightarrow$ H + H + e$^{-}$ & 8  \\
$9$ & H$_{2}$ + H $\rightarrow$ H + H + H & 9, 10 \\
$10$ & H$_{2}$ + H$_{2}$ $\rightarrow$ H$_{2}$ + H + H & 11 \\
$11$ & H + e$^{-}$ $\rightarrow$ H$^{+}$ + e$^{-}$ + e$^{-}$ & 12 \\
$13$\footnote{We switch from the case A to case B recombination rates for hydrogen and helium when $\tau_{HI} > 1$ and $\tau_{HeI} > 1$ respectively.} & H$^{+}$ + e$^{-}$ $\rightarrow$ H + $\gamma$ & 13 \\
$15$ & H$^{-}$ + e$^{-}$ $\rightarrow$ H + e$^{-}$ + e$^{-}$ & 12 \\
$16$ & H$^{-}$ + H $\rightarrow$ H + H + e$^{-}$ & 12 \\
$17$ & H$^{-}$ + H$^{+}$ $\rightarrow$ H$_{2}^{+}$ + e$^{-}$ & 14 \\
$24$ & He + e$^{-}$ $\rightarrow$ He$^{+}$ + e$^{-}$ + e$^{-}$ & 12 \\
$25^{c}$ & He$^{+}$ + e$^{-}$ $\rightarrow$ He + $\gamma$ & 15, 16 \\
$26$ & He$^{+}$ + H $\rightarrow$ He + H$^{+}$ & 17 \\
$27$ & He + H$^{+}$ $\rightarrow$ He$^{+}$ + H & 18 \\
$51$ & H$^{-}$ + $\gamma$ $\rightarrow$ H + e$^{-}$ & 19, 20 \\
$52$ & H$_{2}^{+}$ + $\gamma$ $\rightarrow$ H + H$^{+}$ & 21 \\
$53$ & H$_{2}$ + $\gamma$ $\rightarrow$ H + H & This work \\
$54$ & H$_{2}$ + $\gamma$ $\rightarrow$ H$_{2}^{+}$ + e$^{-}$ & 22, 23 \\
$60$ & H + H $\overrightarrow{_{dust}}$ H$_{2}$ & 24 \\
$61$ & H$^{+}$ + e$^{-}$ $\overrightarrow{_{dust}}$ H & 25 \\
$63$ & He$^{+}$ + e$^{-}$ $\overrightarrow{_{dust}}$ He & 25 \\
$67$ & H + cr $\rightarrow$ H$^{+}$ + e$^{-}$ & 26, 27 \\
$69$ & He + cr $\rightarrow$ He$^{+}$ + e$^{-}$ & 27, 28 \\
$70$ & H$_{2}$ + cr $\rightarrow$ H$_{2}^{+}$ + e$^{-}$ & 28 \\
$83$ & H$_{2}^{+}$ + H$^{-}$ $\rightarrow$ H + H + H & 29 \\
$84$ & H$_{2}$ + e$^{-}$ $\rightarrow$ H$^{-}$ + H & 30 \\
$85$ & H$_{2}$ + He$^{+}$ $\rightarrow$ He + H + H$^{+}$ & 31 \\
$86$ & H$_{2}$ + He$^{+}$ $\rightarrow$ H$_{2}^{+}$ + He & 31 \\
$87$ & H$_{2}^{+}$ + H$^{-}$ $\rightarrow$ H$_{2}$ + H & 29 \\
$88$ & H$_{2}$ + He $\rightarrow$ H + H + He & 32 \\
$97$ & H$_{2}^{+}$ + H$_{2}$ $\rightarrow$ H$_{3}^{+}$ + H & 33 \\
$100$ & CO + H$_{3}^{+}$ $\rightarrow$ HCO$^{+}$ + H$_{2}$ & 34 \\
$101$ & He$^{+}$ + CO $\rightarrow$ C$^{+}$ + O + He & 35 \\
$106$ & H$_{3}^{+}$ + e$^{-}$ $\rightarrow$ H$_{2}$ + H & 36 \\
$107$ & HCO$^{+}$ + e$^{-}$ $\rightarrow$ CO + H & 37 \\
$108$ & H$_{3}^{+}$ + Fe $\rightarrow$ Fe$^{+}$ + H$_{2}$ + H & 38 \\
$111$ & HCO$^{+}$ + $\gamma$ $\rightarrow$ CO + H$^{+}$ & 39 \\
$112$ & H$_{3}^{+}$ + H $\rightarrow$ H$_{2}^{+}$ + H$_{2}$ & 40 \\
$113$ & CO + He$^{+}$ $\rightarrow$ C + O$^{+}$ + He & 35 \\
$114$ & H$_{3}^{+}$ + e$^{-}$ $\rightarrow$ H + H + H & 36 \\
$115$ & H$_{2}$ + H$^{+}$ $\rightarrow$ H$_{3}^{+}$ + $\gamma$ & 41 \\
$116$ & C + O $\rightarrow$ CO + $\gamma$ & 42, 43 \\
$117$ & OH + H $\rightarrow$ O + H + H & 28 \\
$118$ & HOC$^{+}$ + H$_{2}$ $\rightarrow$ HCO$^{+}$ + H$_{2}$ & 44 \\
$119$ & HOC$^{+}$ + CO $\rightarrow$ HCO$^{+}$ + CO & 45 \\
$120$ & C + H$_{2}$ $\rightarrow$ CH + H & 46 \\
$121$ & CH + H $\rightarrow$ C + H$_{2}$ & 47 \\
$122$ & CH + H$_{2}$ $\rightarrow$ CH$_{2}$ + H & 28 \\
$123$ & CH + C $\rightarrow$ C$_{2}$ + H & 48 \\
$124$ & CH + O $\rightarrow$ CO + H & 49, 50 \\
$125$ & CH$_{2}$ + H $\rightarrow$ CH + H$_{2}$ & 51 \\
$126$ & CH$_{2}$ + O $\rightarrow$ CO + H  + H & 52 \\
$127$ & CH$_{2}$ + O $\rightarrow$ CO + H$_{2}$ & 53 \\
$128$ & C$_{2}$ + O $\rightarrow$ CO + C & 54, 55, 56 \\
$129$ & O + H$_{2}$ $\rightarrow$ OH + H & 57 \\
$130$ & OH + H $\rightarrow$ O + H$_{2}$ & 58 \\
$131$ & OH + H$_{2}$ $\rightarrow$ H$_{2}$O + H & 59 \\
$132$ & OH + C $\rightarrow$ CO + H & 48 \\
$133$ & OH + O $\rightarrow$ O$_{2}$ + H & 28, 56, 60 \\
$134$ & OH + OH $\rightarrow$ H$_{2}$O + O & 48 \\
\hline
\end{tabular}
\end{minipage}
\end{table}

\begin{table}
\centering
\begin{minipage}{84mm}
\contcaption{}
\centering
\begin{tabular}{cll}
\hline
$135$ & H$_{2}$O + H $\rightarrow$ H$_{2}$ + OH & 61 \\
$136$ & O$_{2}$ + H $\rightarrow$ OH + O & 28 \\
$137$ & O$_{2}$ + H$_{2}$ $\rightarrow$ OH + OH & 62 \\
$138$ & O$_{2}$ + C $\rightarrow$ CO + O & 28, 48 \\
$139$ & CO + H $\rightarrow$ C + OH & 28 \\
$140$ & C + H$_{2}^{+}$ $\rightarrow$ CH$^{+}$ + H & 28 \\
$141$ & C + H$_{3}^{+}$ $\rightarrow$ CH$^{+}$ + H$_{2}$ & 28 \\
$142$ & C$^{+}$ + H$_{2}$ $\rightarrow$ CH$^{+}$ + H & 63 \\
$143$ & CH$^{+}$ + H $\rightarrow$  C$^{+}$ + H$_{2}$ & 64 \\
$144$ & CH$^{+}$ + H$_{2}$ $\rightarrow$  CH$_{2}^{+}$ + H & 64 \\
$145$ & CH$^{+}$ + O $\rightarrow$  CO$^{+}$ + H & 65 \\
$146$ & CH$_{2}$ + H$^{+}$ $\rightarrow$  CH$^{+}$ + H$_{2}$ & 28 \\
$147$ & CH$_{2}^{+}$ + H $\rightarrow$  CH$^{+}$ + H$_{2}$ & 28 \\
$148$ & CH$_{2}^{+}$ + H$_{2}$ $\rightarrow$  CH$_{3}^{+}$ + H & 66 \\
$149$ & CH$_{2}^{+}$ + O $\rightarrow$ HCO$^{+}$ + H & 28 \\
$150$ & CH$_{3}^{+}$ + H $\rightarrow$ CH$_{2}^{+}$ + H$_{2}$ & 28 \\
$151$ & CH$_{3}^{+}$ + O $\rightarrow$  HCO$^{+}$ + H$_{2}$ & 67 \\
$152$ & C$_{2}$ + O$^{+}$ $\rightarrow$ CO$^{+}$ + C & 28 \\
$153$ & O$^{+}$ + H$_{2}$ $\rightarrow$ OH$^{+}$ + H & 68, 69 \\
$154$ & O + H$_{2}^{+}$ $\rightarrow$ OH$^{+}$ + H & 28 \\
$155$ & O + H$_{3}^{+}$ $\rightarrow$ OH$^{+}$ + H$_{2}$ & 70 \\
$156$ & OH + H$_{3}^{+}$ $\rightarrow$ H$_{2}$O$^{+}$ + H$_{2}$ & 28 \\
$157$ & OH + C$^{+}$ $\rightarrow$ CO$^{+}$ + H & 71 \\
$283$ & OH + C$^{+}$ $\rightarrow$ CO + H$^{+}$ & 71 \\
$158$ & OH$^{+}$ + H$_{2}$ $\rightarrow$ H$_{2}$O$^{+}$ + H & 72 \\
$159$ & H$_{2}$O$^{+}$ + H$_{2}$ $\rightarrow$ H$_{3}$O$^{+}$ + H & 73 \\
$160$ & H$_{2}$O + H$_{3}^{+}$ $\rightarrow$ H$_{3}$O$^{+}$ + H$_{2}$ & 74 \\
$161$ & H$_{2}$O + C$^{+}$ $\rightarrow$ HCO$^{+}$ + H & 75 \\
$162$ & H$_{2}$O + C$^{+}$ $\rightarrow$ HOC$^{+}$ + H & 75 \\
$163$ & H$_{3}$O$^{+}$ + C $\rightarrow$ HCO$^{+}$ + H$_{2}$ & 28 \\
$164$ & O$_{2}$ + C$^{+}$ $\rightarrow$ CO$^{+}$ + O & 66 \\
$165$ & O$_{2}$ + C$^{+}$ $\rightarrow$ CO + O$^{+}$ & 66 \\
$166$ & O$_{2}$ + CH$_{2}^{+}$ $\rightarrow$ HCO$^{+}$ + OH & 66 \\
$167$ & O$_{2}^{+}$ + C $\rightarrow$ CO$^{+}$ + O & 28 \\
$168$ & CO + H$_{3}^{+}$ $\rightarrow$ HOC$^{+}$ + H$_{2}$ & 34 \\
$169$ & HCO$^{+}$ + C $\rightarrow$ CO + CH$^{+}$ & 28 \\
$170$ & HCO$^{+}$ + H$_{2}$O $\rightarrow$ CO + H$_{3}$O$^{+}$ & 76 \\
$171$ & CH + H$^{+}$ $\rightarrow$ CH$^{+}$ + H & 28 \\
$172$ & CH$_{2}$ + H$^{+}$ $\rightarrow$ CH$_{2}^{+}$ + H & 28 \\
$173$ & CH$_{2}$ + He$^{+}$ $\rightarrow$ C$^{+}$ + He + H$_{2}$ & 28 \\
$174$ & C$_{2}$ + He$^{+}$ $\rightarrow$ C$^{+}$ + C + He & 28 \\
$175$ & OH + H$^{+}$ $\rightarrow$ OH$^{+}$ + H & 28 \\
$176$ & OH + He$^{+}$ $\rightarrow$ O$^{+}$ + He + H & 28 \\
$177$ & H$_{2}$O + H$^{+}$ $\rightarrow$ H$_{2}$O$^{+}$ + H & 77 \\
$178$ & H$_{2}$O + He$^{+}$ $\rightarrow$ OH + He + H$^{+}$ & 78 \\
$179$ & H$_{2}$O + He$^{+}$ $\rightarrow$ OH$^{+}$ + He + H & 78 \\
$180$ & H$_{2}$O + He$^{+}$ $\rightarrow$ H$_{2}$O$^{+}$ + He & 78 \\
$181$ & O$_{2}$ + H$^{+}$ $\rightarrow$ O$_{2}^{+}$ + H & 77 \\
$182$ & O$_{2}$ + He$^{+}$ $\rightarrow$ O$_{2}^{+}$ + He & 79 \\
$183$ & O$_{2}$ + He$^{+}$ $\rightarrow$ O$^{+}$ + O + He & 79 \\
$184$ & O$_{2}^{+}$ + C $\rightarrow$ O$_{2}$ + C$^{+}$ & 28 \\
$185$ & CO$^{+}$ + H $\rightarrow$ CO + H$^{+}$ & 80 \\
$186$ & C$^{-}$ + H$^{+}$ $\rightarrow$ C + H & 28 \\
$187$ & O$^{-}$ + H$^{+}$ $\rightarrow$ O + H & 28 \\
$188$ & He$^{+}$ + H$^{-}$ $\rightarrow$ He + H & 81 \\
$189$ & CH$^{+}$ + e$^{-}$ $\rightarrow$ C + H & 82 \\
$190$ & CH$_{2}^{+}$ + e$^{-}$ $\rightarrow$ CH + H & 83 \\
$191$ & CH$_{2}^{+}$ + e$^{-}$ $\rightarrow$ C + H + H & 83 \\
$192$ & CH$_{2}^{+}$ + e$^{-}$ $\rightarrow$ C + H$_{2}$ & 83 \\
$193$ & CH$_{3}^{+}$ + e$^{-}$ $\rightarrow$ CH$_{2}$ + H & 82 \\
$194$ & CH$_{3}^{+}$ + e$^{-}$ $\rightarrow$ CH + H$_{2}$ & 82 \\
$195$ & CH$_{3}^{+}$ + e$^{-}$ $\rightarrow$ CH + H + H & 28 \\
\hline
\end{tabular}
\end{minipage}
\end{table}

\begin{table}
\centering
\begin{minipage}{84mm}
\contcaption{}
\centering
\begin{tabular}{cll}
\hline
$196$ & OH$^{+}$ + e$^{-}$ $\rightarrow$ O + H & 82 \\
$197$ & H$_{2}$O$^{+}$ + e$^{-}$ $\rightarrow$ O + H + H & 84 \\
$198$ & H$_{2}$O$^{+}$ + e$^{-}$ $\rightarrow$ O + H$_{2}$ & 84 \\
$199$ & H$_{2}$O$^{+}$ + e$^{-}$ $\rightarrow$ OH + H & 84 \\
$200$ & H$_{3}$O$^{+}$ + e$^{-}$ $\rightarrow$ H + H$_{2}$O & 85 \\
$201$ & H$_{3}$O$^{+}$ + e$^{-}$ $\rightarrow$ OH + H$_{2}$ & 85 \\
$202$ & H$_{3}$O$^{+}$ + e$^{-}$ $\rightarrow$ OH + H + H & 85 \\
$203$ & H$_{3}$O$^{+}$ + e$^{-}$ $\rightarrow$ O + H + H$_{2}$ & 85 \\
$204$ & O$_{2}^{+}$ + e$^{-}$ $\rightarrow$ O + O & 86 \\
$205$ & CO$^{+}$ + e$^{-}$ $\rightarrow$ C + O & 87 \\
$206$ & HCO$^{+}$ + e$^{-}$ $\rightarrow$ OH + C & 37 \\
$207$ & HOC$^{+}$ + e$^{-}$ $\rightarrow$ CO + H & 28 \\
$208$ & H$^{-}$ + C $\rightarrow$ CH + e$^{-}$ & 28 \\
$209$ & H$^{-}$ + O $\rightarrow$OH + e$^{-}$ & 28 \\
$210$ & H$^{-}$ + OH $\rightarrow$ H$_{2}$O + e$^{-}$ & 28 \\
$211$ & C$^{-}$ + H $\rightarrow$ CH + e$^{-}$ & 28 \\
$212$ & C$^{-}$ + H$_{2}$ $\rightarrow$ CH$_{2}$ + e$^{-}$ & 28 \\
$213$ & C$^{-}$ + O $\rightarrow$ CO + e$^{-}$ & 28 \\
$214$ & O$^{-}$ + H $\rightarrow$ OH + e$^{-}$ & 28 \\
$215$ & O$^{-}$ + H$_{2}$ $\rightarrow$ H$_{2}$O + e$^{-}$ & 28 \\
$216$ & O$^{-}$ + C $\rightarrow$ CO + e$^{-}$ & 28 \\
$217$ & C + e$^{-}$ $\rightarrow$ C$^{-}$ + $\gamma$ & 88 \\
$218$ & C + H $\rightarrow$ CH + $\gamma$ & 38 \\
$219$ & C + H$_{2}$ $\rightarrow$ CH$_{2}$ + $\gamma$ & 38 \\
$220$ & C + C $\rightarrow$ C$_{2}$ + $\gamma$ & 89 \\
$221$ & C$^{+}$ + H $\rightarrow$ CH$^{+}$ + $\gamma$ & 90 \\
$222$ & C$^{+}$ + H$_{2}$ $\rightarrow$ CH$_{2}^{+}$ + $\gamma$ & 91 \\
$223$ & C$^{+}$ + O $\rightarrow$ CO$^{+}$ + $\gamma$ & 42 \\
$224$ & O + e$^{-}$ $\rightarrow$ O$^{-}$ + $\gamma$ & 28 \\
$225$ & O + H $\rightarrow$ OH + $\gamma$ & 28 \\
$226$ & O + O $\rightarrow$ O$_{2}$ + $\gamma$ & 38 \\
$227$ & OH + H $\rightarrow$ H$_{2}$O + $\gamma$ & 92 \\
$228$ & H + H + H $\rightarrow$ H$_{2}$ + H & 93, 94 \\
$229$ & H + H + H$_{2}$ $\rightarrow$ H$_{2}$ + H$_{2}$ & 95 \\
$230$ & H + H + He $\rightarrow$ H$_{2}$ + He & 96 \\
$231$ & C + C + He $\rightarrow$ C$_{2}$ + He & 56, 97 \\
$232$ & C + O + He $\rightarrow$ CO + He & 35, 49 \\
$233$ & C$^{+}$ + O + He $\rightarrow$ CO$^{+}$ + He & 35 \\
$234$ & C + O$^{+}$ + He $\rightarrow$ CO$^{+}$ + He & 35 \\
$235$ & O + H + He $\rightarrow$ OH + He & 58 \\
$236$ & OH + H + He $\rightarrow$ H$_{2}$O + He & 49 \\
$237$ & O + O + He $\rightarrow$ O$_{2}$ + He & 51 \\
$238$ & O + CH $\rightarrow$ HCO$^{+}$ + e$^{-}$ & 98 \\
$239$ & H$_{3}^{+}$ + $\gamma$ $\rightarrow$ H$_{2}$ + H$^{+}$ & 99 \\
$240$ & H$_{3}^{+}$ + $\gamma$ $\rightarrow$ H$_{2}^{+}$ + H & 99 \\
$241$ & C$^{-}$ + $\gamma$ $\rightarrow$ C + e$^{-}$ & 28 \\
$242$ & CH + $\gamma$ $\rightarrow$ C + H & 100 \\
$243$ & CH + $\gamma$ $\rightarrow$ CH$^{+}$ + e$^{-}$ & 21 \\
$244$ & CH$^{+}$ + $\gamma$ $\rightarrow$ C + H$^{+}$ & 100 \\
$245$ & CH$_{2}$ + $\gamma$ $\rightarrow$ CH + H & 100 \\
$246$ & CH$_{2}$ + $\gamma$ $\rightarrow$ CH$_{2}^{+}$ + e$^{-}$ & 28 \\
$247$ & CH$_{2}^{+}$ + $\gamma$ $\rightarrow$ CH$^{+}$ + H & 100 \\
$248$ & CH$_{3}^{+}$ + $\gamma$ $\rightarrow$ CH$_{2}^{+}$ + H & 28 \\
$249$ & CH$_{3}^{+}$ + $\gamma$ $\rightarrow$ CH$^{+}$ + H$_{2}$ & 28 \\
$250$ & C$_{2}$ + $\gamma$ $\rightarrow$ C + C & 100 \\
$251$ & O$^{-}$ + $\gamma$ $\rightarrow$ O + e$^{-}$ & 28 \\
$252$ & OH + $\gamma$ $\rightarrow$ O + H & 101 \\
$253$ & OH + $\gamma$ $\rightarrow$ OH$^{+}$ + e$^{-}$ & 28 \\
$254$ & OH$^{+}$ + $\gamma$ $\rightarrow$ O + H$^{+}$ & 100 \\
$255$ & H$_{2}$O + $\gamma$ $\rightarrow$ OH + H & 102 \\
$256$ & H$_{2}$O + $\gamma$ $\rightarrow$ H$_{2}$O$^{+}$ + e$^{-}$ & 21 \\
$257$ & H$_{2}$O$^{+}$ + $\gamma$ $\rightarrow$ H$_{2}^{+}$ + O & 103 \\
\hline
\end{tabular}
\end{minipage}
\end{table}

\begin{table}
\centering
\begin{minipage}{84mm}
\contcaption{}
\centering
\begin{tabular}{cll}
\hline
$258$ & H$_{2}$O$^{+}$ + $\gamma$ $\rightarrow$ H$^{+}$ + OH & 103 \\
$259$ & H$_{2}$O$^{+}$ + $\gamma$ $\rightarrow$ O$^{+}$ + H$_{2}$ & 103 \\
$260$ & H$_{2}$O$^{+}$ + $\gamma$ $\rightarrow$ OH$^{+}$ + H & 103 \\
$261$ & H$_{3}$O$^{+}$ + $\gamma$ $\rightarrow$ H$^{+}$ + H$_{2}$O & 103 \\ 
$262$ & H$_{3}$O$^{+}$ + $\gamma$ $\rightarrow$ H$_{2}^{+}$ + OH & 103 \\
$263$ & H$_{3}$O$^{+}$ + $\gamma$ $\rightarrow$ H$_{2}$O$^{+}$ + H & 103 \\
$264$ & H$_{3}$O$^{+}$ + $\gamma$ $\rightarrow$ OH$^{+}$ + H$_{2}$ & 103 \\
$265$ & O$_{2}$ + $\gamma$ $\rightarrow$ O$_{2}^{+}$ + e$^{-}$ & 100 \\
$266$ & O$_{2}$ + $\gamma$ $\rightarrow$ O + O & 100 \\
$267$ & CO + $\gamma$ $\rightarrow$ C + O & 104 \\
$268$ & H$_{2}$ + cr $\rightarrow$ H$^{+}$ + H + e$^{-}$ & 28 \\
$269$ & H$_{2}$ + cr $\rightarrow$ H$^{+}$ + H$^{-}$ & 28 \\
$89$ & H$_{2}$ + cr $\rightarrow$ H + H & 28 \\
$270$ & CO + cr $\rightarrow$ CO$^{+}$ + e$^{-}$ & 28 \\
$272$ & CH + $\gamma_{cr}$ $\rightarrow$ C + H & 105 \\
$273$ & CH$^{+}$ + $\gamma_{cr}$ $\rightarrow$ C$^{+}$ + H & 105 \\
$274$ & CH$_{2}$ + $\gamma_{cr}$ $\rightarrow$ CH$_{2}^{+}$ + e$^{-}$ & 28 \\
$275$ & CH$_{2}$ + $\gamma_{cr}$ $\rightarrow$ CH + H & 28 \\
$276$ & C$_{2}$ + $\gamma_{cr}$ $\rightarrow$ C + C & 105 \\
$277$ & OH + $\gamma_{cr}$ $\rightarrow$ O + H & 105 \\
$278$ & H$_{2}$O + $\gamma_{cr}$ $\rightarrow$ OH + H & 105 \\
$279$ & O$_{2}$ + $\gamma_{cr}$ $\rightarrow$ O + O & 105 \\
$280$ & O$_{2}$ + $\gamma_{cr}$ $\rightarrow$ O$_{2}^{+}$ + e$^{-}$ & 105 \\
$281$ & CO + $\gamma_{cr}$ $\rightarrow$ C + O & 106 \\
$303$ & CO$^{+}$ + H$_{2}$ $\rightarrow$ HCO$^{+}$ + H & 114 \\
$304$ & CO$^{+}$ + H$_{2}$ $\rightarrow$ HOC$^{+}$ + H & 114 \\
$44$ & C$^{+}$ + Si $\rightarrow$ C + Si$^{+}$ & 38 \\
$79$ & Fe + C$^{+}$ $\rightarrow$ Fe$^{+}$ + C & 38 \\
$80$ & Fe + Si$^{+}$ $\rightarrow$ Fe$^{+}$ + Si & 107 \\
$294$ & C$^{+}$ + Mg $\rightarrow$ C + Mg$^{+}$ & 38 \\
$295$ & N$^{+}$ + Mg $\rightarrow$ N + Mg$^{+}$ & 38 \\
$296$ & Si$^{+}$ + Mg $\rightarrow$ Si + Mg$^{+}$ & 38 \\
$297$ & S$^{+}$ + Mg $\rightarrow$ S + Mg$^{+}$ & 38 \\
$64$ & C$^{+}$ + e$^{-}$ $\overrightarrow{_{dust}}$ C & 25 \\
$65$ & O$^{+}$ + e$^{-}$ $\overrightarrow{_{dust}}$ O & 25 \\
$66$ & Si$^{+}$ + e$^{-}$ $\overrightarrow{_{dust}}$ Si & 25 \\
$77$ & Fe$^{+}$ + e$^{-}$ $\overrightarrow{_{dust}}$ Fe & 25 \\
$290$ & Mg$^{+}$ + e$^{-}$ $\overrightarrow{_{dust}}$ Mg & 25 \\
$291$ & S$^{+}$ + e$^{-}$ $\overrightarrow{_{dust}}$ S & 25 \\
$292$ & Ca$^{+}$ + e$^{-}$ $\overrightarrow{_{dust}}$ Ca & 25 \\
$293$ & Ca$^{++}$ + e$^{-}$ $\overrightarrow{_{dust}}$ Ca$^{+}$ & 25 \\
$298$ & He$^{+}$ + e$^{-}$ $\rightarrow$ He$^{++}$ + e$^{-}$ + e$^{-}$ & 108 \\
$299$ & He$^{+}$ + $\gamma$ $\rightarrow$ He$^{++}$ + e$^{-}$ & 108 \\
$300$ & He$^{++}$ + e$^{-}$ $\rightarrow$ He$^{+}$ & 108 \\
$301$ & He$^{++}$ + H $\rightarrow$ He$^{+}$ + H$^{+}$ & 108 \\
$302$ & He$^{+}$ + cr $\rightarrow$ He$^{++}$ + e$^{-}$ & 112, 113 \\
CollisIon\footnote{Elements A include C, N, O, Ne, Mg, Si, S, Ca \& Fe.}\footnote{Ionisation states $i$ run from $0$ to $Z - 1$.} & A$^{+i}$ + e$^{-}$ $\rightarrow$ A$^{+i+1}$ + e$^{-}$ + e$^{-}$ & 108 \\
Recomb.\footnote{Radiative and dielectronic} & A$^{+i+1}$ + e$^{-}$ $\rightarrow$ A$^{+i}$ + $\gamma$ & 108 \\
PhotoIon & A$^{+i}$ + $\gamma$ $\rightarrow$ A$^{+i+1}$ + e$^{-}$ & 109, 110 \\
Auger & A$^{+i}$ + $\gamma$ $\rightarrow$ A$^{+i+n}$ + $n$e$^{-}$ & 109, 110, 11 \\
CTHion & A$^{+i}$ + H$^{+}$ $\rightarrow$ A$^{+i+1}$ + H & 108 \\
CTHrec & A$^{+i+1}$ + H $\rightarrow$ A$^{+i}$ + H$^{+}$ & 108 \\
CTHeion & A$^{+i}$ + He$^{+}$ $\rightarrow$ A$^{+i+1}$ + He & 108 \\
CTHerec & A$^{+i+1}$ + He $\rightarrow$ A$^{+i}$ + He$^{+}$ & 108 \\
CosmicRays & A$^{+i}$ + cr $\rightarrow$ A$^{+i+1}$ + e$^{-}$ & 28, 112, 113 \\
\hline
\end{tabular}
\end{minipage}
\end{table}

\label{lastpage}


\begin{thebibliography}{}
\bibitem[\protect\citeauthoryear{Abel et al.}{1997}]{abel97} Abel T., Anninos P., Zhang Y., Norman M. L., 1997, NewA, 2, 181
\bibitem[\protect\citeauthoryear{Abel et al.}{2002}]{abel02} Abel T., Bryan G. L., Norman M. L., 2002, Sci., 295, 93
\bibitem[\protect\citeauthoryear{Adams \& Smith}{1976a}]{adams76a} Adams N. G., Smith D., 1976a, Int. J. Mass Spectrom. Ion Phys., 21, 349
\bibitem[\protect\citeauthoryear{Adams \& Smith}{1976b}]{adams76b} Adams N. G., Smith D., 1976b, J. Phys. B, 9, 1439
\bibitem[\protect\citeauthoryear{Adams et al.}{1978}]{adams78} Adams N. G., Smith D., Grief D., 1978, Int. J. Mass Spectrom. Ion Phys., 26, 405
\bibitem[\protect\citeauthoryear{Adams et al.}{1980}]{adams80} Adams N. G., Smith D., Paulson J. F., 1980, J. Chem. Phys., 72, 288
\bibitem[\protect\citeauthoryear{Adams et al.}{1984}]{adams84} Adams N. G., Smith D., Millar T. J., 1984, MNRAS, 211, 857
\bibitem[\protect\citeauthoryear{Aldrovandi \& P\'{e}quignot}{1973}]{aldrovandi73} Aldrovandi S. M. V., P\'{e}quignot D., 1973, A\&A, 25, 137
\bibitem[\protect\citeauthoryear{Alge et al.}{1983}]{alge83} Alge E., Adams N. G., Smith D., 1983, J. Phys. B, 16, 1433
\bibitem[\protect\citeauthoryear{Andreazza \& Singh}{1997}]{andreazza97} Andreazza C. M., Singh P. D., 1997, MNRAS, 287, 287
\bibitem[\protect\citeauthoryear{Anicich et al.}{1976}]{anicich76} Anicich V. G., Huntress W. T., Futrell J. H., 1976, Chem. Phys. Lett., 40, 233
\bibitem[\protect\citeauthoryear{Azatyan et al.}{1975}]{azatyan75} Azatyan V. V., Aleksandrov E. N., Troshin A. F., 1975, Kinet. Cat., 16, 306
\bibitem[\protect\citeauthoryear{Bakes \& Tielens}{1994}]{bakes94} Bakes E. L. O., Tielens A. G. G. M., 1994, ApJ, 427, 822
\bibitem[\protect\citeauthoryear{Barinovs \& van Hemert}{2006}]{barinovs06} Barinovs G., van Hemert M. C., 2006, ApJ, 636, 923
\bibitem[\protect\citeauthoryear{Barlow}{1984}]{barlow84} Barlow S. G., 1984, PhD thesis, Univ. Colorado
\bibitem[\protect\citeauthoryear{Baulch et al.}{1992}]{baulch92} Baulch D. L. et al., 1992, J. Phys. Chem. Ref. Data, 21, 411
\bibitem[\protect\citeauthoryear{Bertone et al.}{2013}]{bertone13} Bertone S., Aguirre A., Schaye J., 2013, MNRAS, 430, 3292
\bibitem[\protect\citeauthoryear{Black \& Dalgarno}{1977}]{black77} Black J. H., Dalgarno A., 1977, ApJS, 34, 405
\bibitem[\protect\citeauthoryear{Black}{1987}]{black87} Black J. H., 1987, ASSL, 134, 731
\bibitem[\protect\citeauthoryear{Bruhns et al.}{2010}]{bruhns10} Bruhns H., Kreckel H., Miller K. A., Urbain X., Savin D. W., 2010, Phys. Rev. A, 82, 042708
\bibitem[\protect\citeauthoryear{Burton et al.}{1990}]{burton90} Burton M. G., Hollenbach D. J., Tielens A. G. G. M., 1990, ApJ, 365, 620
\bibitem[\protect\citeauthoryear{Carty et al.}{2006}]{carty06} Carty D., Goddard A., K\"{o}hler S. P. K., Sims I. R., Smith I. W. M., 2006, J. Phys. Chem. A, 110, 3101
\bibitem[\protect\citeauthoryear{Cazaux \& Tielens}{2002}]{cazaux02} Cazaux S., Tielens A. G. G. M., 2002, ApJ, 575, L29
\bibitem[\protect\citeauthoryear{Cen}{1992}]{cen92} Cen R., 1992, ApJS, 78, 341
\bibitem[\protect\citeauthoryear{Cohen \& Westberg}{1979}]{cohen79} Cohen N., Westberg K. R., 1979, J. Phys. Chem., 83, 46
\bibitem[\protect\citeauthoryear{Cohen \& Westberg}{1983}]{cohen83} Cohen N., Westberg K. R., 1983, J. Phys. Chem. Ref. Data, 12, 531
\bibitem[\protect\citeauthoryear{Cox \& Tucker}{1969}]{cox69} Cox D. P., Tucker W. H., 1969, ApJ, 157, 1157
\bibitem[\protect\citeauthoryear{Dalgarno \& Lepp}{1987}]{dalgarno87} Dalgarno A., Lepp S., 1987, in Vardya M. S., Tarafdar S. P., eds., Astrochemistry. Reidel, Dordrecht, p. 109
\bibitem[\protect\citeauthoryear{Dalgarno et al.}{1990}]{dalgarno90} Dalgarno A., Du M. L., You J. H., 1990, ApJ, 349, 675
\bibitem[\protect\citeauthoryear{Dalgarno et al.}{1999}]{dalgarno99} Dalgarno A., Yan M., Liu W., 1999, ApJS, 125, 237
\bibitem[\protect\citeauthoryear{Dean et al.}{1991}]{dean91} Dean A. J., Davidson D. F., Hanson R. K., 1991, J. Phys. Chem., 95, 183
\bibitem[\protect\citeauthoryear{de Jong}{1972}]{dejong72} de Jong T., 1972, A\&A, 20, 263
\bibitem[\protect\citeauthoryear{Dere et al}{1997}]{dere97} Dere K. P., Landi E., Mason H. E., Monsignori Fossi B. C., Young P. R., 1997, A\&AS, 125, 149
\bibitem[\protect\citeauthoryear{De Rijcke et al.}{2013}]{derijcke13} De Rijcke S., Schroyen J., Vandenbroucke B., Jachowicz N., Decroos J., Cloet-Osselaer A., Koleva M., 2013 MNRAS, 433, 3005
\bibitem[\protect\citeauthoryear{Dove et al.}{1987}]{dove87} Dove J. E., Rusk A. C. M., Cribb P. H., Martin P. G., 1987, ApJ, 318, 379
\bibitem[\protect\citeauthoryear{Draine}{1978}]{draine78} Draine B. T., 1978, ApJS, 36, 595
\bibitem[\protect\citeauthoryear{Draine et al.}{1983}]{draine83} Draine B. T., Roberge W. G., Dalgarno A., 1983, ApJ, 264, 485
\bibitem[\protect\citeauthoryear{Dubernet et al.}{1992}]{dubernet92} Dubernet M. L., Gargaud M., McCarroll R., 1992, A\&A, 259, 373
\bibitem[\protect\citeauthoryear{Dunseath et al.}{1997}]{dunseath97} Dunseath K. M., Fon W. C., Burke V. M., Reid R. H. G., Noble C. J., 1997, J. Phys. B., 30, 277 
\bibitem[\protect\citeauthoryear{Fairbairn}{1969}]{fairbairn69} Fairbairn A. R., 1969, Proc. R. Soc. A, 312, 207
\bibitem[\protect\citeauthoryear{Federer et al.}{1984}]{federer84} Federer W., Villinger H., Howorka F., Lindinger W., Tosi P., Bassi D., Ferguson E., 1984, Phys. Rev. Lett., 52, 2084
\bibitem[\protect\citeauthoryear{Fehsenfeld et al.}{1973}]{fehsenfeld73} Fehsenfeld F. C., Howard C. J., Ferguson E. E., 1973, J. Chem. Phys., 58, 5841
\bibitem[\protect\citeauthoryear{Fehsenfeld}{1976}]{fehsenfeld76} Fehsenfeld F. C., 1976, ApJ, 209, 638
\bibitem[\protect\citeauthoryear{Ferland et al.}{1992}]{ferland92} Ferland G. J., Peterson B. M., Horne K., Welsh W. F., Nahar S. N., 1992, ApJ, 387, 95
\bibitem[\protect\citeauthoryear{Ferland et al.}{1998}]{ferland98} Ferland G. J., Korista K.T., Verner D.A., Ferguson J.W., Kingdon J.B., Verner E.M., 1998, PASP, 110, 761
\bibitem[\protect\citeauthoryear{Ferland et al.}{2013}]{ferland13} Ferland G. J., Porter R. L., van Hoof P. A. M., Williams R. J. R., Abel N. P., Lykins M. L., Shaw G., Henney W. J., Stancil P. C., 2013, RMxAA, 49, 137
\bibitem[\protect\citeauthoryear{Field et al.}{1980}]{field80} Field D., Adams N. G., Smith D., 1980, MNRAS, 192, 1
\bibitem[\protect\citeauthoryear{Flower et al.}{2000}]{flower00} Flower D. R., Le Bourlot J., Pineau des For\^{e}ts G., Roueff E., 2000, MNRAS, 314, 753
\bibitem[\protect\citeauthoryear{Frank \& Just}{1984}]{frank84} Frank P., Just Th., 1984, Proc. Int. Symp. Shock Tubes Waves, 14, 706
\bibitem[\protect\citeauthoryear{Frank}{1986}]{frank86} Frank P., 1986, Proc. Int. Symp. Rarefied Gas Dyn., 2, 422
\bibitem[\protect\citeauthoryear{Frank et al.}{1988}]{frank88} Frank P., Bhaskaran K. A., Just Th., 1988, Symp. Int. Combust. Proc., 21, 885
\bibitem[\protect\citeauthoryear{Furlanetto \& Stoever}{2010}]{furlanetto10} Furlanetto S. R., Stoever S. J., 2010, MNRAS, 404, 1869
\bibitem[\protect\citeauthoryear{Galli \& Palla}{1998}]{galli98} Galli D., Palla F., 1998, A\&A, 335, 403
\bibitem[\protect\citeauthoryear{Geppert et al.}{2005}]{geppert05} Geppert W. D. et al., 2005, J. Phys. Conf. Ser., 4, 26
\bibitem[\protect\citeauthoryear{Gerlich \& Horning}{1992}]{gerlich92} Gerlich D., Horning S., 1992, Chem. Rev., 92, 1509
\bibitem[\protect\citeauthoryear{Glover et al.}{2006}]{glover06} Glover S. C. O., Savin D. W., Jappsen A.-K., 2006, ApJ, 640, 553
\bibitem[\protect\citeauthoryear{Glover \& Jappsen}{2007}]{glover07} Glover S. C. O., Jappsen A.-K., 2007, ApJ, 666, 1
\bibitem[\protect\citeauthoryear{Glover \& Abel}{2008}]{glover08} Glover S. C. O., Abel T., 2008, MNRAS, 388, 1627
\bibitem[\protect\citeauthoryear{Glover et al.}{2010}]{glover10} Glover S. C. O., Federrath C., Mac Low M.-M., Klessen R. S., 2010, MNRAS, 404, 2
\bibitem[\protect\citeauthoryear{Glover \& Clark}{2012}]{glover12} Glover S. C. O., Clark P. C., 2012, MNRAS, 421, 116
\bibitem[\protect\citeauthoryear{Gnat \& Sternberg}{2007}]{gnat07} Gnat O., Sternberg A., 2007, ApJS, 168, 213
\bibitem[\protect\citeauthoryear{Gnedin \& Hollon}{2012}]{gnedin12} Gnedin N. Y., Hollon N., 2012, ApJS, 202, 13
\bibitem[\protect\citeauthoryear{Goldsmith \& Langer}{1978}]{goldsmith78} Goldsmith P., Langer W. D., 1978, ApJ, 222, 881
\bibitem[\protect\citeauthoryear{Grassi et al.}{2011}]{grassi11} Grassi T., Krstic P., Merlin E., Buonomo U., Piovan L., Chiosi C., 2011, A\&A, 533, A123
\bibitem[\protect\citeauthoryear{Grassi et al.}{2012}]{grassi12} Grassi T., Bovino S., Gianturco F. A., Baiocchi P., Merlin E., 2012, MNRAS, 425, 1332
\bibitem[\protect\citeauthoryear{Grassi et al.}{2013}]{grassi13} Grassi T., Bovino S., Schleicher D. R. G., Prieto J., Seifried D., Simoncini E., Gianturco F. A., 2013, arxiv:1311.1070
\bibitem[\protect\citeauthoryear{Gredel et al.}{1989}]{gredel89} Gredel R., Lepp S., Dalgarno A., Herbst E., 1989, ApJ, 347, 289
\bibitem[\protect\citeauthoryear{Haardt \& Madau}{2001}]{haardt01} Haardt F., Madau P., 2001, in Neumann D. M., Tran J. T. V., eds, XXIst Moriond Astrophys. Meeting, Clusters of Galaxies and the High Redshift Universe Observed in X-rays Editions Frontieres, Paris, 64
\bibitem[\protect\citeauthoryear{Habing}{1968}]{habing68} Habing H. J., 1968, Bull. Astron. Inst. Netherlands, 19, 421
\bibitem[\protect\citeauthoryear{Harding et al.}{1993}]{harding93} Harding L. B., Guadagnini R., Schatz G. C., 1993, J. Phys. Chem., 97, 5472
\bibitem[\protect\citeauthoryear{Herbst}{1985}]{herbst85} Herbst E., 1985, ApJ, 291, 226
\bibitem[\protect\citeauthoryear{Hollenbach \& McKee}{1979}]{hollenbach79} Hollenbach D., McKee C. F., 1979, ApJS, 41, 555
\bibitem[\protect\citeauthoryear{Hummer \& Storey}{1998}]{hummer98} Hummer D. G., Storey P. J., 1998, MNRAS, 297, 1073
\bibitem[\protect\citeauthoryear{Indriolo et al.}{2007}]{indriolo07} Indriolo N., Geballe T. R., Oka T., McCall B. J., 2007, ApJ, 671, 1736
\bibitem[\protect\citeauthoryear{Janev et al.}{1987}]{janev87} Janev R. K., Langer W. D., Evans K., Post D. E., 1987, Elementary Processes in Hydrogen-Helium Plasmas (Berlin: Springer)
\bibitem[\protect\citeauthoryear{Jappsen et al.}{2007}]{jappsen07} Jappsen A.-K., Glover S. C. O., Klessen R. S., Mac Low M.-M., 2007, ApJ, 660, 1332
\bibitem[\protect\citeauthoryear{Jenkins}{2009}]{jenkins09} Jenkins E. B., 2009, ApJ, 700, 1299
\bibitem[\protect\citeauthoryear{Jensen et al.}{2000}]{jensen00} Jensen M. J., Bilodeau R. C., Safvan C. P., Seiersen K., Andersen L. H., Pedersen H. B., Heber O., 2000, ApJ, 543, 764
\bibitem[\protect\citeauthoryear{Jones et al.}{1981}]{jones81} Jones J. D. C., Birkinshaw K., Twiddy N. D., 1981, Chem. Phys. Lett., 77, 484
\bibitem[\protect\citeauthoryear{Kaastra \& Mewe}{1993}]{kaastra93} Kaastra J. S., Mewe R., 1993, A\&AS, 97, 443
\bibitem[\protect\citeauthoryear{Kafatos}{1973}]{kafatos73} Kafatos M., 1973, ApJ, 182, 433
\bibitem[\protect\citeauthoryear{Karpas et al.}{1979}]{karpas79} Karpas Z., Anicich V., Huntress W. T., 1979, J. Chem. Phys. 70, 2877
\bibitem[\protect\citeauthoryear{Kim et al.}{1974}]{kim74} Kim J. K., Theard L. P., Huntress W. T., 1974, Int. J. Mass Spectrom. Ion Phys., 15, 223
\bibitem[\protect\citeauthoryear{Kim et al.}{1975}]{kim75} Kim J. K., Theard L. P., Huntress W. T., 1975, Chem. Phys. Lett., 32, 610
\bibitem[\protect\citeauthoryear{Kimura et al.}{1993}]{kimura93} Kimura M., Lane N. F., Dalgarno A., Dixson R. G., 1993, ApJ, 405, 801
\bibitem[\protect\citeauthoryear{Krumholz et al.}{2011}]{krumholz11} Krumholz M. R., Leroy A. K., McKee, C. F., 2011, ApJ, 731, 25
\bibitem[\protect\citeauthoryear{Krumholz}{2012}]{krumholz12} Krumholz M. R., 2012, ApJ, 759, 9
\bibitem[\protect\citeauthoryear{Landi et al}{2013}]{landi13} Landi E., Young P. R., Dere K. P., Del Zanna G., Mason H. E., 2013, ApJS, 763, 86
\bibitem[\protect\citeauthoryear{Langer}{1978}]{langer78} Langer W. D., 1978, ApJ, 225, 860
\bibitem[\protect\citeauthoryear{Larson et al.}{1998}]{larson98} Larson \AA. et al., 1998, ApJ, 505, 459
\bibitem[\protect\citeauthoryear{Launay et al.}{1991}]{launay91} Launay J. M., Le Dourneuf M., Zeippen C. J., 1991, A\&A, 252, 842
\bibitem[\protect\citeauthoryear{Lee}{1984}]{lee84} Lee L. C., 1984, ApJ, 282, 172
\bibitem[\protect\citeauthoryear{Lepp \& Shull}{1983}]{lepp83} Lepp S., Shull J. M., 1983, ApJ, 270, 578
\bibitem[\protect\citeauthoryear{Lepp \& Shull}{1984}]{lepp84} Lepp S., Shull J. M., 1984, ApJ, 280, 465
\bibitem[\protect\citeauthoryear{Le Teuff et al.}{2000}]{leteuff00} Le Teuff Y. H., Millar T. J., Marwick A. J., 2000, A\&AS, 146, 157
\bibitem[\protect\citeauthoryear{Linder et al.}{1995}]{linder95} Linder F., Janev R. K., Botero J., 1995, in Janev R. J., ed., Atomic and Molecular Processes in Fusion Edge Plasmas. Plenum Press, New York, p. 397
\bibitem[\protect\citeauthoryear{Lotz}{1967}]{lotz67} Lotz W., 1967, ApJS, 14, 207
\bibitem[\protect\citeauthoryear{MacGregor \& Berry}{1973}]{macgregor73} MacGregor M., Berry R. S., 1973, J. Phys. B, 6, 181
\bibitem[\protect\citeauthoryear{Mac Low \& Shull}{1986}]{maclow86} Mac Low M.-M., Shull J. M., 1986, ApJ, 302, 585
\bibitem[\protect\citeauthoryear{Maio et al.}{2007}]{maio07} Maio U., Dolag K., Ciardi B., Tornatore L., 2007, MNRAS, 379, 963
\bibitem[\protect\citeauthoryear{Maloney et al.}{1996}]{maloney96} Maloney P. R., Hollenbach D. J., Tielens A. G. G. M., 1996, ApJ, 466, 561
\bibitem[\protect\citeauthoryear{Martin et al.}{1998}]{martin98} Martin P. G., Keogh W. J., Mandy M. E., 1998, ApJ, 499, 793
\bibitem[\protect\citeauthoryear{Martinez Jr. et al.}{2009}]{martinez09} Martinez O. Jr., Yang Z., Bettes N. B., Snow T. P., Bierbaum V. M., 2009, ApJ, 705, 172
\bibitem[\protect\citeauthoryear{Mathis et al.}{1977}]{mathis77} Mathis J. S., Rumpl W., Nordsieck K. H., 1977, ApJ, 217, 425
\bibitem[\protect\citeauthoryear{Mauclaire et al.}{1978a}]{mauclaire78a} Mauclaire G., Derai R., Marx R., 1978a, Int. J. Mass Spectrom. Ion Phys., 26, 284
\bibitem[\protect\citeauthoryear{Mauclaire et al.}{1978b}]{mauclaire78b} Mauclaire G., Derai R., Marx R., 1978b, Dyn. Mass Spectrom., 5, 139
\bibitem[\protect\citeauthoryear{McCall et al.}{2003}]{mccall03} McCall B. J. et al., 2003, Nature, 422, 500
\bibitem[\protect\citeauthoryear{McCall et al.}{2004}]{mccall04} McCall B. J. et al., 2004, Phys. Rev. A, 70, 052716
\bibitem[\protect\citeauthoryear{McEwan et al.}{1999}]{mcewan99} McEwan M. J., Scott G. B. I., Adams N. G., Babcock L. M., Terzieva R., Herbst E., 1999, ApJ, 513, 287
\bibitem[\protect\citeauthoryear{Milligan \& McEwan}{2000}]{milligan00} Milligan D. B., McEwan M. J., 2000, Chem. Phys. Lett., 319, 482
\bibitem[\protect\citeauthoryear{Mitchell \& Deveau}{1983}]{mitchell83} Mitchell G. F., Deveau T. J., 1983, ApJ, 266, 646
\bibitem[\protect\citeauthoryear{Mitchell}{1990}]{mitchell90} Mitchell J. B. A., 1990, Phys. Rep., 186, 215
\bibitem[\protect\citeauthoryear{Moseley et al.}{1970}]{moseley70} Moseley J., Aberth W., Peterson J. R., 1970, Phys. Rev. Lett., 24, 435
\bibitem[\protect\citeauthoryear{Murrell \& Rodriguez}{1986}]{murrell86} Murrell J. N., Rodriguez J. A., 1986, J. Mol. Struct. Theochem., 139, 267
\bibitem[\protect\citeauthoryear{Natarajan \& Roth}{1987}]{natarajan87} Natarajan K., Roth P., 1987, Combust. Flame, 70, 267
\bibitem[\protect\citeauthoryear{Nelson \& Langer}{1999}]{nelson99} Nelson R. P., Langer W. D., 1999, ApJ, 524, 923
\bibitem[\protect\citeauthoryear{Neufeld \& Kaufman}{1993}]{neufeld93} Neufeld D. A., Kaufman M. J., 1993, ApJ, 418, 263
\bibitem[\protect\citeauthoryear{Neufeld et al.}{1995}]{neufeld95} Neufeld D. A., Lepp S., Melnick G. J., 1995, ApJS, 100, 132
\bibitem[\protect\citeauthoryear{Oldenborg et al.}{1992}]{oldenborg92} Oldenborg R. C., Loge G. W., Harradine D. M., Winn K. R., 1992, J. Phys. Chem., 96, 8426
\bibitem[\protect\citeauthoryear{Omukai et al.}{2005}]{omukai05} Omukai K., Tsuribe T., Schneider R., Ferrara A., 2005, ApJ, 626, 627
\bibitem[\protect\citeauthoryear{Oppenheimer \& Schaye}{2013a}]{oppenheimer13a} Oppenheimer B. D., Schaye J., 2013a, MNRAS, 434, 1043
\bibitem[\protect\citeauthoryear{Oppenheimer \& Schaye}{2013b}]{oppenheimer13b} Oppenheimer B. D., Schaye J., 2013b, MNRAS, 434, 1063
\bibitem[\protect\citeauthoryear{Orel}{1987}]{orel87} Orel A. E., 1987, J. Chem. Phys., 87, 314
\bibitem[\protect\citeauthoryear{Peart \& Hayton}{1994}]{peart94} Peart B., Hayton D. A., 1994, J. Phys. B, 27, 2551
\bibitem[\protect\citeauthoryear{Peebles \& Dicke}{1968}]{peebles68} Peebles P. J. E., Dicke R. H., 1968, ApJ, 154, 891
\bibitem[\protect\citeauthoryear{P\'{e}quignot \& Aldrovandi}{1976}]{pequignot76} P\'{e}quignot D., Aldrovandi S. M. V., 1976, A\&A, 50, 141
\bibitem[\protect\citeauthoryear{Petuchowski et al.}{1989}]{petuchowski89} Petuchowski S. J., Dwek E., Allen J. E. Jr., Nuth J. A. III, 1989, ApJ, 342, 406
\bibitem[\protect\citeauthoryear{Poulaert et al.}{1978}]{poulaert78} Poulaert G., Brouillard F., Claeys W., McGowan J. W., van Wassenhove G., 1978, J. Phys. B., 11, L671
\bibitem[\protect\citeauthoryear{Prasad \& Huntress}{1980}]{prasad80} Prasad S. S., Huntress W. T. Jr., ApJS, 43, 1
\bibitem[\protect\citeauthoryear{Puy et al.}{1993}]{puy93} Puy D., Akecian G., Le Bourlot J., Leorat J., Pineau Des For\^{e}ts G., 1993, A\&A, 267, 337
\bibitem[\protect\citeauthoryear{Rahmati et al.}{2013}]{rahmati13} Rahmati A., Pawlik A. H., Rai\v{c}evi\`{c} M., Schaye J., 2013, MNRAS, 430, 2427
\bibitem[\protect\citeauthoryear{Raksit \& Warneck}{1980}]{raksit80} Raksit A. B., Warneck P., 1980, J. Chem. Soc. Faraday Trans., 76, 1084
\bibitem[\protect\citeauthoryear{Ramaker \& Peek}{1976}]{ramaker76} Ramaker D. E., Peek J. M., 1976, Phys. Rev. A, 13, 58
\bibitem[\protect\citeauthoryear{R\"{o}llig et al.}{2007}]{rollig07} R\"{o}llig M. et al. 2007, A\&A, 467, 187
\bibitem[\protect\citeauthoryear{Ros\'{e}n et al.}{1998}]{rosen98} Ros\'{e}n S. et al., 1998, Phys. Rev., 57, 4462
\bibitem[\protect\citeauthoryear{Ros\'{e}n et al.}{2000}]{rosen00} Ros\'{e}n S. et al., 2000, Faraday Discuss., 115, 295
\bibitem[\protect\citeauthoryear{Santoro \& Shull}{2006}]{santoro06} Santoro F., Shull, J. M., 2006, ApJ, 643, 26
\bibitem[\protect\citeauthoryear{Saslaw \& Zipoy}{1967}]{saslaw67} Saslaw W. C., Zipoy D., 1967, Nature, 216, 976
\bibitem[\protect\citeauthoryear{Savin et al.}{2004}]{savin04} Savin D. W., Krstic P. S., Haiman Z., Stancil P. C., 2004, ApJ, 606, L167 (erratum 607, L147)
\bibitem[\protect\citeauthoryear{Schaye et al.}{2010}]{schaye10} Schaye J., Dalla Vecchia C., Booth C. M., Wiersma R. P. C., Theuns T., Haas M. R., Bertone S., Duffy A. R., McCarthy I. G., van de Voort F., 2010, MNRAS, 402, 1536
\bibitem[\protect\citeauthoryear{Schmeltekopf et al.}{1967}]{schmeltekopf67} Schmeltekopf A. L., Fehsenfeld F. C., Ferguson E. E., 1967, ApJ, 148, L155
\bibitem[\protect\citeauthoryear{Schneider et al.}{1994}]{schneider94} Schneider I. F., Dulieu O., Giusti-Suzor A., Roueff E., 1994, ApJ, 424, 983 (erratum 486, 580)
\bibitem[\protect\citeauthoryear{Schulz \& Asundi}{1967}]{schulz67} Schulz G. J., Asundi R. K., 1967, Phys. Rev., 158, 25
\bibitem[\protect\citeauthoryear{Scott et al.}{1997}]{scott97} Scott G. B. I., Fairley D. A., Freeman C. G., McEwan M. J., Spanel P., Smith D., 1997, J. Chem. Phys., 106, 3982
\bibitem[\protect\citeauthoryear{Shapiro \& Kang}{1987}]{shapiro87} Shapiro P. R., Kang H., 1987, ApJ, 318, 32
\bibitem[\protect\citeauthoryear{Shaw et al.}{2005}]{shaw05} Shaw G., Ferland G. J., Abel N. P., Stancil P. C., van Hoof P. A. M., 2005, ApJ, 624, 794
\bibitem[\protect\citeauthoryear{Shull \& van Steenberg}{1985}]{shull85} Shull J. M., van Steenberg M. E., 1985, ApJ, 298, 268
\bibitem[\protect\citeauthoryear{Sidhu et al.}{1992}]{sidhu92} Sidhu K. S., Miller S., Tennyson J., 1992, A\&A, 255, 453
\bibitem[\protect\citeauthoryear{Silk}{1970}]{silk70} Silk J., 1970, ApL, 5, 283
\bibitem[\protect\citeauthoryear{Singh et al.}{1999}]{singh99} Singh P. D., Sanzovo G. C., Borin A. C., Ornellas F. R., 1999, MNRAS, 303, 235
\bibitem[\protect\citeauthoryear{Slack}{1976}]{slack76} Slack M. W., 1976, J. Chem. Phys., 64, 228
\bibitem[\protect\citeauthoryear{Smith et al.}{2008}]{smith08} Smith B., Sigurdsson S., Abel T., 2008, MNRAS, 385, 1443
\bibitem[\protect\citeauthoryear{Smith \& Adams}{1977a}]{smith77a} Smith D., Adams N. G., 1977a, Int. J. Mass Spectrom. Ion Phys., 23, 123
\bibitem[\protect\citeauthoryear{Smith \& Adams}{1977b}]{smith77b} Smith D., Adams N. G., 1977b, Chem. Phys. Lett., 47, 383
\bibitem[\protect\citeauthoryear{Smith et al.}{1978}]{smith78} Smith D., Adams N. G., Miller T. M., 1978, J. Chem. Phys., 69, 308
\bibitem[\protect\citeauthoryear{Smith et al.}{1992}]{smith92} Smith D., Spanel P., Mayhew C. A., 1992, Int. J. Mass Spectrom. Ion Proc., 117, 457
\bibitem[\protect\citeauthoryear{Smith et al.}{2004}]{smith04} Smith I. W. M., Herbst E., Chang Q., 2004, MNRAS, 350, 323
\bibitem[\protect\citeauthoryear{Smith et al.}{2002}]{smith02} Smith M. A., Schlemmer S., von Richthofen J., Gerlich D., 2002, ApJ, 578, L87
\bibitem[\protect\citeauthoryear{Spitzer}{1978}]{spitzer78} Spitzer L., 1978, Physical Processes in the Interstellar Medium (New York: Wiley)
\bibitem[\protect\citeauthoryear{Stancil \& Dalgarno}{1998}]{stancil98} Stancil P. C., Dalgarno A., 1998, Faraday Discuss., 109, 61
\bibitem[\protect\citeauthoryear{Sternberg \& Dalgarno}{1995}]{sternberg95} Sternberg A., Dalgarno A., 1995, ApJS, 99, 565
\bibitem[\protect\citeauthoryear{Suchkov et al.}{1993}]{suchkov93} Suchkov A., Allen R. J., Heckman T. M., 1993, ApJ, 413, 542
\bibitem[\protect\citeauthoryear{Sutherland \& Dopita}{1993}]{sutherland93} Sutherland R. S., Dopita M. A., 1993, ApJS, 88, 253
\bibitem[\protect\citeauthoryear{Tielens \& Hollenbach}{1985}]{tielens85} Tielens A. G. G. M., Hollenbach D., 1985, ApJ, 291, 722
\bibitem[\protect\citeauthoryear{Trevisan \& Tennyson}{2002}]{trevisan02} Trevisan C. S., Tennyson J., 2002, Plasma Phys. Control. Fusion, 44, 1263
\bibitem[\protect\citeauthoryear{Tsang \& Hampson}{1986}]{tsang86} Tsang W., Hampson R. F., 1986, J. Phys. Chem. Ref. Data, 15, 1087
\bibitem[\protect\citeauthoryear{Tupper}{2002}]{tupper02} Tupper P. F., 2002, Bit Numerical Mathematics, 42, 447
\bibitem[\protect\citeauthoryear{van Dishoeck \& Dalgarno}{1984}]{vandishoeck84} van Dishoeck E. F., Dalgarno A., 1984, ApJ, 277, 576
\bibitem[\protect\citeauthoryear{van Dishoeck}{1987}]{vandishoeck87} van Dishoeck E. F., 1987, in Vardya M. S., Tarafdar S. P., eds., IAU Symp. Vol. 120, Astrochemistry. Reidel, Dordrecht, p. 51
\bibitem[\protect\citeauthoryear{van Dishoeck}{1988}]{vandishoeck88} van Dishoeck E. F., 1988, in Millar T. J., Williams D. A., eds., Rate Coefficients in Astrochemistry. Kluwer, Dordrecht, p. 49
\bibitem[\protect\citeauthoryear{van Dishoeck et al.}{2006}]{vandishoeck06} van Dishoeck E. F., Jonkheid B., van Hemert M. C., 2006, Faraday Discuss., 133, 231
\bibitem[\protect\citeauthoryear{Vasiliev}{2013}]{vasiliev13} Vasiliev E. O., 2013, MNRAS, 431, 638
\bibitem[\protect\citeauthoryear{Verner \& Yakovlev}{1995}]{verner95} Verner D. A., Yakovlev D. G., 1995, A\&AS, 109, 125
\bibitem[\protect\citeauthoryear{Verner et al.}{1996}]{verner96} Verner D. A., Ferland G. J., Korista K. T., Yakovlev D. G., 1996, ApJ, 465, 487
\bibitem[\protect\citeauthoryear{Viggiano et al.}{1980}]{viggiano80} Viggiano A. A., Howorka F., Albritton D. L., Fehsenfeld F. C., Adams N. G., Smith D., 1980, ApJ, 236, 492
\bibitem[\protect\citeauthoryear{Visser et al.}{2009}]{visser09} Visser R., van Dishoeck E. F., Black J. H., 2009, A\&A, 503, 323
\bibitem[\protect\citeauthoryear{Vogelsberger et al.}{2013}]{vogelsberger13} Vogelsberger M., Genel S., Sijacki D., Torrey P., Springel V., Hernquist L., 2013, MNRAS, 436, 3031
\bibitem[\protect\citeauthoryear{Wagner-Redeker et al.}{1985}]{wagnerredeker85} Wagner-Redeker W., Kemper P. R., Jarrold M. F., Bowers M. T., 1985, J. Chem. Phys., 83, 1121
\bibitem[\protect\citeauthoryear{Walkauskas \& Kaufman}{1975}]{walkauskas75} Walkauskas L. P., Kaufman F., 1975, Symp. Int. Combust. Proc., 15, 691
\bibitem[\protect\citeauthoryear{Warnatz}{1984}]{warnatz84} Warnatz J., 1984, in Gardiner W. C. Jr., ed., Combustion Chemistry. Springer-Verlag, NY, p. 197
\bibitem[\protect\citeauthoryear{Weingartner \& Draine}{2001a}]{weingartner01a} Weingartner J. C., Draine B. T., 2001a, ApJ, 548, 296
\bibitem[\protect\citeauthoryear{Weingartner \& Draine}{2001b}]{weingartner01b} Weingartner J. C., Draine B. T., 2001b, ApJ, 563, 842
\bibitem[\protect\citeauthoryear{Wiebe et al.}{2003}]{wiebe03} Wiebe D., Semenov D., Henning T., 2003, A\&A, 399, 197
\bibitem[\protect\citeauthoryear{Wiersma et al.}{2009}]{wiersma09} Wiersma R. P. C., Schaye J., Smith B. D., 2009, MNRAS, 393, 99
\bibitem[\protect\citeauthoryear{Williams et al.}{1998}]{williams98} Williams J. P., Bergin E. A., Caselli P., Myers P. C., Plume, R., 1998, ApJ, 503, 689
\bibitem[\protect\citeauthoryear{Wilms et al.}{2000}]{wilms00} Wilms J., Allen A., McCray R., 2000, ApJ, 542, 914
\bibitem[\protect\citeauthoryear{Wishart}{1979}]{wishart79} Wishart A. W., 1979, MNRAS, 187, 59\textsc{p}
\bibitem[\protect\citeauthoryear{Wolfire et al.}{1995}]{wolfire95} Wolfire M. G., Hollenbach D., McKee C. F., Tielens A. G. G. M., Bakes E. L. O., 1995, ApJ, 443, 152
\bibitem[\protect\citeauthoryear{Wolfire et al.}{2003}]{wolfire03} Wolfire M. G., McKee C. F., Hollenbach D., Tielens A. G. G. M., 2003, ApJ, 587, 278
\bibitem[\protect\citeauthoryear{Wolniewicz et al.}{1998}]{wolniewicz98} Wolniewicz L., Simbotin I., Dalgarno A., 1998, ApJS, 115, 293
\bibitem[\protect\citeauthoryear{Woodall et al.}{2007}]{woodall07} Woodall J., Ag\'{u}ndez M., Markwick-Kemper A. J., Millar T. J., 2007, A\&A, 466, 1197
\bibitem[\protect\citeauthoryear{Yan et al.}{1998}]{yan98} Yan M., Sadeghpour H. R., Dalgarno A., 1998, ApJ, 496, 1044
\bibitem[\protect\citeauthoryear{Zygelman et al.}{1989}]{zygelman89} Zygelman B., Dalgarno A., Kimura M., Lane N. F., 1989, Phys. Rev. A, 40, 2340

\end{thebibliography}
\end{document}